\documentclass[a4paper,11pt]{article}
\pdfoutput=1

\usepackage{jheppub}
\usepackage[T1]{fontenc}
\usepackage{amsmath,amssymb,mathtools,bm}
\usepackage{graphicx,subcaption,booktabs,microtype}
\usepackage[numbers,sort&compress]{natbib}
\usepackage{tikz}
\usepackage{hyperref}
\hypersetup{colorlinks=true,allcolors=blue}
\usepackage{mathrsfs}
\newcommand{\dd}{\mathrm{d}}
\newcommand{\ii}{\mathrm{i}}
\newcommand{\ee}{\mathrm{e}}
\newcommand{\cO}{\mathcal{O}}

\newcommand{\NN}{\mathbb{N}_0}
\newcommand{\tr}{\operatorname{Tr}}

\newcommand{\bx}{\mathbf{x}}
\newcommand{\bk}{\mathbf{k}}

\newcommand{\eg}{{\it e.g.,}\ }
\newcommand{\ie}{{\it i.e.,}\ }

\newcommand{\mt}[1]{\textrm{\tiny #1}}
\newcommand{\GN}{G_\mt{N}}

\newcommand{\TFD}{\mathrm{TFD}}
\subheader{\today}

\author[a,b]{Mingxuan Liu}
\author[a,b]{Shan-Ming Ruan}
\affiliation[a]{School of Physics, Peking University, Beijing 100871, China}
\affiliation[b]{Center of High Energy Physics, Peking University, Beijing 100871, China}

\emailAdd{mingxuanliu26@stu.pku.edu.cn}
\emailAdd{ruanshanming@pku.edu.cn}

\title{Smoothing the Bouncing-Geodesic Singularity in AdS/CFT}

\abstract{Black hole singularities remain a central open problem in quantum gravity. In the AdS/CFT correspondence, bulk spacelike extremal surfaces can probe the black hole interior, and information about the near-singularity geometry is encoded in analytically continued boundary correlators. As the bouncing geodesic approaches the black hole singularity, it becomes nearly null and gives rise to characteristic divergences at complex times. Importantly, these complex-time singularities emerge only in the strict semiclassical limit ($\GN\to 0$ or $c\to\infty$) and are governed by the asymptotically linear quasinormal mode (QNM) spectrum. At finite $c$, nonlinear deformations of the QNM trajectories produce suppression kernel that smooths out the complex-time singularities. Using Liouville CFT as an analytically solvable example, we explicitly demonstrate the finite-$c$ deformation of the QNM spectrum and its associated pole-zero structure. We also identify a complementary Gaussian smoothing mechanism arising from finite-energy fluctuations in the thermal spectral decomposition. These results show how finite $\GN$ quantum corrections modify the holographic imprints of the classical black hole singularity.}

\numberwithin{equation}{section}

\begin{document}
%\preprint{Report-number}
\maketitle
\flushbottom

%%%%%%%%%%%%%%%%%%%%%%%%%%%%%%%%%%%%%%%%%%%%%%%%%%
\section{Introduction}\label{sec:introduction}
%%%%%%%%%%%%%%%%%%%%%%%%%%%%%%%%%%%%%%%%%%%%%%%%%%

Spacetime singularities constitute one of the sharpest indications of the limitations of classical general relativity.  The singularity theorems of Penrose, Hawking, and others establish, under broad assumptions on the causal structure and matter content, that gravitational collapse or cosmological evolution can lead to geodesic incompleteness  \cite{Penrose:1965Singularity,HawkingPenrose:1970}. Importantly, these theorems do not determine the local structure of the near-singularity region itself. A complementary picture is provided by the Belinski--Khalatnikov--Lifshitz (BKL) conjecture, according to which the approach to a generic spacelike singularity is expected to become ultralocal \footnote{The term ``Ultralocal'' means that, to leading order, the evolution at each spatial point is governed by ordinary diﬀerential equations in time, with spatial-gradient terms subleading.} and to proceed through a sequence of approximately Kasner epochs, with transitions that can exhibit oscillatory and chaotic behavior \cite{Belinsky:1970ew,Belinsky:1982pk,Damour:2002et,Belinski:2017awb}. Although a complete characterization of generic spacetime singularities remains an open problem, these developments make clear that singularities are not merely artifacts of highly symmetric solutions, but are deeply rooted in the nonlinear dynamics of general relativity.

At the same time, a curvature singularity marks precisely a regime where the classical gravitational description cannot be expected to remain complete. Curvature invariants and tidal forces can become arbitrarily large, and eventually probe scales at which quantum gravitational effects can no longer be neglected. Understanding whether and how the classical singularity is modified in quantum gravity is therefore a fundamental problem. A complete theory of quantum gravity is expected to resolve the breakdown of classical general relativity associated with spacetime singularities, by replacing the divergent classical description with a well-defined quantum one. For a black hole, the additional challenge is that this quantum region lies behind an event horizon and is not directly accessible to asymptotic observers. One would nevertheless expect the quantum gravity theory to encode the black-hole interior, together with the quantum resolution of its classical singular behavior, in observables accessible from the asymptotic boundary.

\begin{figure}[t]
    \centering
    \includegraphics[width=0.42\linewidth]{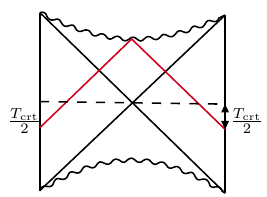}
    \caption{\textit{Bouncing geodesics in the two-sided AdS black hole.} As the conserved geodesic energy is increased, the turning point of a real radial spacelike geodesic moves toward the curvature singularity. In the limiting configuration the geodesic becomes almost null and approaches a trajectory that bounces off the singularity. In terms of the two-sided boundary time, the corresponding critical Lorentzian time is denoted by $T_{\rm crt}$.}
    \label{fig:bounce-geodesics}
\end{figure}

The AdS/CFT correspondence provides a concrete framework in which this question can be addressed. An asymptotically AdS black hole is described by a non-gravitational quantum field theory on the asymptotic boundary, so information about the bulk spacetime geometry, including regions behind the horizon, must ultimately be encoded in boundary observables. A particularly sharp realization of this idea was first developed by Fidkowski, Hubeny, Kleban, and Shenker (FHKS) in \cite{Fidkowski:2003nf}, where the black hole singularity was probed through thermal two-point correlation functions.

For a scalar primary operator $\cO$ with large conformal dimension $\Delta\gg 1$, the holographic dictionary identifies the boundary two-point function with the asymptotic limit of the corresponding bulk propagator. In the large-mass, or equivalently large-$\Delta$, limit, the bulk propagator admits a Wentzel–Kramers–Brillouin (WKB) expansion whose leading saddle points are classical spacelike geodesics. The resulting geodesic approximation therefore gives \cite{Balasubramanian:1999zv,Louko:2000tp,Aparicio:2011zy,Balasubramanian:2012tu}  
\begin{equation}
G_{\mt{LR}}(t_{\mt{L}}, t_{\mt{R}}) \simeq \sum_{\gamma}A_{\gamma} \, \ee^{-\Delta\mathcal{L}_{\gamma}(t_{\mt{L}}, t_{\mt{R}})} \,, 
\label{eq:intro-geodesic-approximation}
\end{equation}
where $\mathcal{L}_{\gamma}$ denotes the renormalized length of the geodesic $\gamma$ joining the left and right boundary insertion points at times $t_{\mt{L}}$ and $t_{\mt{R}}$, respectively, and $A_{\gamma}$ contains the corresponding fluctuation prefactor. In higher-dimensional AdS-Schwarzschild black holes, with bulk spacetime dimension $D=d+1 \ge 4$, the family of real symmetric spacelike geodesics exhibits a characteristic limiting behavior. As the associated conserved energy is increased, the turning point of the geodesic moves progressively deeper into the black-hole interior and approaches the spacelike singularity behind the horizon.  In this limit, the geodesic becomes almost null, except within a shrinking neighborhood of its turning point.  The limiting trajectory may therefore be regarded as a null ray that travels from one asymptotic boundary, effectively bounces off the singularity, and returns to the opposite boundary.  It is thus referred to as the \emph{bouncing geodesic}, as illustrated in figure \ref{fig:bounce-geodesics}. With the boundary time evolution, the limiting bouncing configuration is reached at a finite critical boundary time denoted as $T_{\rm crt}=t_{\mt{L,crt}}+t_{\mt{R,crt}}$. Near this critical time, the corresponding renormalized geodesic length develops a logarithmic divergence, namely 
\begin{equation}
\mathcal{L}_{\rm bounce}(T) \sim 2\log\!\left(T-T_{\rm crt}\right), \qquad T\rightarrow T_{\rm crt}\,.
\label{eq:intro-bounce-length}
\end{equation}
If this geodesic were to determine the semiclassical correlator, the geodesic approximation would consequently lead to 
\begin{equation}
G_{\mt{LR}}^{\rm bounce}(T) \sim \frac{1}{(T-T_{\rm crt})^{2\Delta}} \,, 
\label{eq:intro-bounce-divergence}
\end{equation}
producing a lightcone-like singularity at a finite boundary time. Thus, within the geodesic approximation, the black hole singularity behind the horizon would leave a direct singular imprint on the holographic boundary correlator.

As stressed in \cite{Fidkowski:2003nf}, this conclusion, however, immediately leads to a puzzle. The exact two-sided thermal correlator admits the spectral representation
\begin{equation}
G_{\mt{LR}}(T) = \frac{1}{Z(\beta)} \sum_{m,n} \ee^{-\frac{\beta}{2}(E_m+E_n)} \ee^{-\ii(E_n-E_m)T} \left|\cO_{mn}\right|^2 \,. 
\label{eq:intro-g12-spectral}
\end{equation}
For real Lorentzian time, the oscillating phase factor in the spectral representation implies $\left|G_{\mt{LR}}(T)\right| \leq G_{\mt{LR}}(0)$. Provided that the initial two-sided correlator $G_{\mt{LR}}(0)$ is finite, the physical Lorentzian correlator is therefore bounded \footnote{Recent progress over the past decade has shown that the late-time dynamics of holographic systems is governed by universal random-matrix statistics for quantum-chaotic systems. Suitable spectral diagnostics, such as the spectral form factor, as well as correlation functions of simple operators, exhibit characteristic signatures of this universality. For example, the Gaussian unitary ensemble displays the known slope--ramp--plateau structure, in which an initial decay is followed by a universal ramp generated by long-range spectral correlations and, eventually, by a late-time plateau associated with the discreteness and finiteness of the spectrum \cite{Cotler:2016fpe,Saad:2018bqo,Saad:2019lba}.} for all real Lorentzian time and cannot develop the additional divergence suggested by eq.~\eqref{eq:intro-bounce-divergence}. The naive real bouncing geodesic therefore cannot be the saddle that determines the physical Lorentzian correlator. This apparent paradox was resolved in \cite{Fidkowski:2003nf} by recognizing that the geodesic saddle-point problem is intrinsically multi-valued.  In the AdS$_5$ Schwarzschild example, three geodesic branches coalesce at a branch point in the complexified boundary time. The physical Lorentzian correlator is not determined by the real bouncing-geodesic branch that approaches the black-hole singularity, but instead by a pair of complex-conjugate geodesics selected by continuation from the Euclidean correlator. Schematically, the physical Lorentzian correlator is therefore dominated by 
\footnote{The appearance and relative contributions of the complex saddles in eq.~\eqref{eq:physical-complex-geodesics} can be analyzed more systematically using Picard--Lefschetz theory. In a forthcoming companion work \cite{nextpaper2026}, we apply this framework to a WKB formulation of the path integral for correlation functions, in which the relevant Lefschetz thimbles and their intersection numbers determine which complex saddles, \ie extremal surfaces or geodesics, contribute to the path integral.}
\begin{equation}
G_{\mt{LR}}^{\rm phys}(T) \simeq A_+\,\ee^{-\Delta\mathcal{L}_+(T)} + A_-\,\ee^{-\Delta\mathcal{L}_-(T)},
\qquad \mathcal{L}_-(T)=\mathcal{L}_+^*(T) \,, 
\label{eq:physical-complex-geodesics}
\end{equation}
within the geodesic approximation. These complex saddles remain regular at the bouncing time $T=T_{\rm crt}$, thereby resolving the apparent contradiction with the finiteness of the physical Lorentzian correlator. The real bouncing branch is nevertheless not lost, but remains encoded in the multi-valued analytic structure of the correlator.  By analytically continuing the physical branch further around the branch point, one reaches a secondary sheet containing the real bouncing geodesic, where the time singularity at $T=T_{\rm crt}$ reappears. Thus, information about the black hole singularity remains encoded in the boundary theory, even though the physical Lorentzian correlator itself remains finite.

The basic picture initiated in \cite{Fidkowski:2003nf} has subsequently been refined through complementary analyses of bulk propagation and boundary operator expansions. A direct boundary description was also developed in \cite{Ceplak:2024bja} using the thermal operator product expansion (OPE). The contributions from the identity, the stress tensor, and its composites develop singularities at the locations associated with bouncing geodesics, and their resummation in the large-$\Delta$ limit reproduces the branch structure obtained from the bulk geodesic analysis. Related bouncing singularities also appear in thermal correlators of displacement operators on Wilson lines \cite{Giombi:2026kdz}. These results demonstrate that this diagnostic is not restricted to heavy local scalar operators. A complementary geometric perspective was developed in \cite{Grozdanov:2026cut}. Using Hadamard theory and the propagation of singularities, this work established a connection between null connectivity and singularities of bulk and boundary retarded propagators beyond the strict large-$\Delta$ approximation. Other observables, such as thermal one-point functions whose large-mass dependence encodes the proper time to the singularity, provide new probes with different analytic structures \cite{Grinberg:2020fdj}. The analytic continuations used to select interior trajectories have also been investigated directly within the WKB framework \cite{Afkhami-Jeddi:2025wra, AliAhmad:2026wem,Chakravarty:2026fdm}. For rotating and charged black holes, suitable analytic continuations of thermal correlators can also probe timelike geodesics in the interior of the maximally extended geometry \cite{AliAhmad:2026wem,Chakravarty:2026fdm}.

\begin{figure}[t]
\centering
\includegraphics[width=5.9in]{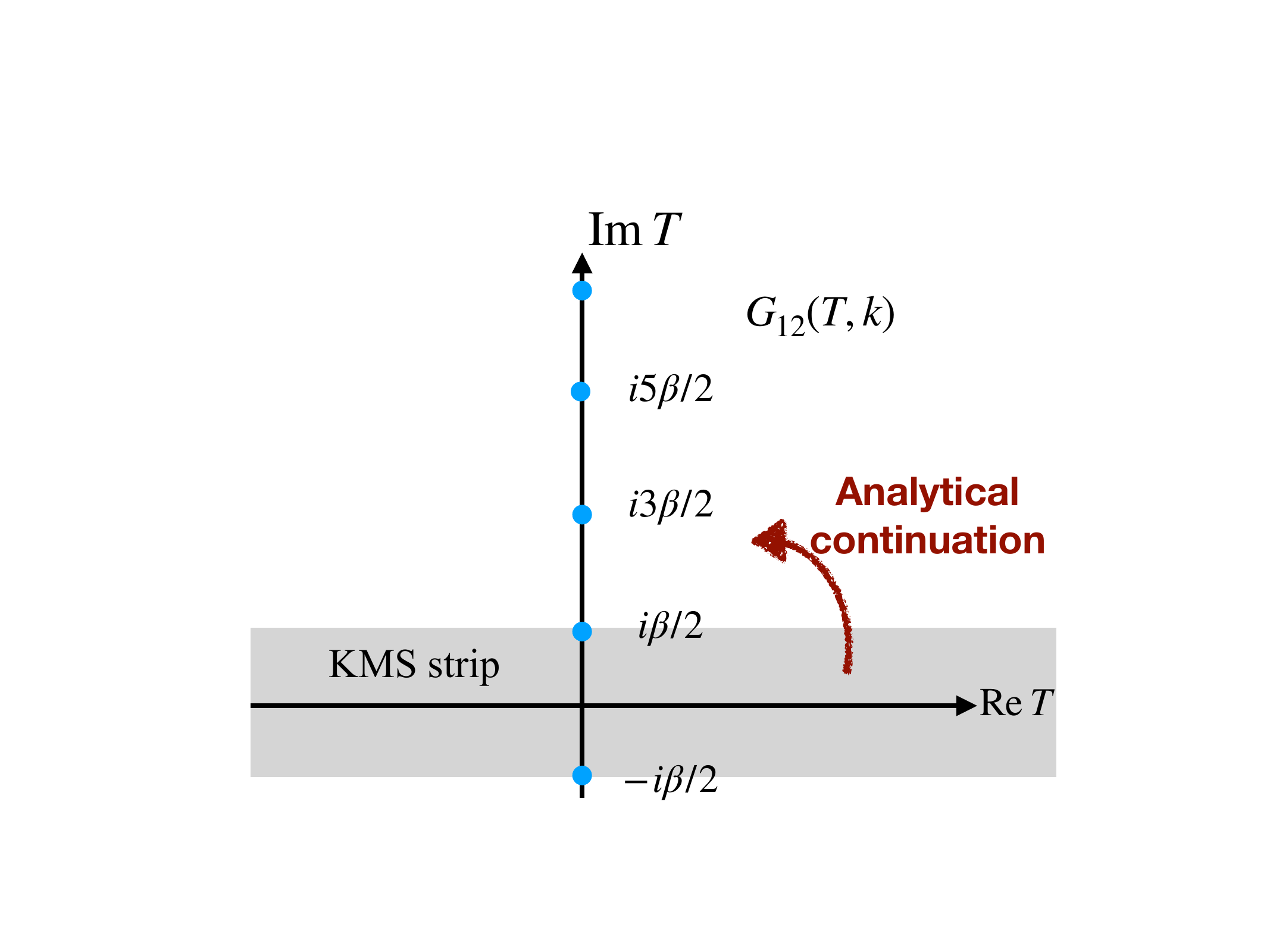}
\caption{\textit{Analytic continuation as a probe of the black hole interior.}
\textbf{Left:} In the maximally extended AdS black-hole geometry, the different exterior and interior regions can be represented using complexified Schwarzschild time $t$, with horizon crossings accompanied by characteristic imaginary-time shifts. Analytic continuation therefore provides a natural way to move inside the black hole interior. \textbf{Right:} The same structure is reflected in the dual boundary theory. The two-sided thermal correlator $G_{\mt{LR}}(T,k)$ is defined in the physical Kubo–Martin–Schwinger (KMS) strip, while its analytic continuation to complex $T$ probes additional singularities outside this strip. In holographic theories, the geometric information about the black hole interior, particularly the near-singularity region and the bouncing-geodesic structure, is encoded in the analytically continued thermal correlators.}
\label{fig:analyticalcontinuation}
\end{figure}

The common theme underlying these developments is the role of analytic continuation, as schematically illustrated in figure~\ref{fig:analyticalcontinuation}. In the bulk, different regions of the maximally extended black hole geometry can be conveniently described using complexified Schwarzschild time, with horizon crossings accompanied by characteristic imaginary-time shifts. In particular, the half-period continuation relating the two asymptotic boundaries has a natural interpretation in terms of the thermofield-double (TFD) state \cite{Maldacena:2001kr}. On the boundary, thermal correlation functions are holomorphic within the KMS strip, while analytic continuation into the complex-time plane reveals additional sheets and singularities that are invisible in the KMS strip. Although the analytical continuation of bulk coordinates and that of boundary correlators play distinct roles, their connection provides a useful holographic framework for extracting information about the spacetime geometry behind the horizon.

A complementary perspective on the bouncing-geodesic singularities also naturally emerges from the analytic structure of thermal correlators in frequency space. The connection between black hole singularities and frequency-space observables was first explored in \cite{Festuccia:2005pi,Festuccia:2008zx}, relating the geodesic description to the high-frequency asymptotics and pole structure of Wightman functions in momentum space. In particular, the poles of retarded Green functions in the semiclassical holographic limit coincide with quasinormal modes (QNMs) of the black hole \cite{Nunez:2003eq,Kovtun:2005ev}. Furthermore, the thermal product formula derived in \cite{Dodelson:2023vrw} shows that two-sided holographic correlators can be reconstructed from QNM data. This nice formula has inspired several further developments; see \eg \cite{Horowitz:2023ury,Grozdanov:2024wgo,Bhattacharya:2025vyi,Grozdanov:2025ulc,Grozdanov:2026ktq}. Related WKB analyses have also identified exponentially suppressed high-frequency contributions that encode the signatures of the black hole singularity in thermal correlators \cite{Afkhami-Jeddi:2025wra,Jia:2025jbi}.

The motivation of our present work starts from the observation that the bouncing-geodesic singularity emerges in the strict semiclassical limit. This raises a natural question: how do quantum effects modify the imprints of the black hole singularity on holographic observables? Rather than attempting to construct a quantum-corrected geometry directly near the black hole singularity, we address this question by investigating how finite-$\GN$, or equivalently finite-$c$, corrections modify the analytic structure of holographic thermal correlators. In particular, we aim to understand the fate of the complex-time singularities associated with bouncing geodesics. 
%\textcolor{blue}{Related recent investigations of bouncing-geodesic singularities can be found in \cite{Buric:2026sit,Batra:2026gbq,Coviello:2026hdc,Komargodski:2026oob,DellaRocca:2026vmu,Ceplak:2026ard,Oh:2026dky}.} 
Related recent studies have investigated the fate of these singularities at finite $\GN$ \cite{Komargodski:2026oob}, their smoothing through gravitational backreaction in 2d gravity \cite{Batra:2026gbq}, and their modification or resolution by stringy effects \cite{Dodelson:2025jff,Buric:2026sit} or in regular AdS black hole geometries \cite{Chen:2026efa}.

From the viewpoint of the Fourier transform, the complex-time singularities of the correlator $G(T)$ are indeed encoded in its corresponding behaviour in frequency space. In particular, a power-law divergence at a generic complex time $T=T_\alpha$ arises from 
\begin{equation}
G(T)\sim \int_0^\infty \dd\omega\,\omega^\nu
\ee^{-\ii\omega(T-T_\alpha)}
\propto \frac{1}{(T-T_\alpha)^{\nu+1}}\,.
\end{equation}
This singular behaviour is therefore determined by the characteristic large-frequency asymptotics of the frequency-space correlator, \ie
\begin{equation}
G(\omega)\sim \omega^\nu\ee^{\ii\omega T_\alpha},
\qquad \beta\omega\gg 1\,.
\end{equation}
The exponential phase factor thus determines the location of the complex-time singularity, while the power-law prefactor $\omega^\nu$ controls its local divergent behaviour. Remarkably, for holographic thermal correlators, this characteristic large-frequency behaviour is determined by the poles of the analytically continued correlator $G_{\mt{LR}}(\omega)$, which appear as QNMs of the dual black hole spacetime. Through the Hadamard product representation \cite{Dodelson:2023vrw}, the distribution of these QNM poles reconstructs the frequency-space thermal correlator. As a result, asymptotically linear QNM trajectories generate an infinite lattice of complex-time singularities in $G_{\mt{LR}}(T)$ \cite{Dodelson:2025jff}. The QNM spectrum thus provides a direct connection between the asymptotic resonance structure of the black hole and the holographic signatures of singularity.

This connection thus suggests a natural smoothing mechanism for the semiclassical singularities. If the QNM trajectories are deformed from their asymptotically linear form, the associated large-frequency contributions (and hence the complex time singularities) are expected to be modified. At finite $c$, quantum corrections would in general be expected to deform the asymptotically linear QNM trajectories, leading to nonlinear corrections of the schematic form:
\begin{equation}\label{eq:intro-qnm-hierarchy}
\omega_n=a_0+a_1 n+\frac{a_2}{c}n^2+\cdots
+O\!\left(\frac{1}{c^2}\right),
\qquad n\in\NN\,.
\end{equation}
Within the intermediate-frequency regime $1\ll\beta|\omega|\ll c$ we show that such nonlinear deformations modify the Stokes sectors of the corresponding Hadamard product. For instance, the leading quadratic correction produces a Gaussian kernel in the corresponding sector of the thermal correlator, \ie  
\begin{equation}
G_{\alpha}(\omega)\sim
A_\alpha\,\omega^\nu
\exp\left[
\ii\omega T_\alpha
-\kappa_\alpha\omega^2
\right],
\qquad
\kappa_\alpha\sim\frac{\beta^2}{c}\,.
\end{equation}
For $\operatorname{Re}\kappa_\alpha>0$, the Gaussian suppression regulates the large-frequency contribution responsible for the original divergence, replacing the sharp complex-time singularity with a smooth profile of characteristic width $\mathcal O(\beta/\sqrt c)$. We further illustrate the finite-$c$ deformation using Liouville CFT, which provides an analytically tractable theory for deriving pole-zero distributions. In particular, the semiclassical linear QNM trajectory is deformed into a quadratic trajectory at finite $c$, providing an explicit CFT realization of the QNM deformation.

Furthermore, we identify a complementary smoothing mechanism arising from the spectral decomposition of holographic thermal correlators. In the strict semiclassical limit, the canonical ensemble of black hole microstates is dominated by a relatively narrow energy window around a characteristic black hole energy, with the thermal two-sided correlator $G_{\mt{LR}}$ to be approximated by that of a typical black hole microstate. At finite $\GN$, however, fluctuations around the canonical saddle produce a Gaussian suppression factor in the transition frequency. For the relevant semiclassical singular sectors $G_{\mt{LR},\alpha}$, this leads to
\begin{equation}
G_{\mt{LR},\alpha}(\omega)
\simeq
\exp\!\left[-\frac{\omega^2}{4(\Delta E)^2}\right]
G_{\mt{LR},\alpha}^{(0)}(\omega),
\qquad
\Delta E=\mathcal O\left(\frac{1}{\sqrt{\GN}}\right)\,,
\label{eq:intro-spectral-gaussian}
\end{equation}
where $\Delta E$ characterizes the energy fluctuations around the canonical saddle. In the time domain, this Gaussian factor corresponds to a convolution that smooths the original singularity over a characteristic time width $\Delta T\sim 1/\Delta E\sim\mathcal O(\sqrt{\GN})$. Unlike the intrinsic finite-$c$ deformation of the QNM spectrum, this mechanism originates from averaging over the black hole microstates in the thermal ensemble. Together, these two mechanisms provide complementary descriptions of how the quantum corrections smear out the holographic signature of the semiclassical black hole singularity.

%\par\noindent\rule{\textwidth}{1pt}
%The boundary description also provides a framework for investigating departures from the semiclassical limit. In two dimensions, heavy--light Virasoro blocks reproduce effective thermal behavior at large central charge \cite{Fitzpatrick:2015zha}, while the analytic structure of heavy--light OPE coefficients can encode Ruelle resonances matching the QNM spectrum of BTZ black hole \cite{Turiaci:2016cvo}. An instructive example of finite-$c$ effects is furnished by the thermal ``forbidden singularities'' of semiclassical heavy--light Virasoro blocks. Their resolution at finite central charge involves contributions nonperturbative in the semiclassical expansion, and numerical studies reveal Stokes phenomena and a breakdown of the naive approximation near the would-be singularities \cite{Fitzpatrick:2016ive,Chen:2017yze}. Although these singularities are distinct from the higher-dimensional bouncing singularity, they illustrate how singular features of an effective black-hole background can be absent from exact conformal blocks. 

%\textcolor{red}{See also \cite{Buric:2026sit,Batra:2026gbq,Coviello:2026hdc,Ceplak:2026ard,Komargodski:2026oob,DellaRocca:2026vmu} for some recent discussion on related topics and questions.} 
%\par\noindent\rule{\textwidth}{1pt}

\emph{Note added:} Several recent works have also investigated the fate of bouncing-geodesic singularities beyond the semiclassical approximation. Ref.~\cite{Komargodski:2026oob} established the absence of these singularities in thermal CFT correlators at finite $\GN$ based on analyticity arguments. This provides non-perturbative constraints without determining the detailed smoothing profile. Another approach considers regular AdS black-hole geometries, where the original divergence is replaced by a smooth bump associated with complex null geodesics \cite{Chen:2026efa}. More closely related to our work, Gaussian smoothing of the bouncing-geodesic singularity has been demonstrated in two-dimensional dilaton gravity by including the gravitational backreaction of the probe worldline \cite{Batra:2026gbq}. The surviving information about the black hole interior is encoded in Stokes phenomena near the coincident-point singularities. By contrast, our present work focuses on generic AdS black holes without assuming a regularized background geometry. We investigate how finite-$c$ deformations of their asymptotic QNM spectra can smooth bouncing-geodesic singularities through the Hadamard product representation of holographic correlators. We also identify a complementary Gaussian smoothing mechanism from the spectral decomposition of thermal correlators. Similar Gaussian corrections arising from canonical energy fluctuations were also discussed in \cite{Batra:2026gbq}.

\textbf{Outline:} The remainder of this paper is organized as follows. In section \ref{sec:semiclassical}, we review the analytic structure of thermal correlators, the semiclassical bouncing-geodesic singularity, and its relation to the asymptotic QNM spectrum. In section \ref{sec:qnm-deformation}, we study finite-$c$ deformations of the QNM trajectories and use Hadamard products and their Stokes structure to explain how the complex-time singularities are smoothed out. We also discuss why OPE singularities remain protected from the smoothing. In section \ref{sec:finite-c-qnm-liouville}, we analyze the pole-zero structure of Liouville CFT as an explicit finite-$c$ example.  In section \ref{sec:spectral-smoothing}, we derive the complementary Gaussian smoothing that follows from the canonical spectral decomposition.  We discuss several related topics and future directions in section \ref{sec:discussion}.

%%%%%%%%%%%%%%%%%%%%%%%%%%%%%%%%%%%%%%%%%%%%%%%%%%%%%%%%%%%%%%%%%

\section{Analytic Continuation, Singularities, and Quasinormal Modes}\label{sec:semiclassical}

Unlike vacuum two-point functions in a CFT, thermal correlators involve an intrinsic scale set by the inverse temperature $\beta$. Their analytic structure is therefore nontrivial and is constrained by the KMS condition, causality, and the OPE. In holographic CFTs with a semiclassical AdS black hole dual, the analytic continuation of these correlators can reveal additional structures associated with bulk propagation. In particular, one encounters time singularities related to bouncing geodesics that probe the black hole interior.

In this section, we first review the complex-time analytic structure of thermal correlators and the origin of the bouncing-geodesic singularities. We then turn to the complementary frequency-space description, where the semiclassical correlator is naturally organized by the QNM spectrum of the black hole. In the holographic semiclassical limit, the QNM spectrum in the large-frequency regime in turn determines the singularity structure of the correlator in the complex-time plane. This establishes a direct connection between three complementary descriptions: the analytic continuation of boundary correlators, bouncing geodesics probing the black hole interior, and the asymptotic spectrum of black hole QNMs.
%%%%%%%%%%%%%%%%%%%%%%%%%%%%%%%%%%%%%%%%%%%%%%%%%%%%%%%%%%%%%%%%%%%%%%%%
\subsection{Analytic structure of thermal correlators}
\label{subsec:thermal-analytic-structure}
%%%%%%%%%%%%%%%%%%%%%%%%%%%%%%%%%%%%%%%%%%%%%%%%%%%%%%%%%%%%%%%%%%%%%%%%
The analytic continuation of thermal correlation functions to complex time is central to our discussion. We therefore begin by reviewing the analytic structure implied by the Kubo--Martin--Schwinger (KMS) condition and its relation to the two Wightman orderings. This structure is completely general and does not rely on holography. We will subsequently discuss the additional singularities associated with holographic CFTs and the black hole interior. Throughout this paper, we denote the bulk spacetime dimension by $D=d+1$, where $d$ denotes the spacetime dimension of the boundary CFT, with spatial coordinates $\bx\in\mathbb{R}^{d-1}$.

We begin with the Euclidean thermal two-point function. Consider a Hermitian bosonic scalar operator $\cO$ in a thermal state at inverse temperature $\beta$. One can introduce the two Euclidean Wightman functions
\begin{equation}
G_{\mt{E}}^>(\tau,\bx)=\frac{1}{Z(\beta)}\tr\!\left[\ee^{-\beta H}\cO(\tau,\bx)\cO(0)\right],
\qquad 0<\tau<\beta ,
\label{eq:euclidean-greater}
\end{equation}
and
\begin{equation}
G_{\mt{E}}^<(\tau,\bx)=\frac{1}{Z(\beta)}\tr\!\left[\ee^{-\beta H}\cO(0)\cO(\tau,\bx)\right],
\qquad -\beta<\tau<0 ,
\label{eq:euclidean-less}
\end{equation}
where
\begin{equation}
\cO(\tau,\bx)=\ee^{\tau H}\cO(0,\bx)\ee^{-\tau H},
\qquad
Z(\beta)=\tr\,\ee^{-\beta H}.
\label{eq:euclidean-heisenberg}
\end{equation}
The two orderings are not independent. Their relation follows directly from cyclicity of the thermal trace. For a bosonic operator one finds 
\begin{equation}
G_{\mt{E}}^>(\tau,\bx)=G_{\mt{E}}^<(\tau-\beta,\bx), \qquad 0<\tau<\beta \,,
\label{eq:euclidean-kms}
\end{equation}
which is the Euclidean form of the KMS condition. Since $\tau-\beta\in(-\beta,0)$, the KMS relation identifies the positive- and negative-Euclidean-time intervals and expresses the fact that the Euclidean thermal correlator lives on a circle of circumference $\beta$. Thus, translating one insertion once around the thermal circle exchanges the operator ordering.

The origin of these thermal intervals is particularly transparent in the energy basis. Inserting complete sets of eigenstates, $H|E_n\rangle=E_n|E_n\rangle$, gives
\begin{equation}\label{eq:Ggreater-spectral}
\begin{split}
G_{\mt{E}}^>(\tau,\bx)
%&=\frac{1}{Z(\beta)}\sum_{m,n}\ee^{-\beta E_m-\tau(E_n-E_m)}
%\cO_{mn}(\bx)\cO_{nm}(0) \\ 
&=\frac{1}{Z(\beta)}\sum_{m,n}\ee^{-(\beta-\tau)E_m-\tau E_n}
\cO_{mn}(\bx)\cO_{nm}(0)\,, 
\end{split}
\end{equation}
where $\cO_{mn}(\bx)\equiv\langle E_m|\cO(0,\bx)|E_n\rangle$. The two independent energy sums are exponentially suppressed provided
\begin{equation}
0<\tau<\beta.
\label{eq:euclidean-thermal-interval}
\end{equation}
Similarly, the natural Euclidean interval of $G_{\mt{E}}^<(\tau,\bx)$ is $-\beta<\tau<0$. At $\bx=0$, or after Fourier transformation to a fixed momentum channel with $\cO_{\bk}^{\dagger}=\cO_{-\bk}$, the product of matrix elements reduces to the corresponding absolute square.

The Lorentzian Wightman functions are obtained from their Euclidean counterparts by the Wick rotation
\begin{equation}
t=-\ii\tau, \qquad \cO(t,\bx)=\ee^{\ii Ht}\cO(0,\bx)\ee^{-\ii Ht}.
\label{eq:wick-rotation}
\end{equation}
The corresponding Lorentzian Wightman functions are therefore
\begin{equation}
G^>(t,\bx)=\frac{1}{Z(\beta)}\tr\!\left[\ee^{-\beta H}\cO(t,\bx)\cO(0)\right],
\qquad
G^<(t,\bx)=\frac{1}{Z(\beta)}\tr\!\left[\ee^{-\beta H}\cO(0)\cO(t,\bx)\right].
\label{eq:lorentzian-wightman}
\end{equation}
We may further analytically continue the Lorentzian time into the complex plane. To avoid introducing unnecessary time variables, we continue to denote the complexified time coordinate by $t=\operatorname{Re}t+\ii\,\operatorname{Im}t\in\mathbb C$. The analyticity domain follows directly from the same thermal spectral representation discussed above. Under the standard assumptions ensuring the existence of thermal correlation functions, $G^>(t,\bx)$ is holomorphic in the lower thermal strip
\begin{equation}
-\beta<\operatorname{Im}\, t<0,
\label{eq:Ggreater-strip}
\end{equation}
while $G^<(t,\bx)$ is holomorphic in the adjacent upper strip $0<\operatorname{Im}\,t<\beta$. The two analytic functions are thus related by the complexified KMS condition
\begin{equation}
G^>(t,\bx)=G^<(t+\ii\beta,\bx),
\label{eq:complex-kms}
\end{equation}
or equivalently $G^>(t-\ii\beta,\bx)=G^<(t,\bx)$. Thus the two Lorentzian Wightman functions are not independent objects, but arise as different boundary values of a single thermal analytic function. For example, the analytic continuation of $G^>(t,\bx)$ to the strip $-\beta<\operatorname{Im}\,t<0$ interpolates between the two Lorentzian orderings: its boundary value at $\operatorname{Im}\,t=0$ gives $G^>(t,\bx)$, whereas its boundary value at $\operatorname{Im}\,t=-\beta$ gives $G^<$ at the real Lorentzian time through the KMS relation. The KMS condition therefore provides a concise characterization of the complex-time analytic structure of thermal equilibrium states. In particular, the Wightman function admits a holomorphic continuation to a strip of width $\beta$ in the complex-time plane, whose two boundaries reproduce the two opposite Lorentzian operator orderings. %These boundary values are related by an imaginary-time translation $t\rightarrow t+\ii\beta$, corresponding to winding once around the thermal circle and exchanging the operator ordering.

In this paper, we are particularly interested in the two-sided AdS black hole. It is therefore convenient to recenter the thermal strip about the Lorentzian axis when studying the analytic continuation of holographic correlators. Starting from the Euclidean time $\tau$, we first introduce the complexified Lorentzian time $t$ through $\tau=\ii t$. We next shift the complex time coordinate by half a thermal period,
\begin{equation}
T\equiv t+\ii\frac{\beta}{2},
\qquad
t=T-\ii\frac{\beta}{2},
\label{eq:centered-complex-time}
\end{equation}
In terms of this centered coordinate, we define
\begin{equation}\label{eq:defineGLR01}
\boxed{ G_{12}(T,\bx)\equiv G_{\mt{E}}^>\!\left(\frac{\beta}{2}+\ii T,\bx\right) = G^>\!\left( T - \ii \frac{\beta}{2},\bx\right) = \frac{1}{Z(\beta)}\tr\!\left(
 \ee^{-\beta H/2}\cO(T,\bx)\ee^{-\beta H/2}\cO(0)\right) \,. } 
\end{equation}
The corresponding analyticity domain is the centered strip
\begin{equation}
-\frac{\beta}{2}<\operatorname{Im} T<\frac{\beta}{2}\,. 
\label{eq:g12-kms-strip}
\end{equation}
Thus, as shown in figure.~\ref{fig:analyticalcontinuation}, the shift $T=t+\ii\beta/2$ simply recenters the original KMS strip so that the physical two-sided Lorentzian correlator lies on the real $T$ axis.

The symmetric correlation function $G_{12}(T,\bx)$ for real $T$ has a direct holographic interpretation. The thermal density matrix can be canonically purified by introducing an auxiliary copy of the Hilbert space and defining the thermofield-double (TFD) state, 
\begin{equation}
|\TFD_\beta\rangle = \frac{1}{\sqrt{Z(\beta)}} \sum_n e^{-\beta E_n/2} |E_n\rangle_{\mt L}| E_n\rangle_{\mt R}\,, 
\end{equation}
where $|E_n\rangle_{\mt R}$ denotes the CPT-conjugate state associated with $|E_n\rangle_{\mt L}$. Here, we choose $t_{\mt L}$ and $t_{\mt R}$ to be the future-directed Lorentzian times on the left and right boundaries \footnote{Notice that this convention differs from the common Schwarzschild Killing coordinate, whose future direction is opposite in the two asymptotic regions. In terms of the complexified Schwarzschild coordinate one has $t_{\rm Sch}^{(\mt R)}=t_{\mt R}$ and $t_{\rm Sch}^{(\mt L)}=-t_{\mt L}-i\beta/2$.}, respectively. In holographic CFTs, the TFD state is dual, in the semiclassical limit, to the maximally extended two-sided AdS black hole \cite{Maldacena:2001kr}. Correspondingly, the time-evolved TFD state is given by 
\begin{equation}
|\TFD(t_{\mt L},t_{\mt R})\rangle = e^{-i(H_{\mt L}t_{\mt L}+H_{\mt R}t_{\mt R})}
|\TFD_\beta\rangle =\frac{1}{\sqrt{Z(\beta)}} \sum_n e^{-\beta E_n/2-iE_nT} |E_n\rangle_{\mt L}|E_n\rangle_{\mt R}\,,
\label{eq:time-evolved-TFD}
\end{equation}
with 
\begin{equation}
    T\equiv t_{\mt L}+t_{\mt R} \,. 
\end{equation}
Since $(H_{\mt L}-H_{\mt R})|\TFD_\beta\rangle=0$, the state depends only on the sum $T$. Considering the two-sided AdS black hole as the dual bulk spacetime, the holographic two-sided correlator is thus given by 
\begin{equation}
	\boxed{ G_{\mt{LR}}(T):=\langle\TFD(t_{\mt L},t_{\mt R})|\mathcal{O}_{\mt L}\mathcal{O}_{\mt R}|\TFD(t_{\mt L},t_{\mt R})\rangle = \langle\TFD_\beta|\mathcal{O}(t_{\mt L})\mathcal{O}(t_{\mt R})|\TFD_\beta\rangle} \,,
\end{equation}
where $\mathcal{O}_{\mt L}$ and $\mathcal{O}_{\mt R}$ denote the operators inserted on the left and right boundaries, respectively. It is straightforward to find that the spectral decomposition has the following form:
\begin{equation}
	G_{\mt{LR}}(T)
	= \frac{1}{Z(\beta)}\sum_{m,n} e^{-\frac{\beta}{2}(E_m+E_n)}\,e^{-iT(E_n-E_m)}\, |{\mathcal{O}_{mn}}|^2 \,. 
\end{equation}
From this decomposition, it is clear that this two-sided correlator is the two-point function $G_{12}(T,\bx)$ defined in eq.~\eqref{eq:defineGLR01}, which can be obtained from the ordinary thermal Euclidean correlator by the analytic continuation
\begin{equation}
 \tau \to \frac{\beta}{2} + \ii \, T \,. 
\label{eq:GLR-half-period}
\end{equation}
The imaginary time shift by $\beta/2$ appearing in \eqref{eq:defineGLR01} admits a natural geometric interpretation in the bulk. In the \textit{maximally extended} eternal AdS black hole, the different exterior and interior regions can be described as distinct patches of a complexified Schwarzschild time coordinate, with imaginary-time shifts accompanying the continuation across the horizons \cite{Maldacena:2001kr,Fidkowski:2003nf}. In particular, continuation from one asymptotic boundary to the other amounts to a half-period shift around the Euclidean thermal circle, $\tau\rightarrow\tau+\beta/2$, or equivalently to an imaginary Lorentzian-time shift by $\mp\ii\beta/2$ depending on the time convention. This is precisely the continuation appearing in \eqref{eq:defineGLR01}, which identifies the left-right correlator with the analytic continuation of an ordinary thermal correlator. The same bulk-boundary relation is naturally incorporated in the real-time holographic prescription, where different segments of the complex-time contour are filled by the corresponding Euclidean and Lorentzian bulk geometries \cite{Herzog:2002pc,Skenderis:2008dg,Skenderis:2008dh}. As shown in figure~\ref{fig:analyticalcontinuation}, this simple bulk picture supports the basic idea that the analytic continuation of boundary correlators can encode information about the black hole interior.

\begin{figure}[t]
\centering
\includegraphics[width=0.62\linewidth]{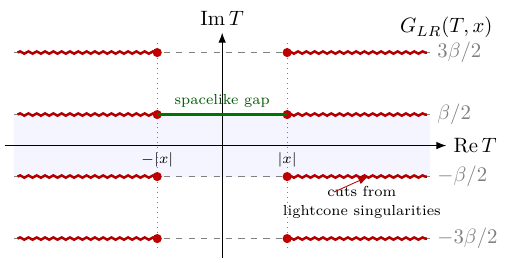}
\caption{\textit{Analytic structure of the two-sided thermal correlator at fixed spatial separation.} The central region denotes the principal KMS strip $-\beta/2<\operatorname{Im}T<\beta/2$. Adjacent KMS strips are analytically connected through the spacelike intervals $-|\bx|<\operatorname{Re}T<|\bx|$ on their shared boundaries, where microcausality makes the two Wightman boundary values coincide. Repeating this gluing across successive thermal boundaries produces a connected principal
thermal sheet, while analytic continuation around the lightcone/OPE branch
points at $T_{\mt{OPE}}^{(\pm,n)}$ leads instead to secondary Riemann sheets.}
\label{fig:g12-kms-analytic-structure}
\end{figure}

The KMS condition determines the domain of analyticity of the thermal correlator, while the singularity structure on the boundaries of this domain is controlled by the local short-distance physics. In particular, as shown in figure~\ref{fig:g12-kms-analytic-structure}, the boundaries of the strip \eqref{eq:g12-kms-strip} contain the lightcone singularities associated with the operator product expansion. Locally, the short-distance OPE gives
\begin{equation}
\cO(\tau,\bx)\cO(0,0) \sim \frac{\mathbf{1}}{\left(\tau^2+|\bx|^2\right)^{\Delta}}+\cdots ,
\label{eq:thermal-ope-local}
\end{equation}
and thermal periodicity generates the corresponding images at $\tau=n\beta$. Although the Euclidean OPE controls only the neighborhood of a coincidence point, the corresponding Lorentzian singularity is associated with null separation. In a local relativistic QFT, such UV singularities propagate along null rays and therefore extend along the full Lorentzian light cone. In terms of the complex time $T$, the resulting lightcone/OPE singularities occur at
\begin{equation}
\boxed{T_{\mt{OPE}}^{(\pm,n)}=\pm|\bx|+\ii\left(n+\frac{1}{2}\right)\beta,
\qquad n\in\mathbb{Z}\,.}
\label{eq:g12-ope-lattice}
\end{equation}

The global continuation of thermal correlators is further constrained by microcausality. Let us consider a fixed nonzero spatial separation $|\bx|$. On the upper and lower boundaries of a KMS strip, the two boundary values correspond to opposite Wightman orderings. However, the operators commute in the spacelike interval
\begin{equation}
-|\bx|<\operatorname{Re}T<|\bx|, \qquad \operatorname{Im}\,T=\pm\frac{\beta}{2} \,.
\label{eq:spacelike-gap}
\end{equation}
As a result, the two Wightman boundary values coincide there. The edge-of-the-wedge theorem then allows the correlator to be analytically continued across this interval, thereby gluing adjacent KMS strips together. 
This gluing terminates at the lightcone branch points \eqref{eq:g12-ope-lattice}. Beyond these points, the common boundaries of neighboring KMS strips lie in the timelike region, where the two Wightman boundary values no longer coincide. These timelike segments therefore form the lightcone branch cuts, as shown in figure~\ref{fig:g12-kms-analytic-structure}. Repeating the same gluing construction over all thermal images produces a connected analytic domain with these cuts removed. The corresponding branch of the correlator is thus referred to as the \emph{principal thermal sheet}. Analytic continuation around a lightcone branch point instead leads to a secondary Riemann sheet.\footnote{Notice that at $\bx=0$ the spacelike interval collapses, and the continuation must instead be specified by a choice of path around the coincident-point singularity.}

\begin{figure}[t]
\centering
\includegraphics[width=5in]{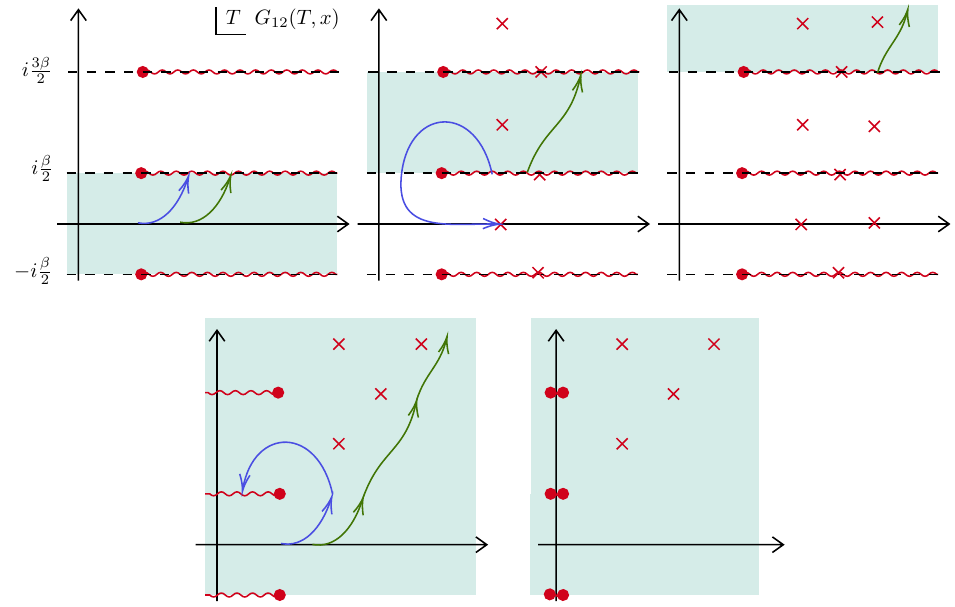}
\caption{\textit{Analytic continuation and sheet structure of $G_{\mt{LR}}(T,\bx)$.} Red dots denote OPE singularities, red wavy lines indicate branch cuts, and red crosses mark possible additional singularities. The upper panels illustrate the first three sheets, with the blue and green curves representing two distinct analytic continuation paths. The cyan regions are continuously connected along the green path, as illustrated in the lower-left panel with a different choice of branch cuts. The lower-right panel shows the limit $|\bx|\to0$, where pairs of OPE singularities merge and the corresponding continuation paths must be reconsidered. In BTZ, analytic continuation introduces no additional singularities beyond the OPE images, whereas higher-dimensional black holes may exhibit additional bouncing-geodesic singularities (see section~\ref{subsec:btz vs high dim bh}).}
\label{fig:BTZAdSBH}
\end{figure}

This distinction between the principal thermal sheet and its secondary Riemann sheets will be central to our discussion of the black hole interior. The KMS condition, microcausality, and the short-distance OPE strongly constrain the analytic structure on the principal thermal sheet, whereas analytic continuation around the lightcone/OPE branch points can access additional structures that are not fixed by these general principles. In a holographic CFT, such additional analytic data may admit a geometric interpretation in the dual bulk spacetime. As illustrated schematically in figure~\ref{fig:BTZAdSBH}, continuation through a lightcone cut can lead to secondary sheets containing a richer singularity structure. The bouncing-geodesic singularity provides a particularly sharp example of such additional information about the black hole interior.

It is useful to illustrate this analytic structure in two-dimensional CFT, where the finite-temperature two-point function is fixed by conformal symmetry. For a scalar primary operator of conformal dimension $\Delta$ on the spatial line, the two-sided thermal correlator is 
\begin{equation}
G_{\mt{LR}}^{\mt{BTZ}}(T,x) \propto 
\left(\frac{\pi}{\beta}\right)^{2\Delta}
\frac{1}{ \left[\cosh\!\left(\frac{\pi(T-x)}{\beta}\right)\cosh\!\left(\frac{\pi(T+x)}{\beta}\right)\right]^{\Delta}}\,,
\label{eq:btz-two-sided-correlator}
\end{equation}
For a holographic CFT$_2$, the same expression is reproduced by the non-winding sector of the non-rotating BTZ geometry. After analytic continuation to complex $T$, this correlator has an infinite lattice of singularities located precisely at $T=T_{\mt{OPE}}^{(\pm,n)}$ in \eqref{eq:g12-ope-lattice}. The singularities nearest to the physical Lorentzian axis lie at $T=\pm x\pm\ii\beta/2$, on the two boundaries of the centered KMS strip. All remaining singularities are obtained by thermal translations by integer multiples of $\ii\beta$ and can therefore be regarded as KMS images of the ordinary lightcone/OPE singularities. Thus, at fixed spatial separation, the complex-time singularity structure of $G_{\mt{LR}}^{\mt{BTZ}}(T,x)$ is generated entirely by the ordinary OPE singularity and its thermal images. Analytic continuation to secondary sheets changes the phase but does not generate an additional family of singularities. This situation changes qualitatively for higher-dimensional AdS black holes. There, nearly null trajectories can probe arbitrarily close to the spacelike curvature singularity and, after analytic continuation to appropriate secondary sheets, give rise to additional bouncing-geodesic singularities, as illustrated in figure~\ref{fig:BTZAdSBH}. We will explain this structure in more detail in subsection \ref{subsec:bouncing-singularity}.

The discussion above has been formulated in position space at fixed spatial separation, in terms of the thermal correlator $G_{12}(T,\bx)\equiv G_{\mt{LR}}(T,\bx)$.  For the analysis in the remainder
of this paper, however, we will primarily work at fixed spatial momentum and make frequent use of both $G_{\mt{LR}}(T,\bk)$ and its frequency-space representation $G_{\mt{LR}}(\omega,\bk)$. We therefore define our Fourier transformation conventions by
\begin{equation}
G_{\mt{LR}}(T,\bk) = \int_{\mathbb R^{d-1}}\dd^{d-1}\bx\, \ee^{-\ii\bk\cdot\bx}\, G_{\mt{LR}}(T,\bx),
\qquad
G_{\mt{LR}}(\omega,\bk)
=
\int_{\mathbb R}\dd T\,
\ee^{\ii\omega T}\,
G_{\mt{LR}}(T,\bk),
\label{eq:fourier-conventions}
\end{equation}
We will thus encounter four closely related representations of the same thermal correlator,
\begin{equation}
G_{\mt{LR}}(T,\bx),\qquad G_{\mt{LR}}(T,\bk),\qquad G_{\mt{LR}}(\omega,\bx),\qquad G_{\mt{LR}}(\omega,\bk).
\label{eq:g12-four-representations}
\end{equation}
The position-space representation is most directly connected to locality, causality, and bulk geodesics, whereas the correlator at fixed spatial momentum provides the natural setting for spectral decompositions and the subsequent description in terms of QNMs. Although these are different representations of the same correlation function, Fourier transformations do not preserve the singularity structure point by point. In particular, under the spatial Fourier transform
$\mathbf{k}\leftrightarrow \mathbf{x}$, moving poles or branch points in position space can collide, pinch the continued integration contour, or cancel, thereby modifying the complex-time singularity structure of $G_{12}(T,\bk)$. We postpone a detailed discussion of this mechanism to appendix~\ref{app:four-representations}. In the remainder of the main text, we focus primarily on the temporal Fourier transform $T\leftrightarrow\omega$.

%%%%%%%%%%%%%%%%%%%%%%%%%%%%%%%%%%%%%%%%%%%%%%%%%%%%%%%%%%%%%%%%%%%%%%%%
%%%%%%%%%%%%%%%%%%%%%%%%%%%%%%%%%%%%%%%%%%%%%%%%%%%%%%%%%%%%%%%%%%%%%%%%
\subsection{Heavy-state correlators and the HHLL representation}
\label{subsec:hhll}
%%%%%%%%%%%%%%%%%%%%%%%%%%%%%%%%%%%%%%%%%%%%%%%%%%%%%%%%%%%%%%%%%%%%%%%%
The thermal correlators discussed above describe expectation values in a canonical ensemble. A closely related but conceptually distinct observable is obtained by keeping the background state fixed. By the state--operator correspondence, a heavy scalar primary operator $\cO_{\mt H}$ creates the pure state
\begin{equation}
|\mathrm{H}\rangle=\cO_{\mt H}(0)|0\rangle,\qquad
\langle \mathrm{H} |=\lim_{r\rightarrow\infty}r^{2\Delta_{\mt H}}\langle0|\cO_{\mt H}^{\dagger}(r\mathbf n).
\label{eq:heavy-state-definition}
\end{equation}
The corresponding heavy-heavy-light-light (HHLL) four-point function can therefore be regarded as a two-point function of light operators in the heavy state,
\begin{equation}
G_{\mt H}(\tau, \mathbf{n})\equiv
\frac{\langle\cO_{\mt H}^{\dagger}(\infty)\cO_{\mt L}(\tau,\mathbf n)\cO_{\mt L}(0,\mathbf 1)\cO_{\mt H}(0)\rangle}
{\langle\cO_{\mt H}^{\dagger}(\infty)\cO_{\mt H}(0)\rangle}
=
\frac{\langle \mathrm{H} |\cO_{\mt L}(\tau,\mathbf n)\cO_{\mt L}(0,\mathbf 1)| \mathrm{H} \rangle}{\langle \mathrm{H} | \mathrm{H}\rangle}\,. 
\label{eq:hhll-heavy-state-correlator}
\end{equation}
Here $\tau=\log r>0$ is the Euclidean cylinder-time separation associated with radial quantization, with $r=e^\tau$, and $\cO_{\mt L}(\tau,\mathbf n)=r^{\Delta_{\mt L}}\cO_{\mt L}(r\mathbf n)$. In particular, we are interested in the semiclassical heavy-light regime
\begin{equation}
\Delta_{\mt H}\sim O(c),\qquad \Delta_{\mt L}\sim O(c^0),
\label{eq:heavy-light-scaling}
\end{equation}
with $c\to\infty$ and $\Delta_{\mt H}/c$ fixed. In this limit, the heavy state backreacts on the bulk geometry at leading order, while the light operator acts as a probe. In a holographic CFT, a sufficiently energetic heavy state can therefore describe a black hole microstate. This interpretation is particularly explicit in CFT$_2$: when the heavy dimension lies above the BTZ threshold, $\Delta_{\mt H}>c/12$, the geometry sourced by $|\mathrm{H} \rangle$ is a BTZ black hole with Hawking temperature
\begin{equation}
T_{\mt H}=\frac{1}{2\pi}\sqrt{\frac{12\Delta_{\mt H}}{c}-1}\,,
\label{eq:heavy-state-BTZ-temperature}
\end{equation}
where the AdS radius has been set to unity. The HHLL correlator then describes the propagation of the light probe in this black-hole background.

It is important, however, to distinguish this heavy-state observable from the canonical thermal correlator. The latter averages over energy eigenstates,
\begin{equation}
G_{\mt{E}}^>(\tau)=\frac{1}{Z(\beta)}\sum_n\ee^{-\beta E_n}G_{E_n}(\tau)\simeq\frac{\displaystyle\int\dd E\,\ee^{S(E)-\beta E}G_E(\tau)}{\displaystyle\int\dd E\,\ee^{S(E)-\beta E}},
\label{eq:thermal-as-heavy-average}
\end{equation}
where $G_E(\tau,\bx)$ denotes the corresponding two-point function in an energy eigenstate. At large entropy, the canonical integral is dominated by the saddle $E_\beta$ satisfying
\begin{equation}
\left.\frac{\partial S(E)}{\partial E}\right|_{E=E_\beta}=\beta.
\label{eq:canonical-heavy-saddle}
\end{equation}
If $|\mathrm{H}\rangle$ is a typical heavy eigenstate with $E_{\mt H}=E_\beta$, the eigenstate thermalization hypothesis for simple probes implies, to leading order in the large-entropy limit,
\begin{equation}
\boxed{ G_{\mt H}(\tau)  \underset{\text{typical state} \,  |\mathrm{H} \rangle  }{\simeq}  G_{\mt{E}}^>(\tau) \,. }
\label{eq:hhll-thermalization}
\end{equation}
In the bulk, this expresses the familiar fact that a sufficiently simple probe cannot distinguish a typical black hole microstate from the corresponding thermal black hole geometry at the semiclassical level. Equation~\eqref{eq:hhll-thermalization}, however, is only an approximate large-$c$ relation. At finite $c$, $G_{\mt H}(\tau)$ is a pure-state correlator and does not obey an exact KMS condition.

\begin{figure}[t]
\centering
\includegraphics[width=4.5in]{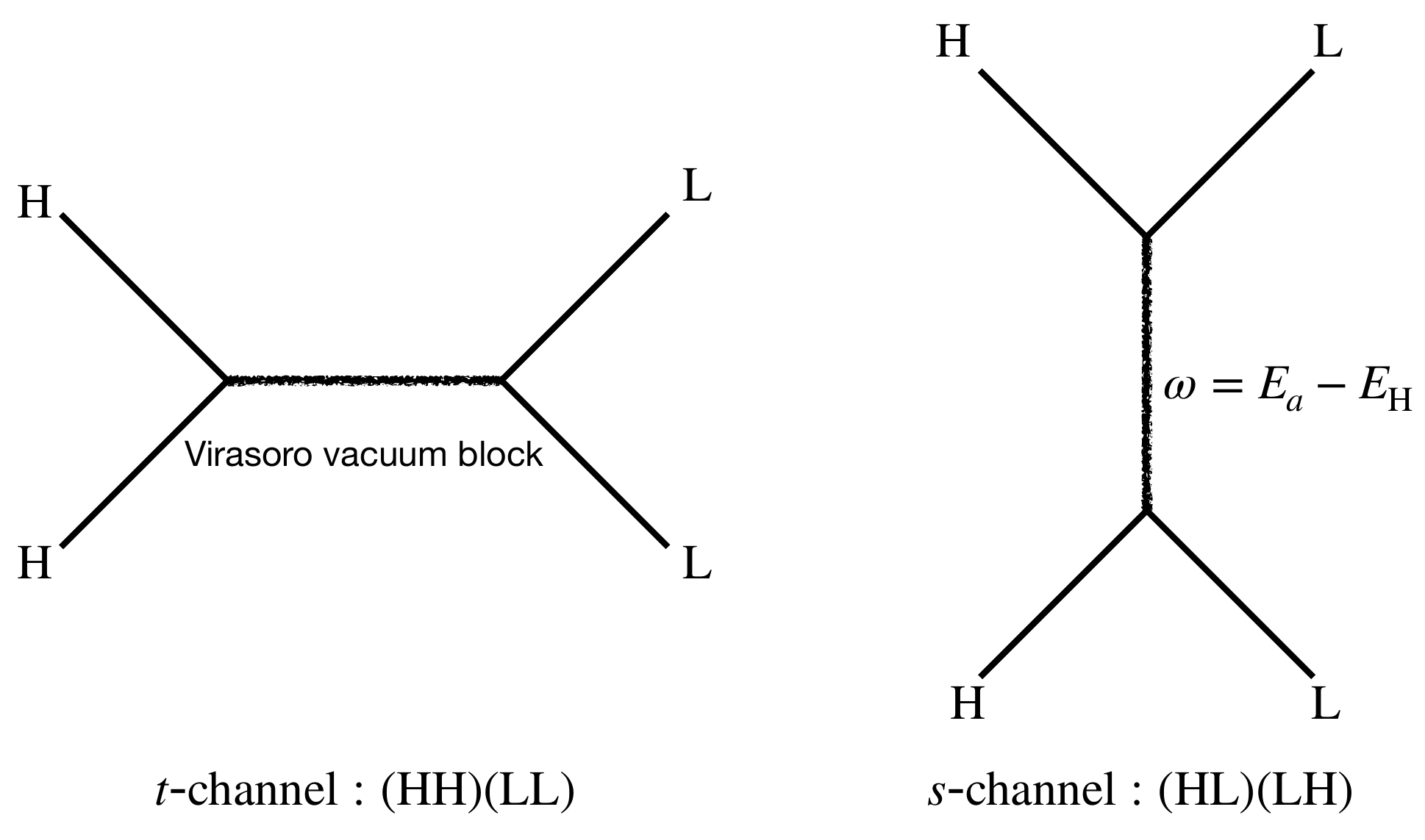}
\caption{\textit{Two complementary OPE channels of the HHLL correlator.} In the $\mathrm{(HH)(LL)}$ channel, the semiclassical heavy-light limit is governed by the vacuum block and exhibits emergent thermal images, including the forbidden singularities. In the crossed $\mathrm{(HL)(HL)}$ channel, the exchanged states are organized by the transition energy $\omega=E_a-E_{\mt H}$, whose continuation is governed by the QNM spectrum in the semiclassical black hole regime.}
\label{fig:hhll-two-channels}
\end{figure}

A particular advantage of the HHLL formulation is that the heavy-state correlator $G_{\mt H}$ is an ordinary CFT four-point function and can therefore be decomposed into conformal blocks in different OPE channels. Crossing symmetry relates two complementary decompositions \cite{Fitzpatrick:2015zha,Fitzpatrick:2016ive,Turiaci:2016cvo}, conventionally referred to as the $\mathrm{(HH)(LL)}$ and $\mathrm{(HL)(HL)}$ channels. As shown schematically in figure~\ref{fig:hhll-two-channels}, the two OPE channels of the HHLL correlator emphasize different aspects of the heavy-state dynamics and black hole physics. The $\mathrm{(HH)(LL)}$ channel naturally captures the emergence of semiclassical thermality, whereas the crossed $\mathrm{(HL)(HL)}$ channel resolves the spectrum of intermediate heavy-light states and is therefore more directly connected to the spectral and QNM descriptions that will be important below.

In the $\mathrm{(HH)(LL)}$ channel, the semiclassical heavy-light limit is governed by the Virasoro vacuum block. The vacuum module resums the exchange of stress-tensor descendants and, in the bulk description, captures the gravitational dressing of the light probe by boundary-graviton and multi-graviton excitations of the heavy background. For a scalar heavy state above the BTZ threshold, the holomorphic vacuum block takes the thermal form
\begin{equation}
\mathcal V_{\infty}(\tau)=\left[\frac{\pi T_{\mt H}}{\sin\!\left(\pi T_{\mt H}\tau\right)}\right]^{\Delta_{\mt L}}\,,
\label{eq:hhll-thermal-vacuum-block}
\end{equation}
together with the corresponding antiholomorphic factor. This semiclassical expression exhibits an emergent thermal periodicity and contains singularities at
\begin{equation}
\tau=n\beta_{\mt H},\qquad n\in\mathbb Z\,. 
\end{equation}
These locations coincide precisely with the KMS images of the ordinary OPE singularity in a genuine thermal correlator at inverse temperature $\beta_{\mt H}$. However, their interpretation in the HHLL correlator is sharply different. The first singularity at $n=0$ is the genuine light-light OPE singularity of the exact four-point function, whereas the singularities at $n\neq0$ are only semiclassical thermal images and constitute the familiar \emph{forbidden singularities}. In the strict large-$c$ limit, these additional singularities mimic the images required by KMS periodicity, even though the fixed-heavy-state correlator itself does not satisfy an exact KMS condition at finite $c$. Consequently, the $n\neq0$ singularities are not genuine OPE singularities of the exact four-point function and must be resolved once finite-$c$ effects are included. Their resolution therefore provides a sharp diagnostic of the breakdown of semiclassical thermality and requires quantum corrections beyond the strict large-$c$ vacuum-block approximation \cite{Fitzpatrick:2015dlt,Fitzpatrick:2016ive,Chen:2017yze}.

It is important to distinguish the fixed-heavy-state correlator from the canonical thermal correlator. In the HHLL correlator $G_{\mt H}$, the initial heavy-state energy $E_{\mt H}$ is held fixed, whereas the thermal correlator additionally averages over the initial-state energy with the Boltzmann weight. For a typical heavy state with $E_{\mt H}=E_\beta$, ETH implies that the two descriptions agree at leading order in the semiclassical regime. Nevertheless, only the canonical thermal correlator obeys an exact KMS condition. Consequently, the corresponding thermal images are genuine features of the thermal correlator, whereas the analogous images appearing in the semiclassical HHLL correlator are only emergent and become the forbidden singularities discussed above. This distinction will be important below when we relate finite-$c$ deformations of the effective QNM data to the smoothing of complex-time singularities.

The crossed $\mathrm{(HL)(HL)}$ channel provides a complementary description that is particularly well suited to a spectral representation. In this channel, each light operator is fused with a heavy operator, so that the correlator can be resolved directly in terms of the corresponding transition energies and matrix elements. For a definite spatial mode of the light operator, collectively denoted by $\lambda$, inserting a complete set of energy eigenstates gives
\begin{equation}
G_{\mt H}(t;\lambda)=\frac{1}{\langle \mathrm{H}|\mathrm{H}\rangle}\sum_a
\ee^{-\ii(E_a-E_{\mt H})t}
\left|\langle \mathrm{H} |\cO_{\mt L,\lambda}| E_a\rangle\right|^2\,.
\label{eq:heavy-state-spectral-time}
\end{equation}
%\smr{\begin{equation}G_{\mt H}(t;\lambda)=\frac{1}{\langle H|H\rangle}\sum_{a,\lambda'}e^{-i(E_{a,\lambda'}-E_{\mt H})t}\langle H|\cO_{\mt L,\lambda}^{\dagger}|a,\lambda'\rangle\langle a,\lambda'|\cO_{\mt L,\lambda}|H\rangle .\end{equation}}
Here $E_a-E_{\mt H}$ is the transition energy induced by the light probe. The matrix elements $\langle E_a|\cO_{\mt L,\lambda}|\mathrm{H}\rangle$ encode the corresponding heavy-light transition amplitudes and, equivalently, the relevant CFT data, including the heavy-light OPE coefficients of the exchanged primaries and the associated descendant contributions. After Fourier transformation to frequency space, the corresponding spectral decomposition thus takes the form\footnote{More precisely, the label $\lambda$ in the previous expression only denotes the discrete spherical-harmonic quantum numbers appropriate to the radial-quantization cylinder $\mathbb R\times S^{d-1}$. To make contact with the continuous momentum description, we instead work in the planar, or equivalently decompactification, limit. In this limit, the radius of the spatial sphere is taken large compared with the length scales probed by the correlator, so that the discrete angular modes become a continuous spatial momentum $\bk$.}
\begin{equation}
G_{\mt H}(\omega,\bk)=\frac{2\pi}{\langle \mathrm{H}|\mathrm{H}\rangle}\sum_a
\left|\langle \mathrm{H} |\cO_{\mt L,\bk}|E_a\rangle\right|^2
\delta\!\left(\omega-(E_a-E_{\mt H})\right),
\label{eq:heavy-state-spectral-frequency}
\end{equation}
where the natural frequency variable at fixed spatial momentum is the transition energy $\omega=E_a-E_{\mt H}$. At finite $c$ and finite volume, the exact heavy-state spectral function consists only of a discrete set of contributions at real transition frequencies $\omega$. However, in the semiclassical black hole regime, we take $|\mathrm{H}\rangle$ to be a typical black hole state with $E_{\mt H}=E_\beta$. The density of intermediate states in the relevant energy window around $E_{\mt H}$ then becomes parametrically large, so that the transition energy $\omega$ may be treated as an effectively continuous variable. Upon coarse graining over this dense spectrum, the corresponding retarded Green function (denoted by $G_{\mt R}$) can be analytically continued from real to complex frequency. In the strict $c\to\infty$ limit, this response is governed by linear perturbations of the classical black hole background, and its complex-frequency analytic structure is characterized by the corresponding QNM poles and associated zeros. At large but finite $c$, the underlying heavy-light transition energies and matrix elements receive quantum corrections. Provided that the coarse-grained response in the relevant semiclassical frequency regime continues to admit an effective meromorphic description, these corrections are naturally encoded as deformations of its pole--zero structure away from the strict classical QNM spectrum. The $\mathrm{(HL)(HL)}$ channel therefore provides a natural CFT framework for studying how finite-$c$ effects modify the semiclassical QNM pole--zero structure.

Taken together, the two OPE channels provide complementary descriptions of finite-$c$ effects in the heavy-state correlator. The $\mathrm{(HH)(LL)}$ channel makes manifest the resolution of the semiclassical forbidden singularities, whereas the $\mathrm{(HL)(HL)}$ channel organizes the corresponding quantum corrections through the heavy-light spectral data and the deformation of the effective QNM pole--zero structure. This interplay will be central to our discussion below of how finite-$c$ effects smooth the semiclassical complex-time singularities.

%%%%%%%%%%%%%%%%%%%%%%%%%%%%%%%%%%%%%%%%%%%%%

\subsection{Bouncing geodesics and the secondary-sheet singularity}
\label{subsec:bouncing-singularity}

We now specialize to holographic CFTs in the semiclassical regime, focusing on the analytic structure of the two-sided correlator $G_{\mt{LR}}$. We begin by reviewing the semiclassical bouncing-geodesic analysis of Ref.~\cite{Fidkowski:2003nf}, which provides a particularly transparent example of how the black-hole singularity is encoded in the analytic continuation of a boundary correlator. As motivated in section~\ref{sec:introduction}, in the large-$\Delta$ limit the two-sided correlator is governed by classical spacelike geodesics through the geodesic approximation \eqref{eq:intro-geodesic-approximation}, which we recall here in the form
\begin{equation}
G_{\mt{LR}}(T,\bx)\simeq\sum_{\gamma}A_{\gamma}(T,\bx)\,\ee^{-\Delta\mathcal{L}_{\gamma}(T,\bx)}\,.
\label{eq:geodesic-approximation}
\end{equation}
Following the configuration in \cite{Fidkowski:2003nf}, we restrict to $\bx=0$, for which a limiting spacelike geodesic approaches a null trajectory that effectively bounces off the black hole singularity. The time singularity generated by this limiting bouncing trajectory does not lie on the principal Lorentzian sheet of $G_{\mt{LR}}(T)$, but is instead revealed after analytic continuation to a secondary Riemann sheet of the complex $T$-plane.

For concreteness, consider the planar AdS$_{d+1}$-Schwarzschild geometry
\begin{equation}
\dd s^2=-f(r)\dd t^2+\frac{\dd r^2}{f(r)}+r^2\dd\bx^2,\qquad
f(r)=r^2\left[1-\left(\frac{r_h}{r}\right)^d\right],
\label{eq:ads-black-brane}
\end{equation}
with inverse Hawking temperature $\beta=4\pi/(d r_h)$. Parameterizing a radial spacelike geodesic by its proper length $s$, one has
\begin{equation}
-f(r)\dot t^{\,2}+\frac{\dot r^{\,2}}{f(r)} = 1\,.
\label{eq:spacelike-geodesic-normalization}
\end{equation}
Time-translation invariance gives a conserved quantity $E$, and the radial dynamics can be summarized as
\begin{equation}
E=f(r)\dot t,\qquad \dot r^{\,2}=E^2+f(r),\qquad E^2+f(r_{\min})=0\,, 
\label{eq:radial-geodesic-equation}
\end{equation}
where $r_{\min}$ denotes the radial turning point. For $d>2$, the blackening factor diverges near the curvature singularity as
\begin{equation}
f(r)\sim-r_h^d r^{2-d},\qquad r\rightarrow 0,
\end{equation}
and the turning-point condition therefore gives
\begin{equation}
\frac{r_{\min}}{r_h}\sim\left(\frac{r_h}{E}\right)^{2/(d-2)},\qquad E\rightarrow\infty.
\label{eq:rmin-large-E}
\end{equation}
Thus, as $E$ increases, the turning point approaches the singularity. Correspondingly, the spacelike geodesic becomes asymptotically null away from a shrinking neighborhood of the turning point, yielding the limiting bouncing geodesic \cite{Fidkowski:2003nf}.

The characteristic time of the bouncing geodesic can be read off from radial null trajectories, for which $\dd t/\dd r=\pm1/f(r)$. Owing to the boost symmetry of the eternal black hole, the two-sided configuration depends only on the invariant combination $T=t_{\mt L}+t_{\mt R}$. For the bouncing geodesic reaching the future black hole singularity, the critical boundary time is\footnote{The imaginary contribution associated with the horizon crossings is absorbed into the half-period shift used to define the centered complex time $T=t+\ii\beta/2$.}
\begin{equation}
T_{\rm crt} = t_{\mt{L,crt}} + t_{\mt{R,crt}} = - 2 \times \operatorname{P.V.}\int_0^\infty\frac{\dd r}{f(r)} = - \frac{\beta}{2}\cot\frac{\pi}{d} <0\,. 
\label{eq:tstar}
\end{equation}
The factor of two accounts for the two radial segments connecting the singularity to the two asymptotic boundaries. The time-reversed trajectory reaching the past white hole singularity occurs at $T=-T_{\rm crt}$. Notice that one has $T_{\rm crt}=0$ for $d=2$, which is consistent with the qualitatively different analytic structure of the BTZ geometry discussed above. However, in the $d=2$ case, the BTZ geometry has no curvature singularity. We will discuss the subtle differences in more detail in section~\ref{sec:discussion}.

To extract the singular behavior near $T_{\rm crt}$, one can regard the boundary time and the renormalized geodesic length parametrically as functions of the conserved energy $E$,
\begin{equation}
T(E)=2\operatorname{Re}\!\left[-\int_{r_{\min}(E)}^\infty\dd r\,\frac{E}{f(r)\sqrt{E^2+f(r)}}\right],
\label{eq:T-E}
\end{equation}
and
\begin{equation}
\mathcal{L}_{\rm ren}(E)=\lim_{r_{\max}\rightarrow\infty}\left[2\int_{r_{\min}(E)}^{r_{\max}}\frac{\dd r}{\sqrt{E^2+f(r)}}-2\log r_{\max}\right],
\label{eq:L-E}
\end{equation}
up to an irrelevant additive constant. In the large-$E$ limit, the geodesic approaches the bouncing trajectory and one finds
\begin{equation}
T(E)\rightarrow T_{\rm crt},\qquad \left|T(E)-T_{\rm crt}\right|\sim E^{-1},\qquad \mathcal{L}_{\rm ren}(E)\sim-2\log E \,. 
\label{eq:large-E-bounce}
\end{equation}
Near the critical time, the bouncing-branch contribution to the two-sided correlator therefore behaves as
\begin{equation}
G_{\mt{LR}}^{\rm bounce}(T,\bx=0)\sim\frac{1}{(T-T_{\rm crt})^{2\Delta}}.
\label{eq:bounce-correlator-singularity}
\end{equation}
Thus, as the spacelike geodesic approaches the null trajectory probing the black hole singularity, the analytically continued holographic two-sided correlator develops a time singularity at $T=T_{\rm crt}$.

As anticipated above, the crucial point is that \eqref{eq:bounce-correlator-singularity} is \emph{not} a singularity of the physical Lorentzian correlator on its principal sheet. The geodesic saddle problem is intrinsically multi-valued, with different solutions of the geodesic equations defining distinct branches of $\mathcal{L}(T)$. In the AdS$_5$ Schwarzschild example studied in \cite{Fidkowski:2003nf}, three geodesics coalesce at the symmetric point $T=0$. In the strict boundary limit, their local behavior takes the form
\begin{equation}
T\sim E^3, \qquad \mathcal{L}(E)-\mathcal{L}(0)\sim-E^4,
\end{equation}
and therefore
\begin{equation}
\mathcal{L}(T)-\mathcal{L}(0)\sim-T^{4/3}.
\label{eq:cubic-geodesic-branch}
\end{equation}
The geodesic approximation consequently defines a multi-sheeted function $E(T)$ of complex $T$. Analytic continuation from the Euclidean correlator selects a pair of complex-conjugate geodesics on the physical branch, which remains regular at $T=\pm T_{\rm crt}$. By contrast, the branch containing the real bouncing geodesic is reached only after analytic continuation around the relevant branch point and develops the singularity at real Lorentzian time $T=T_{\rm crt}$. The bouncing-geodesic singularity located at $T=T_{\rm crt}$ therefore resides on \textit{a secondary Riemann sheet} \footnote{Here the secondary-sheet singularity at $T=T_{\rm crt}$ should not be confused with the ramification points through which the principal and secondary Riemann sheets are connected. These ramification points arise from the multi-valued saddle structure of the geodesic approximation, whereas the singularity at $T=T_{\rm crt}$ is encountered only after analytically continuing onto the bouncing-geodesic branch, \ie the relevant secondary Riemann sheet. Locally, $G_{\mt{LR}}^{(\rm II)}(T,\bx=0)\sim (T-T_{\rm crt})^{-2\Delta}$, while the principal branch remains regular at the same projected value of $T$, with $G_{\mt{LR}}^{(\rm I)}(T_{\rm crt},\bx=0)$ finite on the principal sheet. A simple toy example is $f(z)=\frac{1}{1+\sqrt{z}}$. The branch point at $z=0$ connects the two Riemann sheets. By contrast, there is a second-sheet singularity at $z=1$ due to $f^{(\mathrm{II})}(z)=\frac{1}{1-\sqrt z} =-\frac{2}{z-1}+O(1)$. However, it is still regular with $f^{(\rm I)}(1)=1/2$ on the principal sheet.} of the semiclassical correlator. A detailed analysis of the associated saddle-selection rules and their Picard--Lefschetz interpretation will be presented in a forthcoming companion work \cite{nextpaper2026}.

The absence of the bouncing singularity on the principal Lorentzian sheet is required by the analytic structure of the physical thermal correlator, while its appearance after analytic continuation shows that the imprint of the black hole singularity remains encoded in the thermal correlator. From the boundary perspective, this interpretation has been sharpened by an OPE analysis: the multi-stress-tensor sector develops singularities precisely at the loci associated with bulk bouncing geodesics, and in the large-$\Delta$ limit their resummation reproduces the branch structure predicted by the geodesic analysis \cite{Ceplak:2024bja}. More generally, the relation between null propagation and singularities of bulk and boundary correlators extends beyond the strict geodesic approximation \cite{Grozdanov:2026cut}.

In this work, we keep the heavy-probe limit sufficiently sharp that the geodesic saddle structure remains well defined, and instead ask how the above imprint of the black hole singularity is modified once the strict semiclassical large-$c$ limit is relaxed. Before addressing this question, it is useful to reformulate the analytic structure directly in frequency space. As first emphasized in \cite{Festuccia:2005pi}, information about the black hole singularity is encoded in the high-frequency behavior of thermal correlators. In the next subsection, we make this connection explicit by relating the complex-time singularities to the asymptotic distribution of black hole QNMs.

%%%%%%%%%%%%%%%%%%%%%%%%%%%%%%%%%%%%%%%%%%%%%%%%
\subsection{Linear QNMs and complex-time singularities}
\label{subsec:qnm-singularities}
%%%%%%%%%%%%%%%%%%%%%%%%%%%%%%%%%%%%%%%%%%%%%%%%
We now turn to the frequency-space analytic structure of holographic thermal correlators, which makes clear why black hole quasinormal modes (QNMs) encode information about the black hole singularity through the analytic properties of thermal correlators. Throughout this subsection, we work primarily in the strict semiclassical limit $c\to\infty$, in which the two-sided thermal correlator $G_{\mt{LR}}$ and the heavy-state (HHLL) correlator $G_{\mt H}$ agree at leading order, as discussed in eq.~\eqref{eq:hhll-thermalization}. We therefore do not distinguish between them unless otherwise specified.

\subsubsection{Thermal Green functions and QNM poles}
To explicitly establish this connection, we first recall the relation between the two-sided correlator $G_{\mt{LR}}$ and the retarded Green function $G_{\mt R}$ through their common thermal spectral density, denoted by $\rho(\omega,\bk)$. For a scalar operator $\cO$, the two thermal Wightman functions in momentum space are defined by
\begin{equation}
G^>(t,\bk)=\langle\cO_{\bk}(t)\cO_{-\bk}(0)\rangle_\beta,\qquad G^<(t,\bk)=\langle\cO_{-\bk}(0)\cO_{\bk}(t)\rangle_\beta.
\end{equation}
The corresponding retarded Green function is defined as 
\begin{equation}
G_{\mt R}(t,\bk)=-\ii\theta(t)\left[G^>(t,\bk)-G^<(t,\bk)\right].
\end{equation}
Performing the Fourier transformation defined in eq.~\eqref{eq:fourier-conventions}, \ie $G(\omega,\bk)=\int\dd t\,\ee^{\ii\omega t}G(t,\bk)$, we define the commutator spectral density by
\begin{equation}
\rho(\omega,\bk)\equiv G^>(\omega,\bk)-G^<(\omega,\bk)=-2\operatorname{Im}G_{\mt R}(\omega+\ii0^+,\bk).
\label{eq:retarded-spectral-density}
\end{equation}
The retarded Green function is then determined by the dispersion relation\footnote{More precisely, the spectral density determines the retarded Green function only up to local contact terms, and the necessary UV subtractions must be included to make the dispersion integral convergent.}
\begin{equation}
G_{\mt R}(\omega,\bk)=\int_{-\infty}^{\infty}\frac{\dd\omega'}{2\pi}\,\frac{\rho(\omega',\bk)}{\omega-\omega'+\ii 0^+},\qquad \operatorname{Im} \omega >0,
\label{eq:retarded-dispersion-relation}
\end{equation}
The retarded Green function in momentum space may therefore be regarded as a function of complex frequency. Causality guarantees that $G_{\mt R}(\omega,\bk)$ is holomorphic in the upper half of the complex $\omega$-plane. It can then be analytically continued beyond this domain into the lower half-plane by crossing any real-frequency branch cuts. However, this may introduce isolated poles in the lower half-plane with $\operatorname{Im}\omega<0$. In the semiclassical black hole regime, these poles occur precisely at the QNMs, thereby relating the analytic structure of Green functions to the QNM pole structure studied below.

On the other hand, the same spectral density also determines the two-sided thermal correlator. The key point is that thermal equilibrium relates the two Wightman functions through the KMS condition. In frequency space, one can rewrite the KMS condition \eqref{eq:complex-kms} as
\begin{equation}
G^<(\omega,\bk)=\ee^{-\beta\omega}G^>(\omega,\bk).
\label{eq:kms-frequency}
\end{equation}
Consequently, the spectral density can be expressed as
\begin{equation}
\rho(\omega,\bk)=(1-\ee^{-\beta\omega})G^>(\omega,\bk)\,.
\end{equation}
Recall that the two-sided thermal correlator defined in section~\ref{subsec:thermal-analytic-structure} satisfies
\begin{equation}
G_{\mt{LR}}(T,\bk)=G^>\left(T-\frac{\ii\beta}{2},\bk\right),
\qquad
G_{\mt{LR}}(\omega,\bk)=\ee^{-\beta\omega/2}G^>(\omega,\bk).
\label{eq:GLR-Wightman-frequency}
\end{equation}
It is thus straightforward to obtain the following relation:
\begin{equation}
\boxed{G_{\mt{LR}}(\omega,\bk) = \frac{\rho(\omega,\bk)}{2\sinh\left(\frac{\beta\omega}{2}\right)} \,.}
\label{eq:GLR-spectral-density}
\end{equation}
Thus, the two-sided correlator and the retarded Green function are governed by the same spectral density. In particular, the real-frequency data of $G_{\mt{LR}}(\omega,\bk)$ determine the retarded Green function through eqs.~\eqref{eq:retarded-dispersion-relation} and \eqref{eq:GLR-spectral-density}.

The thermal factor appearing in this relation will play a very important role below. This universal thermal factor $1/\sinh(\beta\omega/2)$ provides the basic seed structure for the complex-time singularities of the thermal correlator $G_{\mt{LR}}(T,\bk)$. In the two half-planes of complex frequency, it admits the following series expansions:
\begin{equation}
\frac{1}{2\sinh(\beta\omega/2)}=
\begin{cases}
\quad\sum_{k=0}^{\infty}\ee^{-(k+\frac12)\beta\omega},
&
\operatorname{Re}\omega>0,
\\[3mm]
-\sum_{k=0}^{\infty}\ee^{(k+\frac12)\beta\omega},
&
\operatorname{Re}\omega<0.
\end{cases}
\label{eq:thermal-factor-expansion}
\end{equation}
The crucial point is that, after inverse Fourier transformation, each such exponential factor generates a genuine singularity of the analytically continued correlator at a corresponding complex time. The simplest illustrative example is
\begin{equation}
\int_0^\infty \dd\omega\,\ee^{-\ii\omega T}\ee^{-a_k\omega} = \frac{-\ii}{T-\ii a_k}, \qquad a_k>0,
\label{eq:seed-fourier-example}
\end{equation}
which exhibits a singularity at the complex time $T=\ii a_k$. The negative-frequency sector similarly generates the corresponding singularities in the lower half of the complex $T$ plane. Consequently, the two frequency sectors of the universal thermal factor naturally generate the infinite paired set of complex-time singularities located at  
\begin{equation}
\boxed{
T_{\rm seed}^{(\pm,k)} = \pm\ii\left(k+\frac12\right)\beta, \qquad k=0,1,2,\ldots
}
\label{eq:thermal-seed-singularities}
\end{equation}
These locations coincide with the thermal/OPE singularity lattice $T_{\mt{OPE}}^{(\pm,n)}$ in \eqref{eq:g12-ope-lattice} for the position-space correlator $G_{\mt{LR}}(T,\bx=0)$. As we will see below, these thermal singularities provide the seed locations from which additional complex-time singularities are generated by the asymptotic QNM structure.

The analytic structure of $G_{\mt R}(\omega,\bk)$ and the two-sided correlator $G_{\mt{LR}}(\omega,\bk)$ has a natural holographic interpretation. In the semiclassical bulk description, the retarded Green function is obtained by solving the linearized bulk wave equation with ingoing boundary conditions at the black hole horizon and the standard asymptotic boundary condition at the AdS boundary \cite{Nunez:2003eq,Kovtun:2005ev}. The QNM frequencies are conventionally written as
\begin{equation}
\text{QNMs}: \qquad \omega_n(\bk) = \Omega_n(\bk) - \ii\Gamma_n(\bk) \,,  \qquad \Gamma_n >0 \,,
\end{equation}
and correspond to the complex frequencies for which the associated homogeneous problem admits a nontrivial solution satisfying both boundary conditions. For simplicity, we label the infinite set of QNMs by a single index $n$. The real part $\Omega_n$ fixes the oscillation frequency, while the negative imaginary part $-\Gamma_n$ determines the decay rate.

The relation between QNMs and the pole structure of Green functions is particularly transparent in the time domain. Suppose that a semiclassical perturbation decays as 
\begin{equation}
G_{\mt R}(t,\bk)\sim\theta(t)\,\ee^{-\ii\Omega_n t}\ee^{-\Gamma_n t}.
\end{equation}
Fourier transformation then produces a singular contribution proportional to $(\omega-\omega_n)^{-1}$. Thus, the QNM frequencies appear as complex poles of the analytically continued retarded Green function,
\begin{equation}
\boxed{G_{\mt R}(\omega,\bk)\sim\frac{R_n(\bk)}{\omega-\omega_n(\bk)} \qquad \text{as}\qquad \omega\rightarrow\omega_n(\bk)\,.}
\label{eq:retarded-qnm-pole}
\end{equation}
Since $G_{\mt{LR}}(\omega,\bk)$ and $G_{\mt R}(\omega,\bk)$ are related through the same real-frequency spectral data, the analytic continuation that gives rise to the QNM poles of $G_{\mt R}$ also constrains the nontrivial complex-frequency structure relevant for $G_{\mt{LR}}$. As we show next, the asymptotic distribution of these QNM poles can be reorganized into a product representation whose large-frequency expansion generates the additional off-axis singularities of the analytically continued two-sided correlator.

%%%%%%%%%%%%%%%%%%%%%%%%%%%%%%%%%%%%%%%%%%%%%%%%%%%%%%%%%%%%%%
\subsubsection{From linear QNMs to complex-time singularities}
\label{subsubsec:linear-qnm-complex-time}

With the preceding ingredients in place, we can now make the connection between the asymptotic QNM spectrum and the complex-time singularities of the two-sided correlator more explicit. A particularly useful starting point is the thermal product formula derived in \cite{Dodelson:2023vrw}. A central result of \cite{Dodelson:2023vrw} is that, for holographic theories with a classical gravity dual, the analytically continued correlator $G_{\mt{LR}}(\omega,\bk)$ has two remarkable properties: it is a meromorphic function of the complex frequency $\omega$, whose poles are determined by the black hole QNMs, and it has no zeros. Equivalently, $1/G_{\mt{LR}}(\omega,\bk)$ is an entire function whose zeros are precisely the QNM frequencies. Moreover, its large-frequency behavior implies that this entire function is of order one. By the Hadamard factorization theorem,\footnote{The Hadamard factorization theorem states that any entire function of finite order can be represented as a canonical product over its zeros, multiplied by the exponential of a polynomial.} these properties strongly constrain $G_{\mt{LR}}(\omega,\bk)$ and allow it to be reconstructed from its QNM spectrum, up to an overall normalization prefactor. More explicitly, the symmetries of the thermal correlator imply that generic QNM poles occur in quartets,
\begin{equation}
\left(\omega_n(\bk),-\omega_n(\bk),\omega_n^\ast(\bk),-\omega_n^\ast(\bk)\right).
\end{equation}
Choosing one representative $\omega_n(\bk)$ from each quartet, the thermal product formula takes the form \cite{Dodelson:2023vrw}
\begin{equation}
G_{\mt{LR}}(\omega,\bk)=
\frac{G_{\mt{LR}}(0,\bk)}{\displaystyle
\prod_{\omega_n \in \rm{QNMs}}
\left(1-\frac{\omega^2}{\omega_n^2(\bk)}\right)
\left(1-\frac{\omega^2}{[\omega_n^\ast(\bk)]^2}\right)}\,.
\label{eq:thermal-product-formula}
\end{equation}
Here $n$ collectively denotes the discrete labels of the QNM spectrum at fixed $\bk$, and the product runs over one representative from each quartet. Purely imaginary QNMs can be incorporated separately by including only a single factor $\left(1-\omega^2/\omega_n^2\right)^{-1}$.

It is worth emphasizing that the continuation of $G_{\mt{LR}}(\omega,\bk)$ to a meromorphic function in the complex $\omega$-plane, as encoded in \eqref{eq:thermal-product-formula}, contains more information than the real-frequency spectral relation in eq.~\eqref{eq:GLR-spectral-density}. In particular, the thermal factor
$1/\sinh(\beta\omega/2)$ appearing in eq.~\eqref{eq:GLR-spectral-density} has apparent poles at the bosonic Matsubara frequencies
\begin{equation}
\omega=\ii \,\frac{2\pi n}{\beta},
\qquad n\in\mathbb Z.
\end{equation}
However, for the holographic correlator, the corresponding numerator vanishes at the same Matsubara frequencies. As a result, these apparent Matsubara poles are removable (this phenomenon is called pole-skipping). Consequently, the nontrivial poles of $G_{\mt{LR}}(\omega,\bk)$ are the QNM poles entering the thermal product formula above.

The thermal product formula makes clear that the complex-frequency structure of
$G_{\mt{LR}}(\omega,\bk)$ is encoded in the QNM spectrum. To determine the
corresponding complex-time singularities of $G_{\mt{LR}}(T,\bk)$, however, the relevant information is not the location of any individual QNM, but rather the collective asymptotic distribution of the QNMs at large overtone number. In the semiclassical holographic regime, the high-overtone QNM spectrum for a broad class of semiclassical AdS black holes typically organizes into asymptotically linear families,
\begin{equation}
    \omega_{n,a}(\bk) = a_{1,a}(\bk)\,n + a_{0,a}(\bk) + \mathcal O(n^{-1}),
    \qquad n\gg 1,
    \label{eq:asymptotic-linear-QNM}
\end{equation}
where $n$ denotes the overtone number and $a$ labels different QNM families. As shown in \cite{Dodelson:2025jff}, when combined with the thermal product formula, it is precisely this asymptotically linear distribution that, after Fourier transformation, gives rise to the characteristic complex-time singularities of the two-sided thermal correlator.

To make this relation explicit, let us first focus on a single exactly linear QNM
family,
\begin{equation}\label{eq:linearQNMs}
\boxed{\omega_n=a_1 n+a_0, \qquad a_0,a_1\in\mathbb C\,.}
\end{equation}
together with its symmetry-related counterparts $-\omega_n$, $\omega_n^\ast$, and $-\omega_n^\ast$. In particular, for a scalar operator of dimension $\Delta$ in an
$\mathrm{AdS}_{d+1}$ black brane, the asymptotic QNM spectrum is \cite{Dodelson:2025jff}
\begin{equation}
    \omega_n \simeq  \frac{4\pi}{\beta} \sin\left(\frac{\pi}{d}\right) e^{-\ii\pi/d}
    \left[ n+ \left(\frac{\Delta}{2}-\frac{d-2}{4} + \frac{\ii\log 2}{2\pi} \right) \right],
    \qquad n\gg1.
    \label{eq:AdS-linear-QNM}
\end{equation}
Following the general connection established in \cite{Dodelson:2025jff}, we now present a closely related derivation adapted to our later analysis, in which the Hadamard product of linear QNMs is reduced explicitly to Gamma functions and the resulting complex-time singularities are extracted from the Stokes sectors of these Gamma functions.

Applying the thermal product formula \eqref{eq:thermal-product-formula}, we first isolate the contribution associated with the QNM family $\omega_n$,
\begin{equation}
    \prod_{n=0}^{\infty}
    \left( 1-\frac{\omega}{a_1 n+a_0} \right)
    =
    \prod_{n=0}^{\infty} \frac{n-z(\omega)}{n+a_0/a_1},
    \label{eq:linear-QNM-product-z}
\end{equation}
where
\begin{equation}\label{eq:linear-z}
    z(\omega) \equiv \frac{\omega-a_0}{a_1}\,,
\end{equation}
is the continuous overtone variable obtained by inverting the linear QNM relation \eqref{eq:linearQNMs}. The infinite product can be evaluated by taking its logarithm, thereby converting it into an infinite sum. After absorbing the $\omega$-independent contribution into the overall normalization, the remaining sum can be evaluated using Hurwitz-zeta regularization,
\begin{equation}
    %\Sigma_0(z) \equiv
    \sum_{n=0}^{\infty}\log\!\left[n-z(\omega)\right]
    =-\left. \partial_s\zeta\!\left(s,-z(\omega)\right) \right|_{s=0}
    = -\log\Gamma\!\left[-z(\omega)\right] +\frac{1}{2}\log(2\pi),
    \label{eq:Sigma0-Hurwitz}
\end{equation}
where $\zeta(s,z)$ denotes the Hurwitz zeta function. Up to an $\omega$-independent normalization, the infinite product associated with one linear QNM family is therefore just proportional to $\Gamma[-z(\omega)]^{-1}$. Including the three symmetry-related QNM families and recalling that the thermal correlator contains the inverse of the corresponding QNM product, we then find
\begin{equation}
    \boxed{ G_{\mt{LR}}(\omega,\bk) \propto
    \Gamma[-z(\omega)] \Gamma[-z(-\omega)] \Gamma[-z^\ast(\omega)] \Gamma[-z^\ast(-\omega)],}
    \label{eq:GLR-linear-QNM-Gamma}
\end{equation}
with $z^\ast(\omega) =\frac{\omega-a_0^\ast}{a_1^\ast}$. Here $z^\ast(\omega)$ only denotes the function obtained by complex conjugating the QNM parameters $a_0$ and $a_1$; for generic complex $\omega$, it should not be confused with the complex conjugate of $z(\omega)$ itself. Equation~\eqref{eq:GLR-linear-QNM-Gamma} provides a useful bridge between the asymptotically linear QNM spectrum and the large-frequency structure of the thermal correlator. As we show next, the latter determines the locations of the singularities after Fourier transformation to complex time.

The connection between the linear QNM spectrum and the complex-time singularities is in fact encoded in the asymptotic behavior of the Gamma functions in \eqref{eq:GLR-linear-QNM-Gamma}. It is known that the Gamma function for large $|z|$ in the right half-plane is described by the standard Stirling expansion, namely 
\begin{equation}
    \Gamma(z) \sim \ee^{z\log z-z+\mathcal O(\log z)}\,, \qquad \operatorname{Re} z>0.
\end{equation}
For $z$ in the left half-plane, one may instead use the reflection formula\footnote{A simple but precise example for later illustration is 
\begin{equation*}
 \prod_{n=1}^{\infty} \left(1-\frac{z^2}{n^2}\right) = \frac{\sin\pi z}{\pi z } = \frac{1}{z \, \Gamma(z)\Gamma(1-z)} \,.
\end{equation*}
},
\begin{equation}
    \Gamma(z)\Gamma(1-z) = \frac{\pi}{\sin(\pi z)} \,, 
\end{equation}
and apply the Stirling expansion to $\Gamma(1-z)$. Up to overall constants and algebraic prefactors, this gives
\begin{equation}
    \Gamma(z) \sim \frac{\ee^{z\log(-z)-z+\mathcal O(\log(-z))}}{\ee^{\ii\pi z}-\ee^{-\ii\pi z}} \,, \qquad \operatorname{Re} z<0 \,. 
\end{equation}
Similar to the universal thermal factor appearing in the two-sided thermal correlator, the trigonometric prefactor plays an important role in the following analysis. For example, when $\operatorname{Im}z>0$, it admits the convergent expansion
\begin{equation}
    \frac{1}{\ee^{\ii\pi z}-\ee^{-\ii\pi z}} = -\ee^{\ii\pi z} \sum_{m=0}^{\infty} \ee^{2\pi\ii m z}\,, 
\end{equation}
where the terms with $m\geq1$ are exponentially suppressed and provide the relevant nonperturbative contributions to the large-$z$ asymptotics.

This structure also admits an interesting geometric interpretation in terms of Picard--Lefschetz theory, which will be particularly relevant for the analysis in our companion paper \cite{nextpaper2026}. Applying Picard--Lefschetz theory to an integral representation of the Gamma function, one finds that the leading Stirling expansion corresponds to the dominant saddle contribution. As $z$ crosses the Stokes line $\operatorname{Re}z=0$, the thimble decomposition of the corresponding integration cycle undergoes a Stokes jump, with infinitely many additional saddle-point thimbles entering the decomposition. The contributions of successive saddles differ by factors of $\ee^{\pm2\pi\ii z}$, giving rise to an infinite series of exponentially suppressed sectors. Their resummation reproduces the trigonometric factor appearing in the reflection formula. For detailed explanations see \cite{Harlow:2011ny}.

For concreteness, let us focus on the large positive-frequency limit $\omega\rightarrow+\infty$ along the real axis and choose the representative QNM family such that $\operatorname{Re}z(\omega)>0$. Up to an $\omega$-independent normalization, the four Gamma functions in \eqref{eq:GLR-linear-QNM-Gamma} can then be organized as
\begin{equation}
G_{\mt{LR}}(\omega,\bk)\sim
\frac{G_0(\omega,\bk)}
{\left[\ee^{-\ii\pi z(\omega)}-\ee^{\ii\pi z(\omega)}\right]
\left[\ee^{-\ii\pi z^\ast(\omega)}-\ee^{\ii\pi z^\ast(\omega)}\right]},
\qquad \omega\rightarrow+\infty,
\label{eq:Gamma-Stokes-prefactor}
\end{equation}
where $G_0(\omega,\bk)$ collects the ordinary Stirling asymptotics of the four Gamma functions. For the branch with $\operatorname{Im}z(\omega)>0$, the two trigonometric factors admit convergent geometric expansions, yielding
\begin{equation}
G_{\mt{LR}}(\omega,\bk)\sim
-\ee^{-2\pi\operatorname{Im}z(\omega)}G_0(\omega,\bk)
\sum_{n_+,n_-=0}^{\infty}
\exp\left\{2\pi\ii\left[n_+z(\omega)-n_-z^\ast(\omega)\right]\right\}.
\label{eq:Stokes-geometric-series}
\end{equation}
The crucial observation is that $z(\omega)$, defined by the linear QNM spectrum \eqref{eq:linearQNMs}, is also linear in $\omega$. Therefore, each term in this double series contains an exponential with a linear phase. Consequently, we can write
\begin{equation}
G_{\mt{LR}}(\omega,\bk)\sim
-\ee^{-2\pi\operatorname{Im}z(\omega)}G_0(\omega,\bk)
\sum_{n_+,n_-=0}^{\infty}
C_{n_+,n_-}\,
\ee^{\ii\omega(n_+v_++n_-v_-)},
\label{eq:Stokes-phase-vpm}
\end{equation}
where
\begin{equation}
v_+\equiv\frac{2\pi}{a_1},\qquad
v_-\equiv-\frac{2\pi}{a_1^\ast}=-v_+^\ast\,.
\label{eq:vpm-from-QNM}
\end{equation}
Here, $C_{n_+,n_-}$ denotes an $\omega$-independent constant and the prefactor $\ee^{-2\pi\operatorname{Im}z(\omega)}G_0(\omega,\bk)$ controls the overall large-frequency behavior.

\begin{figure}[t]
    \centering
    \includegraphics[width=5in]{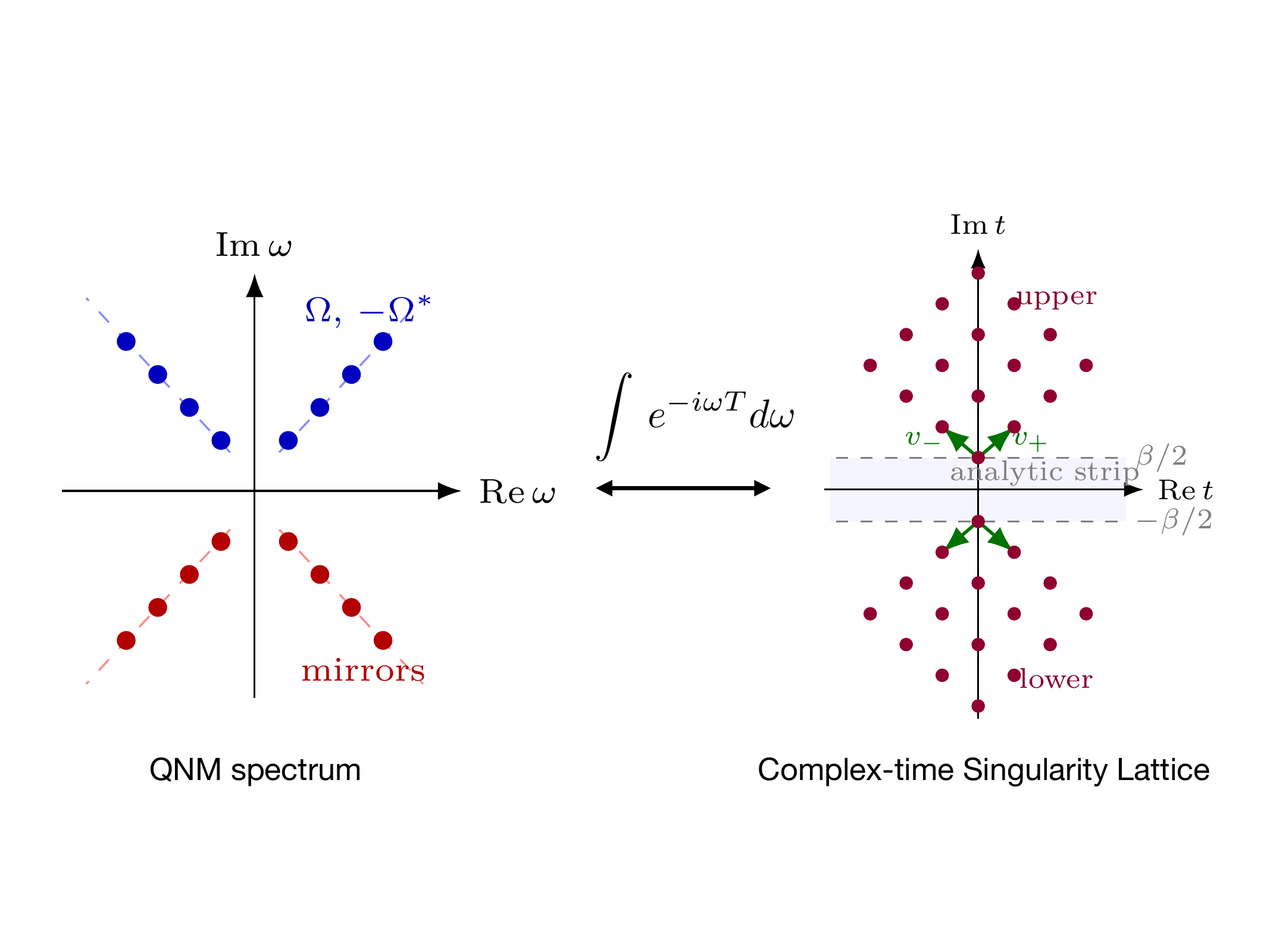}
    \caption{\textbf{Left:} linear QNM spectrum, \ie poles of $G_{\mt{LR}}(\omega,\bk)$ in the complex $\omega$-plane; \textbf{Right:} complex-time singularities of the analytically continued thermal correlator $G_{\mt{LR}}(T,\bk)$ in the complex $T$-plane.}
    \label{fig:fromQNMtosingularties}
\end{figure}

Having determined the large-frequency asymptotics of $G_{\mt{LR}}(\omega,\bk)$, we can now translate this frequency-space structure into the complex-time plane using the inverse Fourier transform
\begin{equation}
G_{\mt{LR}}(T,\bk)=\int\frac{\dd\omega}{2\pi}\,\ee^{-\ii\omega T}G_{\mt{LR}}(\omega,\bk).
\end{equation}
A basic feature of Fourier transforms is that the large-frequency behavior of a function controls its local structure in the conjugate time domain. First, let us recall that the universal thermal factor in the two-sided correlator \eqref{eq:GLR-spectral-density} generates an infinite sequence of seed singularities in the upper half of the complex $T$-plane,
\begin{equation}
T=T_{\rm seed}^{(+,k)}=\ii\left(k+\frac12\right)\beta,\qquad k=0,1,2,\ldots\,.
\end{equation}
Combining the thermal factor with the linear phase factors induced by the QNM spectrum, the contribution associated with a given pair $(n_+,n_-)$ takes the large-frequency form
\begin{equation}\label{eq:large-frequency-phase}
G_{\mt{LR}}(\omega,\bk)\supset \,\omega^\nu
\ee^{\ii\omega T_{\rm seed}^{(+,k)}}
\ee^{\ii\omega(n_+v_++n_-v_-)} ,
\end{equation}
where the power $\nu=2\Delta-d$ can be fixed by matching the dimension of the probe operator.\footnote{We also note that the power governing the bouncing-geodesic singularity of $G^{(\rm II)}_{\mt{LR}}(T,\bx=0)$ differs from that at a fixed spatial momentum. Considering the Fourier transformation 
\begin{equation*}
    G(\omega,\bx=0)=\int\frac{\dd^{d-1}k}{(2\pi)^{d-1}} G(\omega,\bk)\,,
\end{equation*}
and noting that the large-frequency correlator at fixed $\bk$ has the scaling form $G(\omega,\bk)\sim |\omega|^{2\Delta-d}F(\bk/\omega)$, one may rescale $\bk$ such that $\dd^{d-1}k=|\omega|^{d-1}\dd^{d-1}q$. The spatial momentum integration therefore contributes an additional factor of $|\omega|^{d-1}$, yielding
\begin{equation*}
    |\omega|^{2\Delta-d}\times|\omega|^{d-1}= |\omega|^{2\Delta-1}\,.
\end{equation*}
}
Taking each thermal singularity as a seed, the additional exponential phase shifts its location in the complex $T$-plane under the inverse Fourier transform. We therefore find that the analytically continued thermal correlator $G_{\mt{LR}}(T,\bk)$ contains a family of complex-time singularities,
\begin{equation}\label{eq:large-frequency-exp}
G_{\mt{LR}}(T,\bk) \underset{T\to T_{n_+,n_-}}{\sim} \frac{1}{(T-T_{n_+,n_-})^{\nu+1}} \,,
\end{equation}
with 
\begin{equation}
\boxed{T_{n_+,n_-}=T_{\rm seed}^{(+,k)}+n_+v_++n_-v_-\,. }
\label{eq:shifted-complex-time-singularities}
\end{equation}
As summarized in figure~\ref{fig:fromQNMtosingularties}, the universal thermal factor determines the seed singularities, while the asymptotically linear QNM spectrum generates their shifts through the vectors $v_\pm$. By time-reflection and complex-conjugation symmetry, the corresponding singularities in the lower half of the complex $T$-plane are located at $T=(T_{n_+,n_-})^\ast$.

In particular, the leading large-frequency behavior is controlled by the thermal suppression, 
\begin{equation}
   \boxed{ G_{\mt{LR}}(\omega,\bk) \sim  \ee^{-\beta\omega/2}\omega^{\nu}  \,, \qquad  \nu = 2\Delta -d \,.}
    \label{eq:GLR-large-frequency-seed}
\end{equation}
which corresponds to the first KMS singularity at $T_{\rm seed}^{(+,0)}=\frac{\ii\beta}{2}$ with $k=0$. Notice that the ordinary Stirling contributions collected in $G_0(\omega,\bk)$ contain no net exponential dependence on $\omega$. Comparison with \eqref{eq:Stokes-geometric-series} then shows that the thermal condition fixes
\begin{equation}
    \boxed{ 2\pi\,\operatorname{Im} \left(\frac{1}{a_1} \right) = \frac{\beta}{2} \quad \Longrightarrow \quad \operatorname{Im} v_+ = \operatorname{Im} v_- = \frac{\beta}{2} \,. }
\end{equation}
As a result, all higher seeds with $k>0$ are redundant, since they can already be generated from the fundamental seed $T^{(+,0)}_{\rm seed}=\ii\beta/2$ by repeated $v_++v_-$ shifts with $n_+=n_-$.

The connection to the bouncing-geodesic singularity becomes manifest once we specialize to the asymptotic QNM spectrum of the planar AdS black hole. Recalling \eqref{eq:AdS-linear-QNM}, the QNM slope gives rise to 
\begin{equation}
\frac{2\pi}{a_1} =\frac{\beta}{2}\cot\left(\frac{\pi}{d}\right)+ \ii \frac{\beta}{2}
=|T_{\rm crt}|+ \ii \frac{\beta}{2} \,,
\label{eq:vplus-Tstar}
\end{equation}
where $T_{\rm crt}$ is precisely the critical boundary time associated with the future-directed bouncing geodesic. The two elementary translation vectors can therefore be written in the particularly transparent form \cite{Dodelson:2025jff}
\begin{equation}
v_+=-T_{\rm crt}+\ii\frac{\beta}{2},\qquad v_-=T_{\rm crt}+\ii \frac{\beta}{2}.
\label{eq:vpm-Tstar}
\end{equation}
Their real parts are thus fixed directly by the characteristic bouncing time $T_{\rm crt}$, while their imaginary parts are fixed by the thermal scale $\beta$.

Using the fundamental seed singularity $T_{\rm seed}=\ii\beta/2$, the entire upper-half-plane family can then be written as
\begin{equation}
T_{n_+,n_-}=\frac{\ii\beta}{2}+n_+v_++n_-v_- =(n_+-n_-)|T_{\rm crt}| +\ii \frac{\beta}{2}(n_++n_-+1),
\qquad n_\pm\in\mathbb Z_{\geq0} \,, 
\label{eq:complex-time-lattice}
\end{equation}
As illustrated in figure~\ref{fig:fromQNMtosingularties}, we refer to eq.~\eqref{eq:complex-time-lattice} as the complex-time singularity lattice of the holographic correlator $G_{\mt{LR}}(T,\bk)$. The holographic interpretation of this lattice can be made more explicit by reorganizing the two non-negative integers as
\begin{equation}
m\equiv |n_+-n_-|,\qquad p\equiv \min(n_+,n_-).
\end{equation}
The singularity locations can then be written as
\begin{equation}
T_{m,p} =\pm m|T_{\rm crt}|+ \ii \frac{\beta}{2}(2p+m+1).
\label{eq:T_mp}
\end{equation}
In this parametrization, the two integers have a direct geometric interpretation: $m$ counts the number of bounces associated with the black hole singularity, whereas $p$ labels the corresponding thermal images. For $m=0$, the two branches coincide and the singularities lie entirely on the imaginary axis, reproducing the ordinary thermal OPE/KMS images. For $m=1$ and $p=0$, one finds the first pair of off-axis singularities,
\begin{equation}
T_{m=1,p=0}= \pm T_{\rm crt}+\ii\beta \,, 
\end{equation}
which are associated with the past- and future-directed single-bounce geodesics, respectively. More generally, the $p=0$ sequence with $m\geq1$ describes the successive multiple-bounce singularities, while $p>0$ generates their thermal images.

In summary, the QNM spectrum determines the complex-time singularities of the analytically continued two-sided thermal correlator $G_{\mt{LR}}(T,\bk)$, whose horizontal locations are controlled by the characteristic bouncing-geodesic time $T_{\rm crt}$. In this sense, the asymptotic linear QNM spectrum retains an imprint of the black hole interior: the thermal scale $\beta$ fixes the vertical spacing of the complex-time singularities, whereas the bouncing time $T_{\rm crt}$ fixes their horizontal displacement. %BTZ is the degenerate case $T_{\rm crt}=0$: its QNM families have purely imaginary overtone spacing, so the fixed-momentum singularities have no off-axis bouncing locus. Appendix~\ref{app:BTZ vs higher dim BH} compares the two structures in $G_{\mt{LR}}(T,\bk)$ and $G_{\mt{LR}}(\omega,\bk)$.

\iffalse
\begin{figure}[t]
\centering
\includegraphics[width=5in,pagebox=mediabox]{fig/g12_bulk_bouncing_geodesics_2.pdf}
\caption{Schematic relation between bouncing trajectories and the analytic continuation of a two-sided correlator. The upper diagram illustrates repeated bounces and thermal
windings, and the lower-left diagram represents the associated continuation
paths in complex time. The integers $m$ and $n$ label bounce and winding
numbers.}
\label{fig:intro-bouncing-trajectories}
\end{figure}
\fi

\section{Finite-$c$ QNM deformation and smoothing of singularities}
\label{sec:qnm-deformation}

In the previous section, we showed that the sharp complex-time singularities of the semiclassical holographic correlator $G_{\mt{LR}}(T,\bk)$ are encoded in the asymptotic distribution of QNMs. Specifically, an asymptotically linear QNM family produces the complex-time singularity lattice \eqref{eq:complex-time-lattice} through the thermal product formula \eqref{eq:thermal-product-formula} and its large-frequency asymptotic expansion \eqref{eq:large-frequency-exp}. This relationship naturally raises the central question of this paper: how are the imprints of the strict semiclassical black hole singularity modified by quantum effects? A key clue is that the linear QNM spectrum discussed in section~\ref{subsec:qnm-singularities} is a particular property of a classical bulk field propagating on a fixed black hole background, which corresponds to taking the strict semiclassical limit,
\begin{equation}
    c \sim \frac{1}{G_{\rm N}} \rightarrow \infty \,. 
\end{equation}
At finite $c$, or finite $\GN$, the probe is no longer fully decoupled from quantum gravitational fluctuations and the other degrees of freedom of the CFT. The corresponding resonances should therefore be dressed by these quantum corrections. Accordingly, finite-$c$ effects are expected to deform the pole trajectories of the analytically continued holographic correlator away from their strictly linear behavior, while also generating additional zero families. Since the linear QNM spectrum determines the corresponding complex-time singularities, it is natural to expect that finite-$c$ deformations of the QNM spectrum modify these singularities and smooth the boundary signatures associated with bouncing geodesics in the black hole interior.
  
The purpose of this section is to formulate this smoothing mechanism precisely. First, we introduce an effective meromorphic description of the correlator to characterize the finite-$c$ deformation of the semiclassical QNM trajectories. Following the strategy described in the previous section, we use the Hadamard product to determine how nonlinear pole--zero structures modify the large-frequency Stokes sectors. We will see that the leading nonlinear correction generates a Gaussian dressing of the semiclassical exponential sectors, which consequently smooths the corresponding bouncing-geodesic singularities. Finally, we explain why this mechanism does not smooth genuine OPE singularities.

%%%%%%%%%%%%%%%%%%%%%%%%%%%%%%%%%%%%%%%%%%%%%%%%
\subsection{Effective meromorphy and Hadamard factorization}
\label{subsec:deformed-qnm}
In this subsection, we formulate an effective description of the analytic structure of holographic thermal correlators away from the strict semiclassical limit. We begin by identifying the meromorphic structure that emerges at $c\rightarrow\infty$, and then discuss how finite-$c$ corrections modify this structure.

In the strict semiclassical limit $c\to\infty$, or equivalently $\GN\to0$, the holographic correlators considered in the previous section are determined by classical bulk fields propagating on a fixed black hole background. At fixed spatial momentum $\bk$, the bulk equations reduce to radial boundary-value problems in the complex frequency $\omega$. For the non-extremal asymptotically AdS black holes considered here, we impose ingoing boundary conditions at the horizon and normalizability at the asymptotic AdS boundary to select a distinguished bulk solution. Taking a bulk scalar field as an example, its ingoing bulk solution near the asymptotic boundary behaves as
\begin{equation}
    \phi_{\rm in}(r;\omega,\bk) \sim A(\omega,\bk)\,r^{\Delta-d}  + B(\omega,\bk)\,r^{-\Delta} +\cdots\,.
\end{equation}
According to the holographic dictionary, the retarded Green function is then given by
\begin{equation}
    G_{\mt R}(\omega,\bk) \propto \frac{B(\omega,\bk)}{A(\omega,\bk)} \,,
\end{equation}
up to normalization. The radial effective potential for the bulk wave equation, written in tortoise coordinate $z$, admits a special convergent expansion near the horizon $z\to\infty$,
\begin{equation}
    V(z)=\sum_{n=1}^{\infty}c_n\ee^{-n\frac{4\pi}{\beta} z}.
\end{equation}
This guarantees that the connection coefficients $A(\omega,\bk)$ and $B(\omega,\bk)$  admit meromorphic continuation in $\omega$; consequently, so does $G_{\mt R}(\omega,\bk)$.
Its poles occur at the zeros of $A(\omega,\bk)$, which are precisely the QNM frequencies $\{\omega_n(\bk)\}$, \ie
\begin{equation}
    A(\omega_n,\bk)=0
    \qquad\Longleftrightarrow\qquad
    \omega_n(\bk)\ \text{is a QNM}.
\end{equation}
The advanced Green function $G_{\mt A}(\omega,\bk)$ associated with the opposite horizon boundary condition thus admits an analogous meromorphic continuation. As discussed above, the two-sided thermal correlator is related to the same thermal spectral function $\rho(\omega,\bk)$. Specifically, the thermal KMS relation gives
\begin{equation}
    G_{\mt{LR}}(\omega,\bk) = \frac{\rho(\omega,\bk)}{2\sinh(\beta\omega/2)}
     =\frac{\ii\left[G_{\mt R}(\omega,\bk)-G_{\mt A}(\omega,\bk)\right]}{2\sinh(\beta\omega/2)}\,. 
\end{equation}
This in turn provides a natural meromorphic continuation of the two-sided correlator $G_{\mt{LR}}$ on the complex $\omega$-plane. The thermal product formula \eqref{eq:thermal-product-formula} also makes this analytic structure more explicit by reconstructing the analytically continued two-sided correlator directly from the corresponding QNM data.

At finite $c$, the exact thermal correlator in frequency space is no longer naturally described as a globally meromorphic function. Instead, its exact spectral representation is supported on discrete transition frequencies,
\begin{equation}
	\begin{split}
		G_{\mt{LR}}(\omega,\bk) &= \frac{2\pi}{Z(\beta)}\sum_{m,n} \ee^{-\frac{\beta}{2}(E_m+E_n)}\,\delta\!\left(\omega-(E_n-E_m)\right)\, |\mathcal{O}_{mn}(\bk)|^2 \,, 
	\end{split}
\label{eq:GLR-spectral-decomposition}
\end{equation}
where $\mathcal{O}_{mn}(\bk)=\mathcal{O}_{nm}(-\bk)^\ast$ are the momentum-space matrix elements obtained through the spatial Fourier transformation \eqref{eq:fourier-conventions}. An analogous spectral decomposition holds for the heavy-state correlator $G_{\mt H}$ in \eqref{eq:heavy-state-spectral-frequency}, with the pure heavy state taken to be a typical black hole microstate.

Since we focus on the black hole sector, the density of states is parametrically large. After coarse graining over the densely spaced microscopic levels, the exact discrete spectrum can therefore be approximated by a smooth spectral density. However, the analytic continuation of this coarse-grained spectrum may then develop non-meromorphic structures, such as threshold branch points and branch cuts. We therefore reorganize the finite-$c$ correlator in terms of 
\begin{equation}
   \boxed{ G(\omega,\bk) = G_{\rm mero}(\omega,\bk) + R(\omega,\bk) \,,}
    \label{eq:effective-meromorphic-decomposition}
\end{equation}
where $G_{\rm mero}$ denotes the meromorphic component of the analytically continued correlator, while $R$ collects the genuinely non-meromorphic remainder.

In particular, the regime of interest is the following intermediate-frequency window:
\begin{equation}
    1\ll \beta|\omega|\ll c \,.
    \label{eq:intermediate-frequency-window}
\end{equation}
The first condition places us in the large-frequency regime in which the asymptotic behavior relevant for the complex-time singularities becomes accessible. The second keeps the probe energy parametrically below the gravitational energy scale $\sim \mathcal O(c)$, so that its backreaction on the black hole background remains negligible. In this frequency regime, we assume that the meromorphic component $G_{\rm mero}(\omega,\bk)$ captures the leading analytic structure relevant to the semiclassical black hole, while the non-meromorphic remainder gives subleading corrections. Schematically, we expect
\begin{equation}
  |R(\omega,\bk)|\ll |G_{\rm mero}(\omega,\bk)|\,,
    \label{eq:effective-meromorphy-assumption}
\end{equation}
away from small neighborhoods of zeros of $G_{\rm mero}$.

There are several reasons to expect this hierarchy to hold at large $c$. We can recast the dressed correlation function with finite-$c$ corrections in terms of the Dyson form, 
\begin{equation}
    G(\omega,\bk)= \frac{1}{G_{0}^{-1}(\omega,\bk)-\Sigma(\omega,\bk)}\,.
    \label{eq:dressed-propagator-self-energy}
\end{equation}
Here, $G_0$ denotes the leading semiclassical propagator and $\Sigma(\omega,\bk)$ is the self-energy. Meromorphic contributions to $\Sigma$ dress the semiclassical resonances: they shift their locations, modify their residues, and can reorganize the meromorphic structure of the correlator. Such effects are naturally incorporated into $G_{\rm mero}$ (see section \ref{subsec:QNM deformation and non-Hermitian Hamiltonian} for more discussions).  By contrast, non-meromorphic contributions to $\Sigma(\omega,\bk)$, arising, for example, from multiparticle thresholds, generate branch points and associated branch cuts. However, we argue that such contributions are suppressed in the intermediate-frequency regime \eqref{eq:intermediate-frequency-window}. First, some non-meromorphic structures are separated from this regime by an $\mathcal O(c)$ energy scale. This is particularly transparent for the heavy-state correlator. Since the exact spectrum is bounded from below, its spectral support has a lower edge at $\omega_{\rm min}=E_0-E_{\beta}\sim-\mathcal O(c)$ for a typical black hole state. Similarly, thresholds associated with additional genuinely heavy sectors occur at energy differences of order $c/\beta$. As a result, the influence of such singular structures within the intermediate-frequency regime is parametrically suppressed by $1/c\ll1$. 
Second, non-meromorphic structures may also occur at frequencies of order the thermal scale, or even near $\omega=0$. Typical examples are multiparticle thresholds generated by the emission and reabsorption of gravitons or other bulk excitations. Although such effects are not suppressed by their separation in frequency space, they arise beyond the classical fixed-background approximation and are generated by bulk loop effects. Therefore, they are generally suppressed by the gravitational coupling $G_{\rm N}$. These two suppression mechanisms\footnote{More precisely, consider a continuum of intermediate states that generates a non-meromorphic contribution to the self-energy. The contribution can be represented schematically by the dispersive integral, \ie
\begin{equation*}
    \Sigma_{\rm cut}(\omega)\sim \GN \int_{\omega_{\rm th}}^\infty \dd\mu\, \frac{f(\mu)}{\omega-\mu+\ii0}\,.
\end{equation*}
If the relevant intermediate states differ in energy from the black hole state $E_\beta$ by an amount of order $c/\beta$, then $|\beta\omega_{\rm th}|\sim \mathcal O(c)$. Within the intermediate-frequency window, their nontrivial frequency dependence is therefore suppressed by
$|\omega/\omega_{\rm th}|\sim\beta|\omega|/c\ll1$. By contrast, intermediate states consisting of the original black hole together with soft gravitons or other light excitations can have energy differences $\mu=E_a-E_\beta\sim \mathcal O(\beta^{-1})$. Such contributions arising from bulk quantum loops are therefore suppressed by the gravitational coupling $\GN\sim1/c$.}
thus motivate our assumption that the genuinely non-meromorphic remainder $R$ is subleading in the intermediate-frequency regime, while $G_{\rm mero}$ captures the leading analytic structure relevant to the semiclassical black hole background.

Having argued that the non-meromorphic remainder is subleading in the intermediate-frequency regime, we can focus on the meromorphic component $G_{\rm mero}$. The advantage of this effective description is that the analytic structure of a meromorphic function can be organized systematically in terms of its poles and zeros. We further assume that $G_{\rm mero}(\omega,\bk)$ is of \emph{finite order} in $\omega$.\footnote{An entire function is of finite order $\rho$ if its growth at large $|\omega|=r$ is bounded by $\exp(r^{\rho+\epsilon})$ for any $\epsilon>0$. There are two physical reasons to expect such behavior in the semiclassical/EFT regime. First, we expect the relevant pole and zero families to have polynomial asymptotic trajectories in their overtone numbers, without superpolynomial growth of their degeneracies, so that finite-genus canonical products suffice. Second, the UV/OPE behavior of local CFT correlators exhibits large-frequency growth of the form shown in eq.~\eqref{eq:GLR-large-frequency-seed}. This suggests that the residual entire factor $\ee^{Q(\omega)}$ should not exhibit superexponential growth such as $\ee^{\ee^\omega}$.} By the Hadamard factorization theorem,\footnote{If one prefers not to assume finite order, the Weierstrass factorization theorem can instead be used to represent $G_{\rm mero}(\omega,\bk)$ in a similar form.} we can then recast the meromorphic function $G_{\rm mero}$ as a ratio of canonical products,
\begin{equation}
	\boxed{ G_{\rm mero}(\omega,\bk)=\ee^{Q(\omega,\bk)} \omega^{\ell}
	\frac{\displaystyle
          \prod_{m}E_{q}\!\left(\frac{\omega}{\zeta_m(\bk)}\right)}{
		  \displaystyle\prod_{n}
		  E_{p}\!\left(\frac{\omega}{\omega_n(\bk)}\right)} \,.} 
	\label{eq:hadamard-meromorphic}
\end{equation}
Here $\{\zeta_m(\bk)\}$ and $\{\omega_n(\bk)\}$ denote the nonzero zeros and poles, respectively. The integer $\ell\in\mathbb Z$ accounts for a possible zero or pole at $\omega=0$. The canonical factors $E_p(z)$ of genus $p$ are defined by
\begin{equation}
   	E_p(z)= (1-z) \prod_{k=1}^p \ee^{z^k/k} =  (1-z) \exp\!\left(z+\frac{z^2}{2}+\cdots+\frac{z^p}{p} \right),
    \qquad
    p\in\mathbb Z_{\geq 0},
\end{equation}
with $E_0(z)=1-z$. The finite integers $p$ and $q$ are chosen as the smallest nonnegative integers such that the corresponding canonical products over the poles and zeros converge. The exponent $Q(\omega,\bk)$ of the entire prefactor is taken as a polynomial in $\omega$ for a finite-order meromorphic function.

The Hadamard factorization \eqref{eq:hadamard-meromorphic} makes explicit that the pole--zero data and the entire prefactor encode distinct pieces of information about holographic correlators. At finite $c$, the poles $\omega_n(\bk)$ include the resonances continuously connected to the semiclassical QNMs, while additional pole and zero families may also appear. Their distributions determine the nontrivial meromorphic structure of the correlator, whereas the polynomial $Q(\omega,\bk)$ captures the residual entire contribution not fixed by the pole--zero locations alone and is constrained by the normalization and large-frequency asymptotics.

It is worth noting that additional symmetries further constrain the Hadamard representation \eqref{eq:hadamard-meromorphic}. In particular, reality and frequency-reflection properties relate different members of the pole and zero sets. For the two-sided thermal correlator $G_{\mt{LR}}(\omega,\bk)$ of a neutral Hermitian operator defined in eq.~\eqref{eq:GLR-spectral-decomposition}, one has the usual reality condition together with frequency-reflection symmetry. Consequently, its generic poles and zeros should occur in symmetry-related quartets,
\begin{equation}
    \{\omega_n,-\omega_n,\omega_n^\ast,-\omega_n^\ast\},
    \qquad
    \{\zeta_a,-\zeta_a,\zeta_a^\ast,-\zeta_a^\ast\}.
\end{equation}
These symmetries constrain not only the pole--zero distributions but also the allowed polynomial $Q(\omega,\bk)$ and the behavior of the correlator at $\omega=0$. For comparison, the HHLL correlator $G_{\mt H}$ satisfies $G_{\mt H}(t)^\ast=G_{\mt H}(-t)$, as follows directly from its spectral representation~\eqref{eq:heavy-state-spectral-time}. Equivalently, its analytic continuation obeys $G_{\mt H}(\omega^\ast)^\ast=G_{\mt H}(\omega)$, implying that its poles and zeros always appear in complex-conjugate pairs. In general, however, $G_{\mt H}$ is not even under either time or frequency reflection. In the leading semiclassical regime, where the heavy-state correlator thermalizes, it is related to the two-sided thermal correlator by
\begin{equation}
    G_{\mt{LR}}^{(0)}(\omega,\bk)= \ee^{-\beta\omega/2}G_{\mt{H}}^{(0)}(\omega,\bk)\,.
\end{equation}
Since the exponential factor is entire and nowhere vanishing, the two correlators have the same pole-zero data in the semiclassical limit. Their different frequency-reflection properties are instead encoded in the entire prefactor of the Hadamard representation.

The strict semiclassical thermal correlator discussed in section~\ref{subsec:qnm-singularities} provides a particularly simple specialization of the general Hadamard representation. In this limit, the relevant meromorphic correlator is of order one, contains no additional zero families, and is regular and non-vanishing at $\omega=0$, so that $\ell=0$. Its poles are therefore determined entirely by the QNM frequencies and their symmetry-related counterparts. A stronger property of the semiclassical holographic QNM spectrum is that each asymptotic QNM trajectory is linear in a single overtone number,
\begin{equation}
    \omega_n^{(0)} \sim a_1 n+ a_0,  \qquad  n\gg1\,.
\end{equation}
For such a linear pole sequence, the genus of the corresponding canonical product is $p=1$, since
\begin{equation}
    \sum_n\frac{1}{|\omega_n^{(0)}|^{p+1}}\sim \sum_n\frac{1}{n^{p+1}}
\end{equation}
diverges for $p=0$ but converges for $p=1$. Furthermore, frequency-reflection symmetry allows opposite poles to be paired, yielding 
\begin{equation}
    E_1\left({\omega}/{\omega_n^{(0)}}\right)E_1\left(-{\omega}/{\omega_n^{(0)}}\right) =1-\frac{\omega^2}{(\omega_n^{(0)})^2}\,,
\end{equation}
where the exponential factors cancel pairwise. A similar pairing applies to the complex-conjugate QNM poles. Finally, an order-one Hadamard representation allows $Q(\omega,\bk)$ to be at most linear in $\omega$, while the frequency-reflection symmetry of $G_{\mt{LR}}$ eliminates its linear part. The remaining part is therefore a constant that can be fixed by normalization at $\omega=0$. The general Hadamard representation of the holographic thermal correlator $G_{\mt{LR}}(\omega,\bk)$ then naturally reduces to the thermal product formula~\eqref{eq:thermal-product-formula}.

In summary, we have shown that, within the intermediate-frequency regime, the analytically continued holographic correlator is determined to leading order by the Hadamard representation of $G_{\rm mero}$ in \eqref{eq:hadamard-meromorphic}. In other words, one can reconstruct the correlator from the following information:
\begin{equation}
\boxed{
\begin{array}{ccl}
\{\omega_n(\bk)\} &:& \text{poles},\\
\{\zeta_m(\bk)\} &:& \text{zeros},\\
\ell &:& \text{zero or pole at }\omega=0,\\
Q(\omega,\bk) &:& \text{asymptotic information}.
\end{array}}
\end{equation}
In particular, in the strict semiclassical limit, the asymptotic QNM trajectories are linear in the overtone number. As shown in section~\ref{subsec:qnm-singularities}, the resulting large-frequency exponential factors generate the sharp lattice of complex-time singularities of the analytically continued correlator. The generic Hadamard representation \eqref{eq:hadamard-meromorphic} then provides a natural framework for extending this analysis away from the strict semiclassical limit. The central question is thereby translated into determining how finite-$c$ effects deform the asymptotic distributions of poles and zeros, and how these deformations modify the associated large-frequency factors. As we will show below, these modifications directly control the fate of the corresponding complex-time singularities.

%%%%%%%%%%%%%%%%%%%%%%%%%%%%%%%%%%%%%%%%%%%%%%%%
%%%%%%%%%%%%%%%%%%%%%%%%%%%%%%%%%%%%%%%%%%%%%%%%
\subsection{Deformed QNM trajectories and singularity smoothing}
\label{subsec:QNM-deformation}
%%%%%%%%%%%%%%%%%%%%%%%%%%%%%%%%%%%%%%%%%%%%%%%%

We now parameterize the generic finite-$c$ deformation of the asymptotic QNM spectrum and investigate its consequences for the complex-time singularities discussed in the previous section. Within the effective meromorphic description developed above, this amounts to determining how the pole/zero trajectories entering the Hadamard representation are deformed away from their strict semiclassical form. 

In the following discussion, we first isolate the contribution of the deformed poles. This is sufficient to exhibit the smoothing mechanism by which a nonlinear QNM trajectory modifies the corresponding Stokes sector. At finite $c$, the complete Hadamard representation also contains additional zero families, whose inverse trajectories contribute in an analogous manner with the opposite sign in $\log G_{\rm mero}$. Their combined effect is essential for the global UV consistency of the correlator and, in particular, for preserving the singularities required by the OPE. We return to these constraints in section~\ref{subsec:protected-singularities}.

For simplicity, we first focus on a single family of QNM poles parametrized by one overtone number $n$. In the strict semiclassical limit, its large-overtone trajectory is asymptotically linear,
\begin{equation}
    \omega_n^{(0)} \sim a_0+a_1 n, \qquad  n\gg 1.
    \label{eq:classical-linear-qnm-leading}
\end{equation}
For the planar AdS black hole, this behavior is realized explicitly by the asymptotic spectrum in eq.~\eqref{eq:AdS-linear-QNM}. At finite $c$, we parameterize the dressed trajectory as
\begin{equation}
    \omega_n = \omega_n^{(0)}  + \delta\omega_n,
    \qquad
    \delta\omega_n\rightarrow0 \quad\text{as}\quad  c\rightarrow\infty.
    \label{eq:finite-c-qnm-deformation}
\end{equation}
The essential new feature is that $\delta\omega_n$ need not remain linear in $n$. Finite-$c$ interactions and gravitational dressing can therefore bend the asymptotic pole trajectory in the complex $\omega$-plane. To make this deformation explicit, let us consider a one-dimensional polynomial trajectory, 
\begin{equation}
   \boxed{ \omega_n=P_c(n)=\sum_{r=0}^{K} A_r(c,\bk) \,n^r\,.} 
    \label{eq:polynomial-qnm-trajectory}
\end{equation}
This trajectory naturally generalizes the linear QNM spectrum discussed in the previous section. Here the complex coefficients $A_r$ may depend on $c$, the spatial momentum $\bk$, and the particular pole family under consideration. The precise complex coefficients, as well as the detailed form of the finite-$c$ deformation, are of course theory dependent. To be consistent with the semiclassical limit, we require that the QNM trajectory reduce to the linear one in eq.~\eqref{eq:classical-linear-qnm-leading} in the strict semiclassical limit $c \to \infty$. In terms of the overtone number, the intermediate-frequency regime corresponds to
\begin{equation}
    1\ll n\ll c,
\end{equation}
where the large-overtone asymptotics relevant for the complex-time singularities are accessible while the nonlinear finite-$c$ corrections remain under perturbative control. We can therefore expand the QNM trajectory as 
\begin{equation}
    \omega_n = P_c(n) = a_0+ a_1 n + \frac{1}{c}\sum_{r=2}^{K}a_r n^r +  \mathcal O(c^{-2}),
    \label{eq:polynomial-qnm-large-c}
\end{equation}
where possible $1/c$ corrections to the constant and linear terms have been absorbed into $a_0$ and $a_1$. The simplest nontrivial deformation is a quadratic bending,
\begin{equation}
    \omega_n = a_0+a_1n+\frac{a_2}{c}n^2+\cdots \,, 
    \label{eq:quadratic-qnm-deformation}
\end{equation}
for which the nonlinear correction remains perturbative relative to the linear semiclassical trajectory as long as $n\ll c$.

For the following analysis, it is more convenient to invert the QNM trajectory and regard the overtone number $n=z(\omega)$ as a function of frequency. Given the polynomial trajectory $P_c(n)$ in eq.~\eqref{eq:polynomial-qnm-trajectory}, we therefore consider the inverse equation
\begin{equation}
    P_c(z)=\omega \,.
    \label{eq:inverse-trajectory-definition}
\end{equation}
For a degree-$K$ polynomial, this equation admits $K$ roots,
\begin{equation}
    z_\mu(\omega), \qquad \mu=1,\ldots,K \,,
    \label{eq:inverse-root-branches}
\end{equation}
which define the corresponding inverse branches. Unlike the linear case, the finite-$c$ polynomial trajectory therefore gives rise to multiple inverse branches. Among them, we distinguish the branch $z_{\rm sc}(\omega)$ that continuously approaches the inverse semiclassical QNM trajectory as $c\rightarrow\infty$. To characterize this branch explicitly, let us consider the finite-$c$ deformation \eqref{eq:polynomial-qnm-large-c} whose perturbative inversion is given by 
\begin{equation}
    z_{\rm sc}(\omega) = z^{(0)}(\omega) - \frac{1}{a_1c} \sum_{r=2}^{K}a_r
    \left[z^{(0)}(\omega)\right]^r
    +\mathcal O(c^{-2})\,,
    \label{eq:inverse-qnm-trajectory}
\end{equation}
where $z^{(0)}(\omega)=(\omega-a_0)/a_1$ denotes the inverse semiclassical trajectory. For the simplest quadratic deformation in eq.~\eqref{eq:quadratic-qnm-deformation}, this reduces to
\begin{equation}\label{eq:zsc-quadratic}
    z_{\rm sc}(\omega)=z^{(0)}(\omega) -  \frac{a_2}{a_1c}   \left[z^{(0)}(\omega)\right]^2 +\cdots \,.
\end{equation}
However, the remaining $K-1$ branches exhibit a qualitatively different large-$c$ behavior. For the polynomial deformation (with $a_1a_K\neq0$), these roots scale as
\begin{equation}
    z_{{\rm UV},\mu}(\omega) \sim \left( -\frac{a_1}{a_K} \,c \right)^{\frac{1}{K-1}} e^{ 2\pi \ii \mu /(K-1)}+\mathcal O(1)\,, 
    \qquad
    \mu=0,\ldots,K-2 \,.
\end{equation}
Their characteristic magnitude corresponds to the overtone scale $n\sim c^{1/(K-1)}$, at which the highest-degree correction becomes comparable to the semiclassical linear term and the perturbative expansion ceases to be reliable. Although the polynomial inverse equation admits $K$ algebraic branches\footnote{For any physical QNM frequency $\omega_n=P_c(n)$, the polynomial equation can be factorized as 
\begin{equation*}
P_c(z)-\omega_n=P_c(z) -P_c(n) = (z-n)R_{K-1}(z;n) \,, 
\end{equation*}
where $R_{K-1}(z;n)$ is a polynomial of degree $K-1$. Therefore, one inverse root is exactly $z_{\rm phys}=n$, while the remaining $K-1$ roots are determined by $R_{K-1}(z;n)=0$. These additional roots are merely algebraic solutions and do not represent independent physical QNM families.}, the physical QNM spectrum defines a unique inverse correspondence $z_{\rm phys}(\omega_n)=n \in\mathbb Z_{\geq0}$. Within the semiclassical regime, this is represented by the distinguished branch $z_{\rm sc}(\omega)$, which continuously reduces to the inverse linear QNM trajectory as $c\to\infty$.

More generally, within the intermediate-frequency regime, the semiclassically connected branch admits a local expansion of the form
\begin{equation}
    z_{\rm sc}(\omega)= z^{(0)}(\omega) + \delta z_{\rm sc}(\omega) =\sigma_0+\sigma_1\omega+\sigma_2\omega^2+\sigma_3\omega^3+\cdots,
    \label{eq:inverse-qnm-local-expansion}
\end{equation}
where the nonlinear coefficients $\sigma_{r\geq2}$ vanish in the strict semiclassical limit. The distinction between the semiclassical and finite-$c$ regimes is thus encoded directly in the frequency dependence of the inverse trajectory. At $c=\infty$, the linear function $z_{\rm sc}(\omega)$ generates pure exponential Stokes factors corresponding to linear shifts in complex time. At finite $c$, nonlinear corrections modify these factors: a quadratic term produces Gaussian-type dressing in frequency space, while higher-order terms generate more general nonlinear exponential contributions.

To establish this connection explicitly, we can now generalize the analysis of section~\ref{subsec:qnm-singularities} to deformed QNM trajectories. First, note that the polynomial trajectory \eqref{eq:polynomial-qnm-trajectory} admits the following factorization: 
\begin{equation}
    P_c(n)-\omega = A_K\prod_{\mu=1}^{K}
    \left[n-z_\mu(\omega)\right]\,.
\end{equation}
The same factorization at $\omega=0$ then allows the product associated with this pole family in the Hadamard representation~\eqref{eq:hadamard-meromorphic} to be expressed as
\begin{equation}
    \prod_{n=0}^{\infty}\left(1-\frac{\omega}{\omega_n}\right)
    =
    \prod_{\mu=1}^{K} \prod_{n=0}^{\infty}\frac{n-z_\mu(\omega)}{n-z_\mu(0)} 
    = 
    \prod_{\mu=1}^{K}\frac{\Gamma[-z_\mu(0)]}{\Gamma[-z_\mu(\omega)]} \,, 
    \label{eq:polynomial-product-Gamma}
\end{equation}
Here, we have applied the same Hurwitz-zeta regularization used in eq.~\eqref{eq:Sigma0-Hurwitz} to obtain the last equality.

The Gamma-function representation~\eqref{eq:polynomial-product-Gamma} allows us to extend the Stokes analysis of section~\ref{subsec:qnm-singularities} to nonlinear QNM trajectories. As in the linear case, the relevant Gamma functions generate exponentially suppressed contributions involving $\exp[\pm 2\pi\ii z_\mu(\omega)]$. In general, each inverse root $z_\mu(\omega)$ can contribute its own Stokes factors, and the complete structure involves combinations of the different branches. 
Our primary interest, however, lies in the finite-$c$ deformation of the semiclassical complex-time singularities, which originate from the semiclassical limit of the distinguished inverse branch $z_{\rm sc}(\omega)$. We therefore focus on the corresponding Stokes contributions, which can be decomposed into their semiclassical and finite-$c$ parts as
\begin{equation}
   \boxed{\frac{1}{\Gamma(-z_{\rm sc}(\omega))} \rightarrow \mathcal S_m(\omega)
    =\ee^{2\pi\ii m z_{\rm sc}(\omega)}
    =C_m \ee^{\ii\omega mv}\ee^{2\pi\ii m\delta z_{\rm sc}(\omega)},
    \qquad m=1,2,\ldots,}
    \label{eq:deformed-single-Stokes}
\end{equation}
where 
\begin{equation}
    C_m=\ee^{-2\pi\ii m a_0/a_1},
    \qquad
    v=\frac{2\pi}{a_1}\,.
    \label{eq:single-Stokes-parameters}
\end{equation}

As discussed in section~\ref{subsec:qnm-singularities}, the linear dependence on $\omega$ generates the complex-time translations of the seed singularities, whereas the nonlinear correction modifies their frequency-space profiles. To illustrate this effect, substituting the leading quadratic deformation~\eqref{eq:zsc-quadratic} into the Stokes factor $\mathcal{S}_m(\omega)$ \eqref{eq:deformed-single-Stokes} yields
\begin{equation}
    \mathcal{S}_m(\omega) =
    C_m \ee^{\ii\omega mv}
    \exp\left[-\kappa_m(\omega-a_0)^2+\mathcal O\left(\frac{m(\omega-a_0)^3}{c^2}\right)\right],
    \qquad
    \kappa_m=\frac{2\pi\ii m a_2}{a_1^3c}\,.
    \label{eq:quadratic-deformed-Stokes}
\end{equation}
The leading quadratic deformation introduces an additional Gaussian-type factor in frequency space, providing a natural mechanism for smoothing out the semiclassical complex-time singularities.

To examine this effect explicitly, we consider the contribution to the analytically continued correlator associated with a particular semiclassical complex-time singularity located at $T=T_m$. Suppressing the overall normalization, its leading behavior can be written as
\begin{equation}
   \boxed{ G(T,\bk)\supset I_\nu(\Delta T_m,\kappa_m)
    =\int_0^\infty\dd\omega\,\omega^\nu
    \exp\left[-\kappa_m\omega^2-\ii\omega\Delta T_m\right],}
    \label{eq:single-Stokes-smoothing-integral}
\end{equation}
where $\Delta T_m=T-T_m^{(c)}$, with
\begin{equation}
    T_m^{(c)}=T_m-2\ii\kappa_m a_0\,.
    \label{eq:finite-c-Stokes-location}
\end{equation}
Here $T_m$ denotes the location of the corresponding semiclassical complex-time singularity, while $T_m^{(c)}$ incorporates the leading finite-$c$ correction arising from the quadratic deformation. Convergence of the integral requires $\operatorname{Re}\kappa_m>0$ and $\operatorname{Re}\nu>-1$. In the absence of the quadratic correction, the large-frequency contribution produces the familiar power-law singularity at $T=T_m$. At finite $c$, however, the Gaussian factor suppresses the large-frequency modes responsible for this divergence, leading to a regular profile near the would-be singularity.

\begin{figure}[t]
    \centering
    \includegraphics[width=4in]{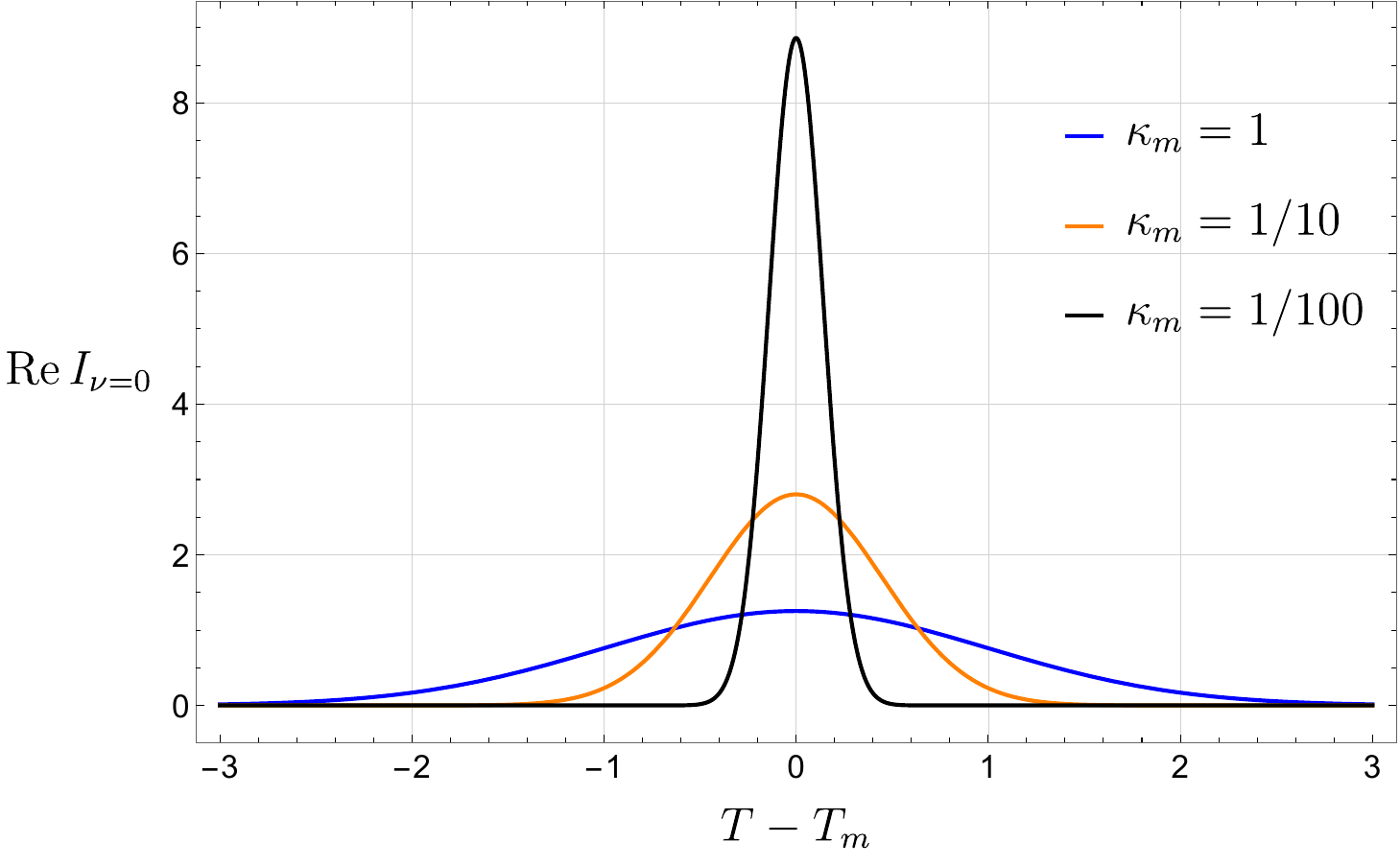}
    \caption{Real part of the Gaussian-regularized integral $I_0(\Delta T_m,\kappa_m)$. This is the simplest example illustrating the smoothing of the semiclassical complex-time singularity for $\operatorname{Re}\kappa_m>0$, with $|\kappa_m| \sim \beta^2/c$.}
    \label{fig:GaussianIntegral}
\end{figure}

In fact, for $\operatorname{Re}\kappa_m>0$, the Gaussian-regulated integral in eq.~\eqref{eq:single-Stokes-smoothing-integral} is not only finite at $\Delta T_m=0$, but also defines an entire function of the complex variable $\Delta T_m$. Thus, the semiclassical singularity is completely removed within this Gaussian approximation. To see this explicitly, for $\nu=0$, the integral can be evaluated as
\begin{equation}
I_0(\Delta T_m,\kappa_m)=\frac{\sqrt{\pi}}{2\sqrt{\kappa_m}}\exp\left[-\frac{\Delta T_m^2}{4\kappa_m}\right]\operatorname{erfc}\left(\frac{\ii\Delta T_m}{2\sqrt{\kappa_m}}\right),
\label{eq:Gaussian-integral-erfc}
\end{equation}
where $\operatorname{erfc}$ denotes the complementary error function. More generally, for $\operatorname{Re}\nu>-1$, the result is expressed in terms of the parabolic cylinder function $D_a(z)$,
\begin{equation}
I_\nu(\Delta T_m,\kappa_m)=\frac{\Gamma(\nu+1)}{(2\kappa_m)^{(\nu+1)/2}}\exp\left[-\frac{\Delta T_m^2}{8\kappa_m}\right]D_{-\nu-1}\left(\frac{\ii\Delta T_m}{\sqrt{2\kappa_m}}\right).
\label{eq:Gaussian-integral-parabolic}
\end{equation}
For nonnegative integer $\nu$, this expression can equivalently be obtained by applying $\ii^\nu\partial_{\Delta T_m}^\nu$ to $I_0$. In particular, at the center of the regularized profile, one finds
\begin{equation}
I_\nu(0,\kappa_m)=\frac{\Gamma\left(\frac{\nu+1}{2}\right)}{2\kappa_m^{(\nu+1)/2}},
\label{eq:Gaussian-integral-center}
\end{equation}
which remains finite for any nonzero $\kappa_m$ satisfying the convergence condition. In contrast, the strict semiclassical limit $\kappa_m\to0$ recovers the original power-law singularity,
\begin{equation}
I_\nu(\Delta T_m,0)=\lim_{\epsilon\to0^+}\frac{\Gamma(\nu+1)}{(\epsilon+\ii\Delta T_m)^{\nu+1}}\sim\frac{1}{(T-T_m)^{\nu+1}}.
\label{eq:Gaussian-integral-semiclassical}
\end{equation}
The Gaussian deformation therefore replaces the sharp semiclassical singularity with a regular profile characterized by the width
\begin{equation}
\boxed{|\Delta T_m|\sim\sqrt{|\kappa_m|}\sim\frac{\beta}{\sqrt c}\,,}
\label{eq:single-Stokes-smoothing-scale}
\end{equation}
for fixed $m$. As illustrated in figure~\ref{fig:GaussianIntegral}, the regularized profile becomes increasingly localized as $\kappa_m$ decreases, recovering the semiclassical singularity in the nonuniform limit $c\to\infty$.

The above analysis assumes $\operatorname{Re}\kappa_m>0$, for which the Gaussian factor suppresses large frequencies along the positive real axis. However, the quadratic coefficient induced by finite-$c$ QNM deformations is generally complex, and its real part need not be positive. When $\operatorname{Re}\kappa_m<0$, the quadratic approximation instead produces exponential growth, making the corresponding real-frequency integral divergent. This does not necessarily imply a singularity in the exact finite-$c$ correlator. Rather, it indicates that the truncated inverse QNM trajectory cannot be extrapolated to arbitrarily large frequencies. Indeed, the UV behavior of the full thermal correlator is constrained by the KMS relation and short-distance OPE, as shown in eq.~\eqref{eq:GLR-large-frequency-seed}. The apparent anti-Gaussian growth must therefore be resolved by the UV completion of the Stokes expansion, including the higher-order frequency dependence and, where necessary, the contributions from other pole-zero sectors.

One possible resolution is that higher-order corrections provide sufficient suppression along the original real-frequency contour. To illustrate this mechanism, suppose that the inverse trajectory admits a finite polynomial representation valid throughout the positive real-frequency regime,
\begin{equation}
z_{\rm sc}(\omega)=\sigma_0+\sigma_1\omega+\sum_{r=2}^{N}\sigma_r\omega^r,\qquad
\mathcal S_m(\omega)=\ee^{2\pi\ii m z_{\rm sc}(\omega)}\,.
\label{eq:Smomega}
\end{equation}
For the $m>0$ sector, the sufficient damping condition
\begin{equation}
\operatorname{Im}\sigma_N>0
\label{eq:highest-order-Stokes-damping}
\end{equation}
ensures that the highest-order term produces superlinear exponential suppression at large positive frequencies. After absorbing the constant and linear contributions into the overall amplitude and the shifted singularity location $T_m$, the corresponding Fourier integral becomes
\begin{equation}
    I_\nu(\Delta T_m) =\int_0^\infty\dd\omega\,\omega^\nu \exp\left[2\pi\ii m\sum_{r=2}^{N}\sigma_r\omega^r -\ii\omega (T-T_m) \right].
    \label{eq:polynomial-Stokes-smoothing-integral}
\end{equation}
For $\operatorname{Re}\nu>-1$, the superlinear damping dominates the Fourier kernel for any finite complex time $T$, so the integral defines an entire function. Thus, within this polynomial model, sufficiently strong higher-order suppression provides a direct mechanism for smoothing the semiclassical complex-time singularity.

However, superlinear damping along the real-frequency axis is not a necessary condition for smoothing. More generally, the exact inverse QNM trajectory need not admit a finite polynomial representation, and its local expansion may contain an inverted Gaussian even when the full Stokes factor has acceptable UV behavior. In such cases, the quadratic approximation can still provide a controlled description of the smoothing mechanism through an appropriate deformation of the integration contour into the complex-frequency plane. The key motivation comes from Picard--Lefschetz theory, which allows the original integration cycle to be decomposed into complex steepest-descent contours associated with the saddle points of the integral action \cite{Witten:2010cx,nextpaper2026}. More precisely, after complexifying the original real integration variable $x$ to $z$, the physical integration cycle can be expressed as a sum of Lefschetz thimbles in the complex $z$-plane,
\begin{equation}
    \mathcal{C}_{\rm phys}\sim\sum_\sigma n_\sigma\mathcal J_\sigma,\qquad
    \int_{\mathcal C_{\rm phys}}\dd x\,\ee^{-S(x)}
    =\sum_\sigma n_\sigma\int_{\mathcal J_\sigma}\dd z\,\ee^{-S(z)},
    \label{eq:UV-contour-PL-decomposition}
\end{equation}
where $\mathcal J_\sigma$ denotes the downward Lefschetz thimble associated with the saddle point $z_\sigma$, and $n_\sigma$ is the corresponding intersection number.\footnote{More precisely, $n_\sigma=\langle\mathcal C_{\rm phys},\mathcal K_\sigma\rangle$, where $\mathcal K_\sigma$ denotes the upward Lefschetz thimble associated with $z_\sigma$.} Thus, even when the original physical contour $\mathcal{C}_{\rm phys}$ lies along the real axis, the equivalent thimble representation generally involves integration along complex contours. This suggests that the appropriate (complex) integration contour of the effective theory could be inherited from the thimble decomposition of the full UV-complete theory. Schematically, we can express the relation between the UV-complete integral and its effective description in terms of 
\begin{equation}
    \int_{\mathcal C_{\mt{UV}}}\dd\omega\,\ee^{-S_{\mt{UV}}(\omega)}
    \simeq
    \int_{\mathcal C_{\mt{EFT}}}\dd\omega\,\ee^{-S_{\mt{EFT}}(\omega)}
    +\text{subleading corrections}\,,
    \label{eq:UV-EFT-contour-matching}
\end{equation}
where $S_{\mt{EFT}}$ is the truncated effective action and $\mathcal C_{\mt{EFT}}$ denotes the corresponding complex integration contour inherited from the Picard--Lefschetz decomposition of the full theory. The approximation is valid when the relevant UV thimble contributions are captured by the effective description, while the omitted higher-order corrections and other saddle contributions remain subleading. In this way, the integration contour of the effective theory can still encode enough information about the UV completion that is not captured by the truncated effective action itself. In particular, if the relevant contribution is captured by a saddle described by the quadratic approximation, while other saddle contributions remain subleading, the effective Gaussian integral can still reproduce the leading smoothing profile along the corresponding complex contour $\mathcal C_{\mt{EFT}}$, even when it diverges along the original real-frequency axis.

To illustrate this idea, we can consider a quartic toy model and its quadratic approximation,
\begin{equation}
\begin{split}
    I_{\mt{EFT}}&=\int_{\mathcal C_{\mt{EFT}}}\dd\omega\, e^{-\ii (T-T_m)\omega}
    \exp\left[\frac{\alpha}{c}\omega^2\right] \,, \\
    I_{\mt{UV}}&=\int_{\mathbb R}\dd\omega\,
    e^{-\ii (T-T_m)\omega} \exp\left[\frac{\alpha}{c}\omega^2
    -\frac{\omega^4}{c^3}\right]\,,\\
\end{split}
    \label{eq:quartic model}
\end{equation}
The quartic term ensures convergence of the full integral along the real axis, whereas the quadratic approximation becomes divergent on the same contour when $\operatorname{Re}\alpha>0$. Nevertheless, for suitable complex values of $\alpha$, the original integration contour (\ie $\mathbb{R}$) can be deformed into complex steepest-descent contours whose leading contributions are captured by the quadratic approximation. By choosing the effective complex contour $\mathcal C_{\mt{EFT}}$ consistently with this deformation, the truncated integral can reproduce the leading smooth profile of the full theory, despite its apparent divergence along the real axis. This provides a natural interpretation of the contour prescription in effective field theory: the integration cycle of the effective theory can retain information about the UV completion that is not captured by the truncated effective action itself. We leave the detailed discussion on the Picard--Lefschetz analysis and its validity within the effective theory in Appendix~\ref{subsec:deformed-integral-EFT}.
%This agreement requires that the relevant saddle contributions are preserved while additional UV saddles remain subleading.

%\par\noindent\rule{\textwidth}{1pt}

More generally, the smoothing mechanism can be interpreted in terms of the large-frequency behavior of the inverse QNM trajectory. Recall that the semiclassical linear QNM spectrum gives rise to Stokes factors that are exponential in frequency, thereby generating displaced complex-time singularities from the original OPE seed. At finite $c$, however, the exact inverse trajectory may exhibit qualitatively different UV behavior. To illustrate this, suppose that the Stokes factor has the asymptotic form $\mathcal S_m(\omega)=\ee^{2\pi\ii m z_{\rm sc}(\omega)}\sim\ee^{a\omega^\gamma}$, with $\gamma>0$. In particular, we require $\operatorname{Re}a<0$, ensuring that each individual Stokes sector is exponentially suppressed at large positive frequencies and remains consistent with the UV constraints imposed by the OPE. The corresponding Fourier contribution can then be modeled by
\begin{equation}
\int_0^\infty\dd\omega\,\omega^\nu
    \ee^{-\ii\omega(T-T_{\rm seed})}
    \ee^{2\pi\ii m z_{\rm sc}(\omega)}
    \ \xrightarrow{\mathrm{UV}}\ 
    \int_0^\infty\dd\omega\,\omega^\nu
    \ee^{-\ii\omega(T-T_{\rm seed})+a\omega^\gamma}\,. 
    \label{eq:general-UV-Stokes-smoothing}
\end{equation}
For $\gamma=1$, the linear exponential generates a finite displacement of the original complex-time singularity. By contrast, when $\gamma>1$, the \emph{superlinear suppression} renders the Fourier transform entire. For $0<\gamma<1$, the sublinear exponent no longer produces a displaced singularity, and the corresponding Fourier transform remains analytic away from the original seed location. Thus, both sublinear and sufficiently suppressive superlinear UV behavior can eliminate the additional complex-time singularities generated by the semiclassical linear QNM spectrum.

These two possibilities can be realized by different deformations of the inverse QNM trajectory. The finite polynomial model discussed above belongs to the superlinear case. On the other hand, a QNM spectrum with asymptotic behavior $\omega_n\sim n^p, (p>1)$ has an inverse trajectory $z_{\rm sc}(\omega)\sim \omega^{1/p}$, corresponding to $\gamma=1/p<1$ (see appendix~\ref{sec:asymptotic-hadamard}). In particular, an exact quadratic QNM spectrum gives $\gamma=1/2$, even though its expansion in the semiclassical regime contains a linear term and a quadratic Gaussian correction. This illustrates how the smooth finite-$c$ profile approaches a sharp displaced singularity in the strict semiclassical limit $c\to\infty$. In the next subsection, we investigate this mechanism explicitly using an exactly solvable quadratic QNM model and show how its leading smoothing profile can be recovered from the intermediate-frequency expansion through an appropriate complex-frequency contour.
%\par\noindent\rule{\textwidth}{1pt}
%\par\noindent\rule{\textwidth}{1pt}
%\newpage

\subsection{An exactly solvable quadratic QNM model}
\label{subsec:Quadratic-QNM-Model}

In the last subsection, we showed that finite-$c$ corrections can deform the semiclassical linear QNM spectrum and modify the corresponding complex-time singularities. To investigate this mechanism beyond the perturbative expansion, we can consider a simple quadratic QNM model,
\begin{equation}
    \omega_n=a_0+a_1n+\frac{a_2}{c}n^2\,,
\end{equation}
where the quadratic trajectory is now taken to be exact. This provides a simple UV completion of the semiclassical QNM trajectory and allows us to investigate the crossover between the intermediate-frequency and ultraviolet regimes. More importantly, it provides an explicit example in which the leading smoothing profile can be recovered from the effective quadratic approximation by choosing an appropriate complex-frequency contour. As we shall see in section~\ref{sec:finite-c-qnm-liouville}, the same quadratic structure also arises naturally from the exact OPE coefficients in Liouville CFT.

The inverse branch continuously connected to the semiclassical trajectory is derived as 
\begin{equation}
    z_{\rm sc}(\omega)=\frac{a_1c}{2a_2}
    \left[\sqrt{1+\frac{4a_2(\omega-a_0)}{a_1^2c}}-1\right]\,. 
    \label{eq:exact-quadratic-inverse}
\end{equation}
In the intermediate-frequency regime, its expansion gives
\begin{equation}
    z_{\rm sc}(\omega)
    =\frac{\omega-a_0}{a_1}
    -\frac{a_2(\omega-a_0)^2}{a_1^3c}
    +O\left(\frac{(\omega-a_0)^3}{c^2}\right),
    \label{eq:exact-quadratic-inverse-IR}
\end{equation}
which reproduces the same Gaussian coefficient $\kappa_m=2\pi\ii m a_2/(a_1^3c)$ introduced in eq.~\eqref{eq:quadratic-deformed-Stokes}. However, the exact inverse trajectory grows as 
\begin{equation}
 z_{\rm sc}(\omega) \xrightarrow{\mathrm{UV}} z_{\rm sc}(\omega) \sim   \sqrt{\omega} \,, 
\end{equation}
in the UV regime with a fixed finite $c$. Its associated Stokes factor therefore no longer contains the linear exponential phase that can generate the semiclassical bouncing-geodesic singularities on the complex $T$ plane. 

To make the following analysis simpler and more explicit, let us consider the following choice of coefficients for one particular Stokes factor,
\begin{equation}
    a_0=0,\qquad a_1=-\frac{2\pi\ii}{\tau_*},\qquad
    a_2=-\frac{\pi^2}{\tau_*^2},\qquad \tau_*>0\,. 
    \label{eq:quadratic-solvable-parameters}
\end{equation}
The corresponding Stokes factor $\mathcal{S}_{m=1}(\omega)$ reduces to 
\begin{equation}
    \begin{aligned}
    \mathcal S_1(\omega)
    &=\exp\left[-2\tau_*c\left(\sqrt{1+\frac{\omega}{c}}-1\right)\right]=\exp\left[-\tau_*\omega+\frac{\tau_*}{4c}\omega^2
    +O\left(\frac{\omega^3}{c^2}\right)\right].
    \end{aligned}
    \label{eq:quadratic-exact-Stokes-action}
\end{equation}
Here, we consider only the $m=1$ case and the higher Stokes sectors follow by replacing $\tau_*$ with $m\tau_*$. The exact Stokes factor is exponentially suppressed along the positive real-frequency axis, consistently with the UV constraint, even though its quadratic approximation contains an inverted Gaussian. In particular, its UV behavior is governed by 
\begin{equation} 
\mathcal S_1(\omega) \xrightarrow{\mathrm{UV}}\  \exp[-2\tau_*\sqrt{c\omega}+O(c)] \,, 
\end{equation}
rather than the pure linear exponential responsible for the semiclassical singularity. 

We can verify the resulting smoothing directly from the exact Fourier transform. For simplicity, we just take $\nu=0$ and define
\begin{equation}
    F_c(s)=\int_0^\infty\dd\omega\,
    \exp\left[-s\omega-2\tau_*c
    \left(\sqrt{1+\frac{\omega}{c}}-1\right)\right],
    \qquad s=\ii(T-T_{\rm seed}) \,. 
    \label{eq:quadratic-exact-transform}
\end{equation}
In the strict semiclassical limit, this reduces to $F_\infty(s)=1/(s+\tau_*)$, with a displaced singularity at $T=T_{\rm seed}+\ii\tau_*$. At finite $c$, however, the integral can be evaluated exactly and gives\footnote{Here, we define the square roots for $\operatorname{Re}s>0$ and then continue them around $s=0$ through the upper half of the $s$-plane.}
\begin{equation}
    F_c(s)=\frac{1}{s}
    -\frac{\tau_*\sqrt{\pi c}}{s^{3/2}}
    \exp\left[\frac{c(s+\tau_*)^2}{s}\right]
    \operatorname{erfc}\left((s+\tau_*)\sqrt{\frac{c}{s}}\right).
    \label{eq:quadratic-exact-transform-erfc}
\end{equation}
Unlike the semiclassical result, this exact expression is locally analytic at the would-be complex-time singularity $T=T_{\rm seed}+\ii \tau_*$ for any finite $c$. In particular, continuation through the upper half-plane gives
\begin{equation}
    F_c(-\tau_*+\ii0)
    =-\frac{1}{\tau_*}-\ii\sqrt{\frac{\pi c}{\tau_*}},
    \label{eq:quadratic-exact-center}
\end{equation}
which remains finite at finite $c$. Thus, the semiclassical complex-time singularity located at $s=-\tau_\ast$ is replaced by a smooth profile and only emerges in the strict $c\to\infty$ limit.

\begin{figure}[t]
    \centering
    \includegraphics[width=2.95in]{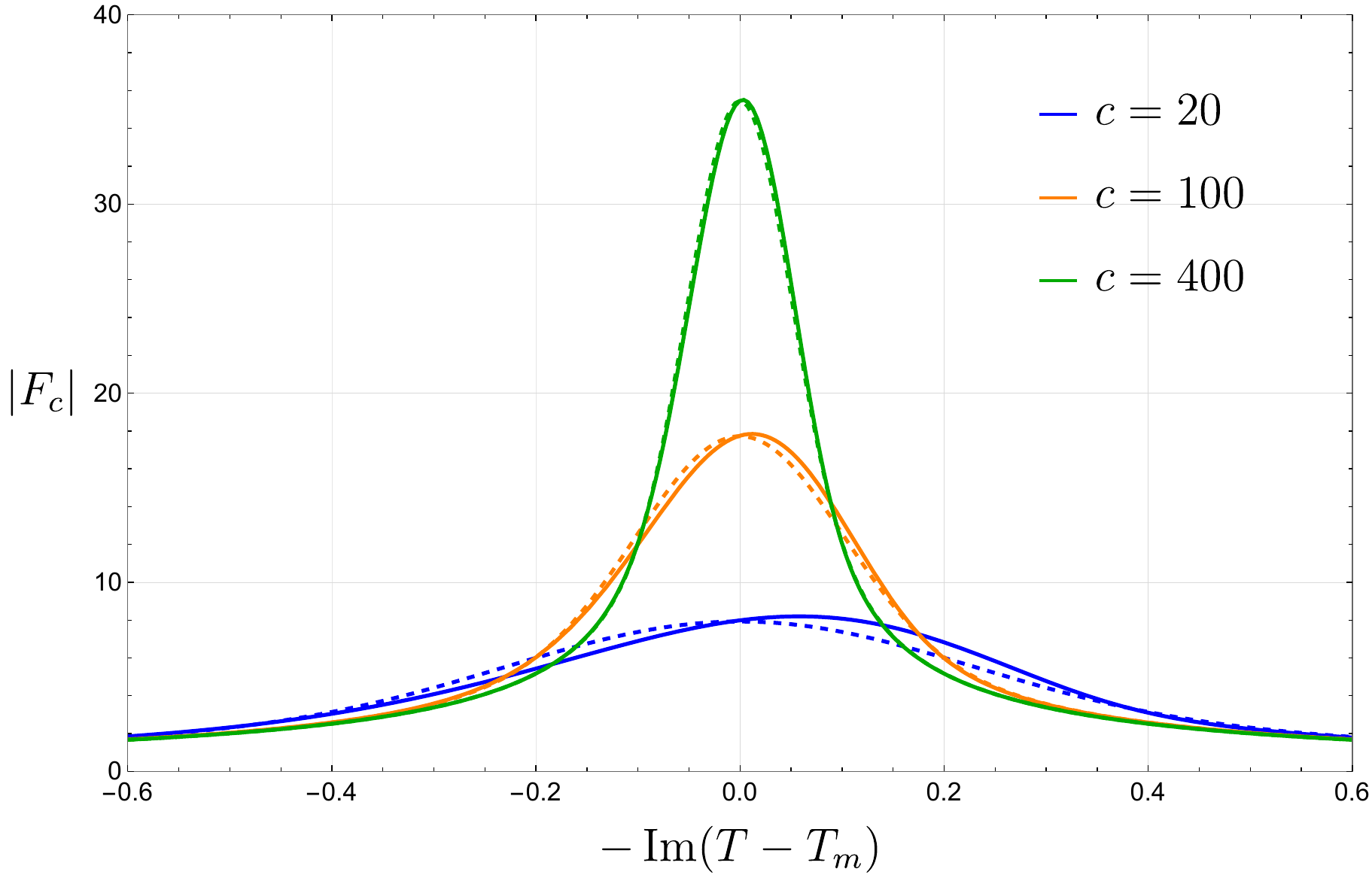}
     \includegraphics[width=2.95in]{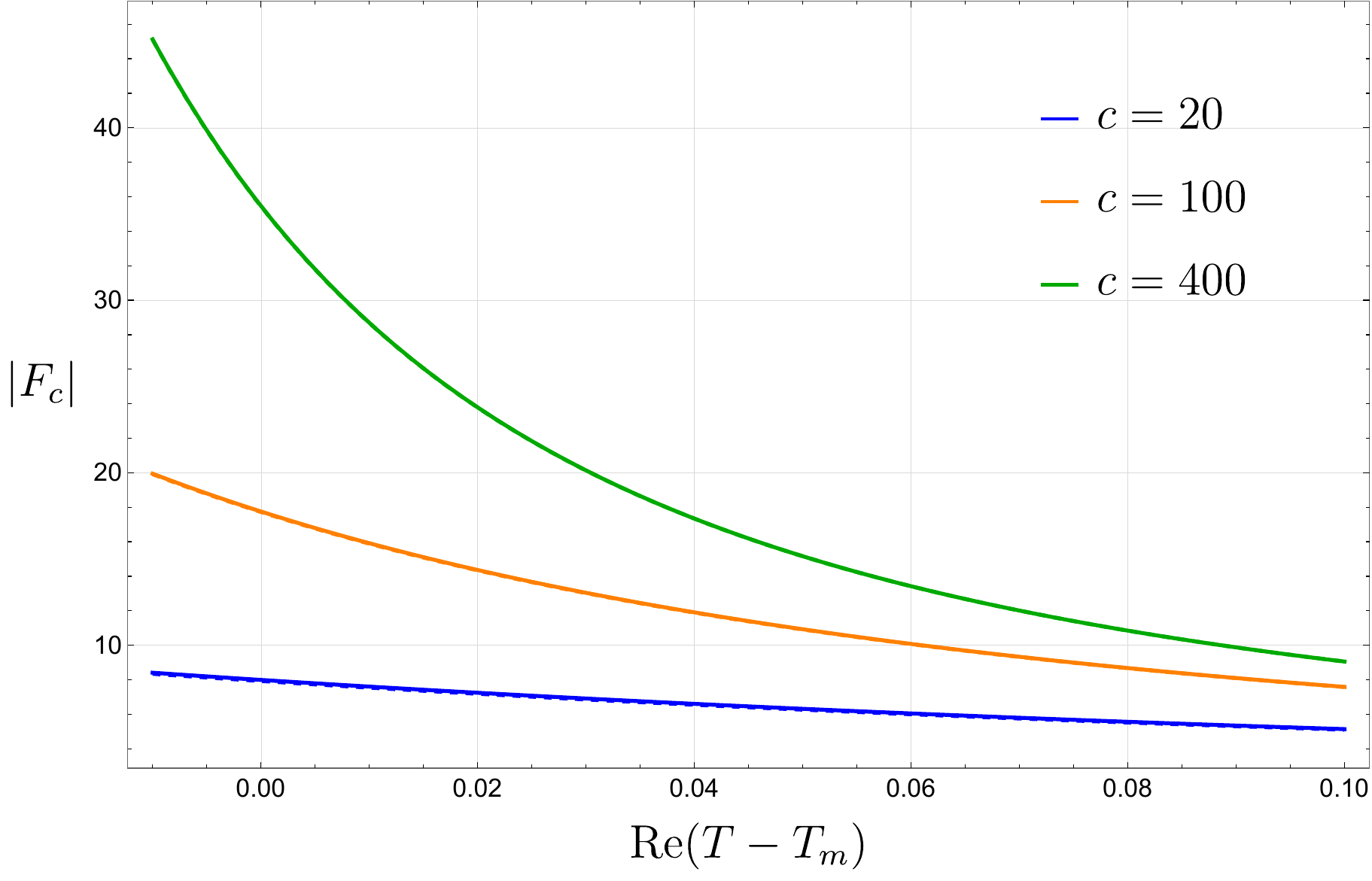}
   \caption{Comparison of the exact Fourier transform $F_c$ (solid lines) and its quadratic approximation $F_c^{\mt{EFT}}$ (dashed lines) for different values of $c$, with $\tau_*=1$. The left and right panels show the profiles along the imaginary-time and real-time directions, respectively, near the semiclassical displaced singularity $T_m=T_{\rm seed}+\ii\tau_*$. The quadratic approximation evaluated along the complex contour accurately reproduces the leading smooth profile in both directions. As $c$ increases, the profile becomes increasingly localized, with a characteristic width $O(c^{-1/2})$ and amplitude $O(c^{1/2})$, recovering the sharp semiclassical singularity in the strict semiclassical limit $c\to\infty$.}
\label{fig:quadratic_z_comparison}
\end{figure}

Interestingly, the leading smoothing profile around the would-be pole $s=-\tau_\ast$ can already be recovered from the quadratic approximation in eq.~\eqref{eq:quadratic-exact-Stokes-action}. Although the resulting inverted-Gaussian integral diverges along the positive real-frequency axis, the analytic continuation of the exact theory selects a different complex contour. Taking the integral contour along the negative imaginary-frequency axis, the leading effective description is obtained as\footnote{Here, we choose the upper continuation while the lower continuation similarly selects the opposite contour and gives the complex-conjugate profile.}
\begin{equation}
    \begin{aligned}
    F_c^{\mt{EFT}}(s)
    &=\int_0^{-\ii\infty}\dd\omega\,
    \exp\left[-(s+\tau_*)\omega+\frac{\tau_*}{4c}\omega^2\right]\\
    &=-\ii\sqrt{\frac{\pi c}{\tau_*}}
    \exp\left[-\frac{c(s+\tau_*)^2}{\tau_*}\right]
    \operatorname{erfc}\left(-\ii(s+\tau_*)\sqrt{\frac{c}{\tau_*}}\right).
    \end{aligned}
    \label{eq:quadratic-IR-matching}
\end{equation}
Remarkably, the quadratic approximation reproduces the leading behavior of the exact result near the semiclassical singularity,
\begin{equation}
    F_c(s)=F_c^{\mt{EFT}}(s)+O(1),\qquad
    |s+\tau_*|=O\left(\sqrt{\frac{\tau_*}{c}}\right).
    \label{eq:quadratic-bump-scaling}
\end{equation}
As illustrated in figure~\ref{fig:quadratic_z_comparison}, the regularized profile remains smooth at the would-be singularity, with a characteristic width of $O(c^{-1/2})$ and an amplitude of $O(c^{1/2})$. The agreement between the exact result and its quadratic approximation can be understood from the relevant frequency scale $|\omega|=O(\sqrt c)$, which lies within the intermediate-frequency regime $|\omega|\ll c$. In this regime, the quadratic correction captures the leading finite-$c$ smoothing effect, while higher-order corrections remain subleading. This model therefore explicitly demonstrates that the quadratic EFT approximation can reproduce the leading smoothing profile when evaluated along a complex contour inherited from the analytic continuation of the exact theory, even though the truncated anti-Gaussian integral diverges along the real-frequency axis. For readers interested in the Picard--Lefschetz analysis, we refer to the discussion in Appendix~\ref{subsec:deformed-integral-EFT} for more details.

Finally, the same conclusion can be easily extended to the original Fourier transform with the power-law prefactor $\omega^\nu$. For nonnegative integer $\nu$, the exact result follows by applying $(-\partial_s)^\nu$ to $F_c(s)$. More generally, for $\operatorname{Re}\nu>-1$, the corresponding quadratic approximation gives the parabolic-cylinder profile as discussed in the previous subsection, with the same characteristic width $\Delta T_m=O(c^{-1/2})$ and an amplitude scaling as $O\left(c^{(\operatorname{Re}\nu+1)/2}\right)$. The branch of $\omega^\nu$ is also fixed by the same analytic continuation as the integration contour.

%\par\noindent\rule{\textwidth}{1pt}
%%%%%%%%%%%%%%%%%%%%%%%%%%%%%%%%%%%%%%%%%%%%%%%%

%%%%%%%%%%%%%%%%%%%%%%%%%%%%%%%%%%%%%%%%%%%%%%%%
\subsection{Protected seed singularities}
\label{subsec:protected-singularities}
%%%%%%%%%%%%%%%%%%%%%%%%%%%%%%%%%%%%%%%%%%%%%%%%

\begin{figure}[t]
    \centering
    \includegraphics[width=5in]{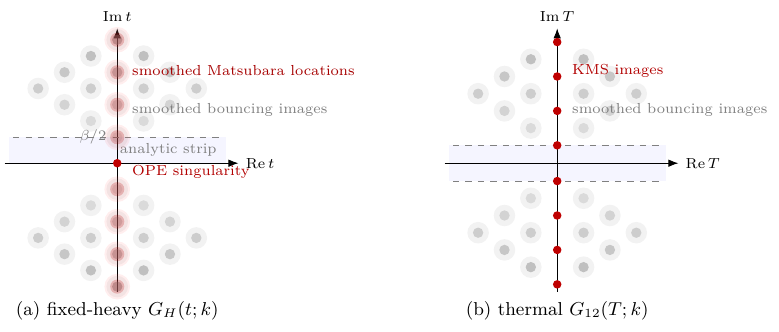}
    \caption{Schematic complex-time singularity structure at finite $c$ and fixed spatial momentum. Left: for the HHLL correlator $G_{\mt H}(t,\bk)$, the red point at $t=0$ denotes the genuine light-light OPE singularity. The additional semiclassical thermal images, indicated by red halos, are smoothed out by finite-$c$ corrections. Right: for the thermal two-sided correlator $G_{\mt{LR}}(T,\bk)$, the red points denote the protected OPE/KMS singularities at $T=T_{\rm seed}^{(\pm,k)}$, while the gray halos represent the smoothing of additional bouncing-geodesic singularities generated by the deformation of linear QNMs.}
    \label{fig:finite-c-time-lattices}
\end{figure}

In previous subsections, we showed how nonlinear finite-$c$ deformations of the QNM spectrum can smooth the additional complex-time singularities generated by the semiclassical linear QNM trajectories. However, this mechanism must preserve the genuine short-distance singularities required by the boundary OPE. The distinction is particularly important when comparing the thermal two-sided correlator $G_{\mt{LR}}$ with the fixed-heavy-state HHLL correlator $G_{\mt{H}}$. For the former, both the OPE singularities and their KMS-related thermal images are protected, whereas for the latter, the additional thermal images arising in the semiclassical approximation are not protected by an exact KMS condition. The resulting finite-$c$ structures are illustrated schematically in figure~\ref{fig:finite-c-time-lattices}.

The protection of the seed singularities on the complex-time plane follows directly from the local OPE and the exact KMS condition, independently of the QNM spectrum. As discussed in section~\ref{subsec:thermal-analytic-structure}, the identity contribution to the OPE fixes the short-distance singularity of the Euclidean thermal correlator. Thermal periodicity then reproduces the same local singular structure whenever the two operator insertions coincide on the Euclidean thermal circle, $\tau=n\beta$, with $n\in\mathbb Z$. Under the continuation $\tau=\beta/2+\ii T$, these coincidence points correspond to the KMS singularities
\begin{equation}
    T_{\rm seed}^{(\pm,k)}
    =\pm\ii\left(k+\frac12\right)\beta,
    \qquad k=0,1,2,\ldots\,,
    \label{eq:protected-KMS-singularities}
\end{equation}
on the corresponding thermal continuation. Their existence is therefore required by the boundary theory and is not affected by finite-$c$ corrections to the semiclassical QNM spectrum. In other words, these seed singularities $T_{\rm seed}^{(\pm,k)}$ should all be protected from the smoothing mechanism described in the last subsection. This protection is also manifest in the exact spectral representation~\eqref{eq:GLR-spectral-density},
\begin{equation}
    G_{\mt{LR}}(\omega,\bk)
    =\frac{\rho(\omega,\bk)}
    {2\sinh(\beta\omega/2)}\,.
\end{equation}
The universal thermal factor contains the full sequence of linear exponential weights, as shown in eq.~\eqref{eq:thermal-factor-expansion}. The short-distance OPE further fixes the leading UV behavior of the spectral density, $\rho(\omega,\bk)\sim C_{\mathcal O}\omega^\nu$, where $\nu=2\Delta-d$. Consequently, the two-sided correlator must retain the large-frequency behavior
\begin{equation}
    G_{\mt{LR}}(\omega,\bk) \sim C_{\mathcal O}\omega^\nu \ee^{-\beta\omega/2},
    \qquad \omega\rightarrow+\infty\,,
    \label{eq:protected-LR-UV}
\end{equation}
with a nonvanishing OPE coefficient $C_{\mathcal O}$. After Fourier transformation to the complex $T$-plane, it follows that the genuine OPE singularity at $T_{\rm seed}^{(+,0)}=\ii \frac{\beta}{2}$ is protected by this universal leading asymptotic behavior. Correspondingly, the remaining KMS images are protected by the exact thermal periodicity. 

By contrast, the HHLL correlator $G_{\mt H}$ describes a two-point function in a fixed heavy state and does not generally satisfy an exact KMS condition at finite $c$. Its short-distance behavior is nevertheless constrained by the light-light OPE, which requires
\begin{equation}
    G_{\mt H}(\omega,\bk) \sim C_{\mt H}\omega^\nu,
    \qquad \omega\rightarrow+\infty\,,
    \label{eq:protected-HHLL-UV}
\end{equation}
in the large-frequency limit. This protects the genuine OPE singularity at $t=0$. However, the additional semiclassical singularities at $t=\ii n\beta_{\mt H}$, with $n\neq 0$, arise from the emergent thermal periodicity of the semiclassical HHLL correlator rather than from an exact KMS relation. These are precisely the forbidden singularities discussed in \cite{Fitzpatrick:2015dlt,Fitzpatrick:2016ive,Chen:2017yze}, which are not protected against finite-$c$ corrections.

The Hadamard representation associated with the QNM spectrum and the distribution of zeros should be reconciled with these protected complex-time singularities. In the appropriate Stokes sector, extracting the contribution associated with the distinguished inverse QNM branch allows us to write the correlator schematically as
\begin{equation}
    G(\omega,\bk) = \mathcal B(\omega,\bk)
    \left[1+\sum_{m=1}^{\infty}\ee^{2\pi\ii m z_{\rm sc}(\omega)}
    \right],
    \label{eq:protected-Stokes-decomposition}
\end{equation}
where $\mathcal{B}(\omega,\bk)$ contains the entire factor in the Hadamard representation~\eqref{eq:hadamard-meromorphic}, the leading asymptotics of pole-zero contributions, and for the thermal case the $\ee^{-\beta\omega/2}$ contribution from the explicit Matsubara factor.  Importantly, the nontrivial Stokes terms are expected to be asymptotically suppressed. Matching the complete correlator to the OPE thus requires that $B(\omega,\bk)$ satisfy the UV behavior required by $G_{\mt{LR}}$ and $G_{\mt{H}}$, respectively. In discussion \ref{subsec:IR UV constraints} and appendix \ref{app:UV-IR-constraints}, we will also show how these UV conditions further constrain the distributions of poles and zeros.

We therefore conclude that finite-$c$ deformations of the QNM spectrum can smooth the additional semiclassical complex-time singularities while preserving those protected by the boundary OPE and, for thermal correlators, the exact KMS condition. The consistency of these two structures imposes nontrivial consistency conditions on the full Hadamard reconstruction, requiring the pole-zero data and the entire prefactor to collectively reproduce the OPE/KMS structure. We further examine these consistency conditions in more detail in appendix~\ref{app:UV-IR-constraints}.

%%%%%%%%%%%%%%%%%%%%%%%%%%%%%%%%%%%%%%%%%%%%%%%%%%%%%%%%%%%%
%%%%%%%%%%%%%%%%%%%%%%%%%%%%%%%%%%%%%%%%%%%%%%%%%%%%%%%%%%%%

\section{Finite-$c$ QNM Deformation in Liouville CFT}
\label{sec:finite-c-qnm-liouville}

In the previous section, we discussed how the complex-time singularities of semiclassical holographic correlators are encoded in the asymptotic distribution of QNMs, and how their finite-$c$ deformation may lead to the smoothing of these singularities. However, an explicit investigation of this mechanism in a generic holographic CFT remains challenging. While the semiclassical QNM spectrum can be obtained from linearized perturbations around a fixed black-hole background, its finite-$c$ corrections generally involve quantum gravitational effects that modify the analytic structure of the boundary correlator. In particular, the exact locations and asymptotic distributions of the corresponding poles and zeros are generally unknown at finite $c$, making it difficult to directly establish their consequences for complex-time singularities.

To gain analytic control over this problem, we instead consider Liouville CFT as an exactly solvable example. Unlike generic two-dimensional CFTs, whose three-point structure constants are generally not known in closed form, Liouville CFT possesses exact three-point coefficients determined by the Dorn--Otto--Zamolodchikov--Zamolodchikov (DOZZ) formula \cite{Dorn:1994xn,Zamolodchikov:1995aa}. In the crossed $(\mathrm{HL})(\mathrm{HL})$ fusion channel, these coefficients govern the dynamical part of the HHLL correlator $G_{\mt H}$ and allow us to investigate the pole--zero structure explicitly. Remarkably, the exact DOZZ formula illustrates how the semiclassical QNM spectrum is deformed at finite $c$. In particular, the relevant DOZZ pole trajectory reproduces the familiar linear QNM spectrum of the BTZ black hole in the semiclassical limit $c \to \infty$, while exhibiting a quadratic deformation at finite $c$. Of course, Liouville CFT is not a conventional holographic CFT, as it lacks several properties expected of a CFT dual to semiclassical AdS$_3$ Einstein gravity. For example, it has no normalizable conformally invariant vacuum, and its physical primary states form a continuous spectrum. Nevertheless, the pole--zero structure of its analytically continued heavy-state correlators exhibits striking similarities to that of thermal correlators in holographic theories. Liouville theory thus provides an analytically controlled example of finite-$c$ QNM deformation and an explicit CFT realization of the smoothing mechanism for complex-time singularities established in the previous section.

\subsection{Liouville CFT and the DOZZ three-point coefficient}
\label{subsec:liouville-dozz}

We begin by briefly reviewing the relevant properties of Liouville CFT and establishing our conventions. The theory is defined by the Liouville action
\begin{equation}
    S_{\mathrm{L}}=\frac{1}{4\pi}\int \dd^2x\,\sqrt{g}\left[g^{ab}\partial_a\phi\partial_b\phi+Q\mathcal{R}\phi+4\pi\mu \ee^{2b\phi}\right],
\end{equation}
where $\mu$ is the cosmological constant, $b$ is the Liouville coupling, and $Q$ denotes the background charge. Quantum conformal invariance requires
\begin{equation}
   Q=b+b^{-1} \,,  \qquad  c=1+6Q^2 = 1 + 6(b+ b^{-1})^2\,. 
    \label{eq:liouville-central-charge}
\end{equation}
The classical limit $c\to\infty$ thus corresponds to $Q\to\infty$ and $b\to0$. The Virasoro primary operators are exponential fields $V_\alpha=\ee^{2\alpha\phi}$, characterized by equal holomorphic and antiholomorphic conformal dimensions,
\begin{equation}
    h(\alpha)=\bar h(\alpha)=\alpha(Q-\alpha).
    \label{eq:liouville-conformal-dimension}
\end{equation}
For real positive $b$, the normalizable primary spectrum is continuous and parameterized by
\begin{equation}
    \alpha=\frac{Q}{2}+\ii P,\qquad P\in\mathbb{R}_{\geq 0},
    \label{eq:liouville-physical-spectrum}
\end{equation}
with conformal dimensions $h(\alpha)=Q^2/4+P^2$. The reflection relation identifies the representations associated with $\alpha$ and $Q-\alpha$, up to a momentum-dependent reflection coefficient \cite{Teschner:1995yf,Teschner:2001rv}. In the semiclassical limit $b\to 0$, light operators with finite conformal dimensions and heavy operators with dimensions of order $c$ are characterized by the following scalings:
\begin{equation}
\text{Light:}\,h(\alpha_\mt{L})\sim \mathcal O(c^0)\,, \, \alpha_\mt{L}\approx h(\alpha_\mt{L})b\sim \mathcal O(b)\,; \quad 
\text{Heavy:}\, h(\alpha_\mt{H})\sim \mathcal O(c)\,, \, \alpha_\mt{H}\sim \mathcal O(b^{-1})\,,
\end{equation}
respectively. Although these light operators lie outside the normalizable spectrum of Liouville CFT, their three-point coefficients can still be obtained by analytic continuation of the DOZZ formula \cite{Harlow:2011ny}.

To connect these exact CFT data from Liouville theory with the pole structure of correlators and the QNM spectrum, we consider the HHLL four-point function in Liouville CFT by identifying the light and heavy operators with the Liouville primaries 
\begin{equation}
\mathcal{O}_{\mt{L}} \to V_{\alpha_{1}}\,, \qquad \mathcal{O}_{\mt{H}} \to V_{\alpha_{2}} \,, 
\end{equation}
respectively. The corresponding HHLL correlator \eqref{eq:hhll-heavy-state-correlator} in the conformal frame can be written as\footnote{The cylinder correlator given by eq.~\eqref{eq:hhll-heavy-state-correlator} and the plane correlator defined here are related by the conformal transformation
\begin{equation}
G_{\mt H}(\tau,\phi)=\ee^{-\Delta_{\mt L}\tau}G_{\mt H}^{\mathrm{plane}}\left(z,\bar z\right)\,.   
\end{equation}
} 
\begin{equation}
    G^{\rm plane}_{\mt H}(z,\bar z)\propto
    \left\langle V_{\alpha_{2}}(\infty)V_{\alpha_{1}}(1)V_{\alpha_{1}}(z,\bar z)V_{\alpha_{2}}(0)\right\rangle\,, \qquad z=\ee^{-\tau+\ii\phi} \,, 
    \label{eq:liouville-hhll-correlator}
\end{equation}
where the overall normalization of the heavy state is suppressed.\footnote{Unlike conventional CFTs with a discrete spectrum, Liouville CFT has a continuous primary spectrum, and its primary states are delta-function normalized, $\langle P'|P\rangle\propto\delta(P-P')$.} As discussed in subsection~\ref{subsec:hhll}, by the operator-state correspondence, this four-point function can be interpreted as a two-point function of light operators in a heavy primary state. In the semiclassical heavy-light limit, such correlators provide a natural CFT description of probe propagation in the background associated with the heavy state. For holographic CFTs with an appropriate sparse light spectrum, the leading heavy-light Virasoro vacuum block exhibits the characteristic thermal behavior of a BTZ black hole \cite{Fitzpatrick:2015zha,Fitzpatrick:2016ive}. 

The essential dynamical information is encoded in the Liouville three-point structure constants, defined by
\begin{equation*}
C_{\mathrm L}(\alpha_1,\alpha_2,\alpha_3) = \left\langle V_{\alpha_1}(0)V_{\alpha_2}(1)V_{\alpha_3}(\infty)\right\rangle \,, 
\end{equation*}
which are symmetric under permutations of their arguments and obey the Liouville reflection relation
\begin{equation*}
C_{\mathrm L}(\alpha_1,\alpha_2,\alpha_3) =R(\alpha_3)\,C_{\mathrm L}(\alpha_1,\alpha_2,Q-\alpha_3)\,.
\end{equation*}
Here, $R(\alpha)$ is the Liouville reflection coefficient, relating the primary operators $V_\alpha$ and $V_{Q-\alpha}$ with identical conformal dimensions \cite{Teschner:1995yf,Teschner:2001rv}. Following the spectral representation in eq.~\eqref{eq:heavy-state-spectral-frequency}, we consider the crossed $(\text{HL})(\text{HL})$ channel, where the intermediate primary states are labeled by the continuous Liouville momentum $\alpha_3=Q/2+\ii P$. The HHLL correlator admits a conformal-block decomposition of the form \cite{Teschner:2001rv,Ponsot:1999uf}
\begin{equation}
\boxed{ G^{\mathrm{plane}}_{\mt H}(z,\bar z)\propto\int_0^\infty \dd P\, C_{\mathrm L}(\alpha_1,\alpha_2,\alpha_3) C_{\mathrm L}(\alpha_1,\alpha_2,Q-\alpha_3) \left|\mathcal{V}_{\mt{HLLH}}(h_3;z)\right|^2 \,,}
\label{eq:liouville-hlhl-channel}
\end{equation}
where $h_3=h(\alpha_3)=Q^2/4+P^2$. Here, the integral runs over the continuous primary spectrum of Liouville theory\footnote{The conformal-block decomposition \eqref{eq:liouville-hlhl-channel} applies directly to the continuous spectrum, while analytic continuation of the external Liouville momenta may require additional discrete residue contributions~\cite{Teschner:2001rv,Harlow:2011ny,Balasubramanian:2017fan}.}, and the product of three-point coefficients determines the corresponding heavy-light transition amplitudes appearing in \eqref{eq:heavy-state-spectral-frequency}. The Virasoro conformal block $\mathcal{V}_{\mt{HLLH}}$ incorporates the contributions from the intermediate primary and its descendants.

Remarkably, unlike generic CFTs, the three-point structure constants of Liouville theory are known exactly through the DOZZ formula \cite{Dorn:1994xn,Zamolodchikov:1995aa},
\begin{equation}
    C_{\mathrm{L}}(\alpha_1,\alpha_2,\alpha_3)=\frac{\left[\pi\mu\gamma(b^2)b^{2-2b^2}\right]^{\frac{Q-\alpha_1-\alpha_2-\alpha_3}{b}} \,\Upsilon_b'(0)\Upsilon_b(2\alpha_1)\Upsilon_b(2\alpha_2)\Upsilon_b(2\alpha_3)}
    {\Upsilon_b(\alpha_1+\alpha_2+\alpha_3-Q)\Upsilon_b(\alpha_1+\alpha_2-\alpha_3)\Upsilon_b(\alpha_1+\alpha_3-\alpha_2)\Upsilon_b(\alpha_2+\alpha_3-\alpha_1)}\,. 
\label{eq:liouville-dozz-formula}
\end{equation}
Here $\gamma(x)=\Gamma(x)/\Gamma(1-x)$, and $\Upsilon_b(x)$ is the Liouville Upsilon function, defined in terms of the Barnes double-gamma function $\Gamma_b(x)$ as
\begin{equation}
    \Upsilon_b(x)=\frac{1}{\Gamma_b(x)\Gamma_b(Q-x)}\,.
    \label{eq:liouville-upsilon-definition}
\end{equation}
The function $\Upsilon_b(x)$ is entire and satisfies the reflection identity $\Upsilon_b(x)=\Upsilon_b(Q-x)$. The DOZZ formula thus determines the three-point structure constants as meromorphic functions of the Liouville momenta, which admit analytic continuation beyond the physical spectrum \cite{Teschner:2001rv,Harlow:2011ny}. In particular, analytically continuing the intermediate Liouville momentum allows us to determine explicitly the pole--zero structure of the heavy-light spectral coefficients.

%\subsection{Pole-zero structure from the DOZZ coefficient}
%\label{subsec:dozz-poles}

\subsection{Pole--zero structure at finite $c$}
\label{subsec:liouville-qnm-deformation}

As discussed in subsection~\ref{subsec:hhll}, the spectral representation \eqref{eq:heavy-state-spectral-frequency} of the HHLL correlator involves the transition energies and matrix elements connecting the initial heavy state to intermediate states through the light operator. In the semiclassical black hole regime, the analytic continuation of these spectral data reveals the resonance structure associated with the QNM spectrum of the effective black hole background. The connection between the analytic structure of heavy-light OPE coefficients and black hole QNMs was previously investigated in \cite{Turiaci:2016cvo}, which showed that the poles of the Liouville DOZZ coefficients reproduce the semiclassical BTZ QNM spectrum. Here, we retain the exact finite-$c$ dependence of the DOZZ coefficients to investigate the finite-$c$ deformation of the pole--zero structure in the complex-frequency plane.

In Liouville CFT, the heavy-light transition amplitudes are encoded in the three-point coefficients $C_{\mathrm{L}}(\alpha_1,\alpha_2,\alpha_3)$ entering the conformal-block decomposition \eqref{eq:liouville-hlhl-channel}. The Virasoro conformal blocks account for the contributions of the exchanged primary operators and all their descendants. However, in the semiclassical heavy-light regime,
\begin{equation}
h_{\mt L}=h(\alpha_1)=\mathcal O(c^0), \quad h_{\mt H}=h(\alpha_2)=\mathcal O(c),
\quad h_3-h_{\mt H}=h(\alpha_3)-h(\alpha_2)=\mathcal O(c^0),
\end{equation}
the contributions of Virasoro descendants in the $\mathrm{(HL)(HL)}$ channel are suppressed at large $c$. Consequently, the conformal block reduces to its leading primary contribution \cite{Fitzpatrick:2015zha,Fitzpatrick:2014vua,Berenstein:2022ico,Grabovsky:2024jwf},
\begin{equation}
    \mathcal{V}_{\mt{HLLH}}(h_3;z) = z^{h_3-h_{\mt H}-h_{\mt L}} \sum_{N=0}^{\infty}f_N(h_3)z^N= z^{h_3-h_{\mt H}-h_{\mt L}} \left[1+\mathcal O(c^{-1})\right]\,. 
\end{equation}
The transition frequency $\omega=E_a-E_{\mt H}$ is thus dominated by the primary-to-primary contribution,
\begin{equation}
   \omega = \Delta(\alpha_3) - \Delta_{\mt{H}} = 2h(\alpha_3)-  2h(\alpha_2) \,, 
    \label{eq:liouville-heavy-frequency-dictionary}
\end{equation}
corresponding to $N=\bar N=0$. After accounting for the conformal factor $\ee^{-\tau \Delta_{\mt L}}$ and Fourier transforming the leading primary contribution with respect to the cylinder time, we find that the heavy-state correlator in frequency space is governed by the product of DOZZ coefficients,
\begin{equation}
 G_{\mt H}(\omega) \propto C_{\mathrm L}(\alpha_1,\alpha_2,\alpha_3) C_{\mathrm L}(\alpha_1,\alpha_2,Q-\alpha_3) \,, 
\end{equation}
where the transition frequency is expressed in terms of the Liouville parameters as
\begin{equation}
\omega =  2(\alpha_3-\alpha_2)(Q-\alpha_3-\alpha_2)\,.
\end{equation}
Analytically continuing $\alpha_3$ into the complex plane then allows us to investigate the meromorphic structure of the corresponding spectral coefficients at finite $c$. Using the exact DOZZ formula \eqref{eq:liouville-dozz-formula}, we can determine the pole--zero structure of the DOZZ coefficients in the complex-frequency plane.

To make the connections to the bulk spacetime and our subsequent QNM analysis more explicit, we adopt the Liouville momentum parameterization for the initial heavy state,
\begin{equation}
\alpha_2=\frac{Q}{2}+\ii\kappa, \qquad \Delta_{\mt H}=2h(\alpha_2)=\frac{Q^2}{2}+2\kappa^2\,,
\end{equation}
which yields  
\begin{equation}\label{eq:transition-frequency}
\boxed{ \omega= 2(P^2-\kappa^2)= - 4 \ii \kappa\,(\alpha_3 - \alpha_2)-2(\alpha_3 - \alpha_2)^2 } \,. 
\end{equation}
Here, $\kappa\geq0$ is the real Liouville momentum characterizing the heavy conformal dimension. The positive imaginary branch is chosen for convenience, since the two branches $\alpha_2=Q/2\pm\ii\kappa$ are related by Liouville reflection symmetry. More importantly, the momentum $\kappa$ is associated with the temperature of a typical black hole state (see subsection~\ref{subsec:hhll}) through
\begin{equation}
T_{\mt H}=\frac{1}{2\pi}\sqrt{\frac{12\Delta_{\mt H}}{c}-1} \quad\longrightarrow\quad \kappa=\sqrt{\frac{1+4\pi^2cT_{\mt H}^2}{24}}\,. 
\end{equation}
In the large-$c$ limit with $b\to0$ and fixed $T_{\mt H}$, the semiclassical black hole temperature is thus related to the Liouville momentum of the heavy state by
\begin{equation}
\kappa \approx \frac{\pi T_{\mt H}}{b}=\frac{\pi}{b\beta_{\mt H}}\,.
\end{equation}

From the exact DOZZ formula \eqref{eq:liouville-dozz-formula}, the poles and zeros are determined by the zeros of the $\Upsilon_b$ functions appearing in its denominator and numerator. Recall that the Liouville Upsilon function $\Upsilon_b(x)$ is entire, with zeros located at
\begin{equation}
    \Upsilon_b(x)=0\quad\Longleftrightarrow\quad 
    x=\begin{cases}
        -\dfrac{m}{b}-nb\,,\\[4pt]
        Q+\dfrac{m}{b}+nb\,,
    \end{cases}
    \qquad m,n\in\mathbb{Z}_{\geq 0}\,.
    \label{eq:liouville-upsilon-zeros}
\end{equation}
These two families of zeros follow directly from the pole structure of the two double-gamma functions $\Gamma_b(z)$ in eq.~\eqref{eq:liouville-upsilon-definition}. 

For fixed external Liouville momenta $\alpha_1$ and $\alpha_2$, the poles of $C_{\mathrm{L}}(\alpha_1,\alpha_2,\alpha_3)$ arise from the four denominator factors in eq.~\eqref{eq:liouville-dozz-formula}, whose arguments are\footnote{Due to the reflection symmetry $\omega(\alpha_3)=\omega(Q-\alpha_3)$, we do not need to consider the other coefficient $C_{\mathrm{L}}(\alpha_1,\alpha_2,Q-\alpha_3)$ appearing in the HHLL decomposition, since it provides only the corresponding reflected pole--zero families.}
\begin{equation}
\begin{aligned}
    x_1&=\alpha_1+\alpha_2+\alpha_3-Q, & x_2&=\alpha_1+\alpha_2-\alpha_3,\\
    x_3&=\alpha_1+\alpha_3-\alpha_2, & x_4&=\alpha_2+\alpha_3-\alpha_1.
\end{aligned}
\label{eq:liouville-dozz-denominator-arguments}
\end{equation}
The pole conditions \eqref{eq:liouville-upsilon-zeros} determine the locations of the poles in the complex $\alpha_3$-plane.

The four denominator factors of the DOZZ coefficient generate eight candidate pole branches in the complex Liouville-momentum plane. Owing to the reflection symmetry $h(\alpha)=h(Q-\alpha)$, with $\omega(\alpha_3)=\omega(Q-\alpha_3)$, these branches correspond to four distinct frequency trajectories. In the semiclassical heavy-light limit, \ie
\begin{equation}
b\to0,\qquad \alpha_1 \approx bh_{\mt{L}}= \frac{b \Delta}{2} \sim \mathcal O(b),\qquad \alpha_2\sim\kappa\sim \mathcal O(b^{-1})\,, 
\end{equation}
only two pole trajectories remain at finite frequency, while the others are pushed to frequencies of order $c$. To track the poles connected to the semiclassical QNMs, we focus specifically on the two pole branches generated by $\Upsilon_b(x_2)$ and $\Upsilon_b(x_3)$, \ie
\begin{equation}
    \alpha_3= \alpha_2 + \alpha_1+ \dfrac{m}{b}+ nb,
    \qquad
    \alpha_3 = \alpha_2  -\alpha_1-\dfrac{m}{b}-nb\,.
    \label{eq:liouville-dozz-pole-families}
\end{equation}
Substituting these expressions into the definition of the transition frequency \eqref{eq:transition-frequency} gives rise to the following poles in the complex-frequency plane\footnote{We emphasize that the pole trajectories are related by complex conjugation but are not invariant under $\omega\to-\omega$. This originates from the fact that the HHLL correlator $G_{\mt H}(\omega)$ generally does not have frequency-reflection symmetry. By contrast, the two-sided thermal correlator $G_{\mt{LR}}(\omega)$ is even in frequency, so its meromorphic pole spectrum exhibits the fourfold symmetry $\{\omega_n,-\omega_n,\omega_n^*,-\omega_n^*\}$.}:
\begin{equation}
 \omega_{m,n}^{\rm pole}=\pm 4\ii\kappa\left(\alpha_1+\frac{m}{b}+bn\right) - 2\left(\alpha_1+\frac{m}{b}+bn\right)^2 \,,  \qquad \{m,n\}\in\mathbb Z_{\geq0}\,,
\end{equation}
which are parametrized by two nonnegative integers $(m,n)$. The other two pole branches from these denominator factors reproduce the same frequency trajectories through Liouville reflection symmetry. The remaining four pole branches generate two additional families,
\begin{equation}
    \widetilde{\omega}_{m,n}^{\rm pole}
    =\pm4\ii\kappa\left(Q-\alpha_1+\frac{m}{b}+bn\right)
    -2\left(Q-\alpha_1+\frac{m}{b}+bn\right)^2\,.
    \label{eq:liouville-additional-pole-families}
\end{equation}
In the semiclassical limit $b\to0$, these additional poles are pushed to frequencies of order $c$ and therefore do not contribute to the finite-frequency QNM spectrum. Nevertheless, they remain part of the exact finite-$c$ pole--zero structure and become relevant in the large-frequency regime.

By contrast, the $m=0$ branches of $\omega_{m,n}^{\rm pole}$, namely 
\begin{equation}\label{eq:QNM-finite-c}
\boxed{ \omega_{m=0,n}^{\rm pole}=\pm 4\ii\kappa\left(\alpha_1+bn\right) - 2\left(\alpha_1+bn\right)^2  \,,  \qquad n\in\mathbb Z_{\geq0} \,. }
\end{equation}
remain at finite frequency in the semiclassical limit. Taking the strict limit $c \to \infty$, \ie $b \to 0$, this pole branch reduces to \cite{Turiaci:2016cvo}
\begin{equation}
\omega_{m=0,n}^{\rm pole} \overset{c \to \infty}{=} \pm 4\ii\kappa\left(\alpha_1+bn\right) 
=
\pm \ii \frac{2\pi}{\beta_{\mt H}} (\Delta + 2n) \,,  \qquad n\in\mathbb Z_{\geq0} \,,
\end{equation}
using $\kappa=\frac{\pi}{b\beta_{\mt H}}$ and $\alpha_1\approx bh=b\Delta/2$. 
These poles in the lower half-plane precisely reproduce the semiclassical BTZ QNM spectrum $\omega_n^{\mt{QNM}}$ \cite{Birmingham:2001pj,Turiaci:2016cvo} of a scalar field with conformal dimension $\Delta$, after identifying the effective temperature $T_{\mt H}$ with the BTZ Hawking temperature. Importantly, the finite-$c$ correction to this semiclassical QNM trajectory $\omega_{m=0,n}^{\rm pole}$ provides an explicit finite-$c$ quadratic deformation of the semiclassical linear QNMs.

\begin{figure}[t]
    \centering
    \includegraphics[width=5.5in]{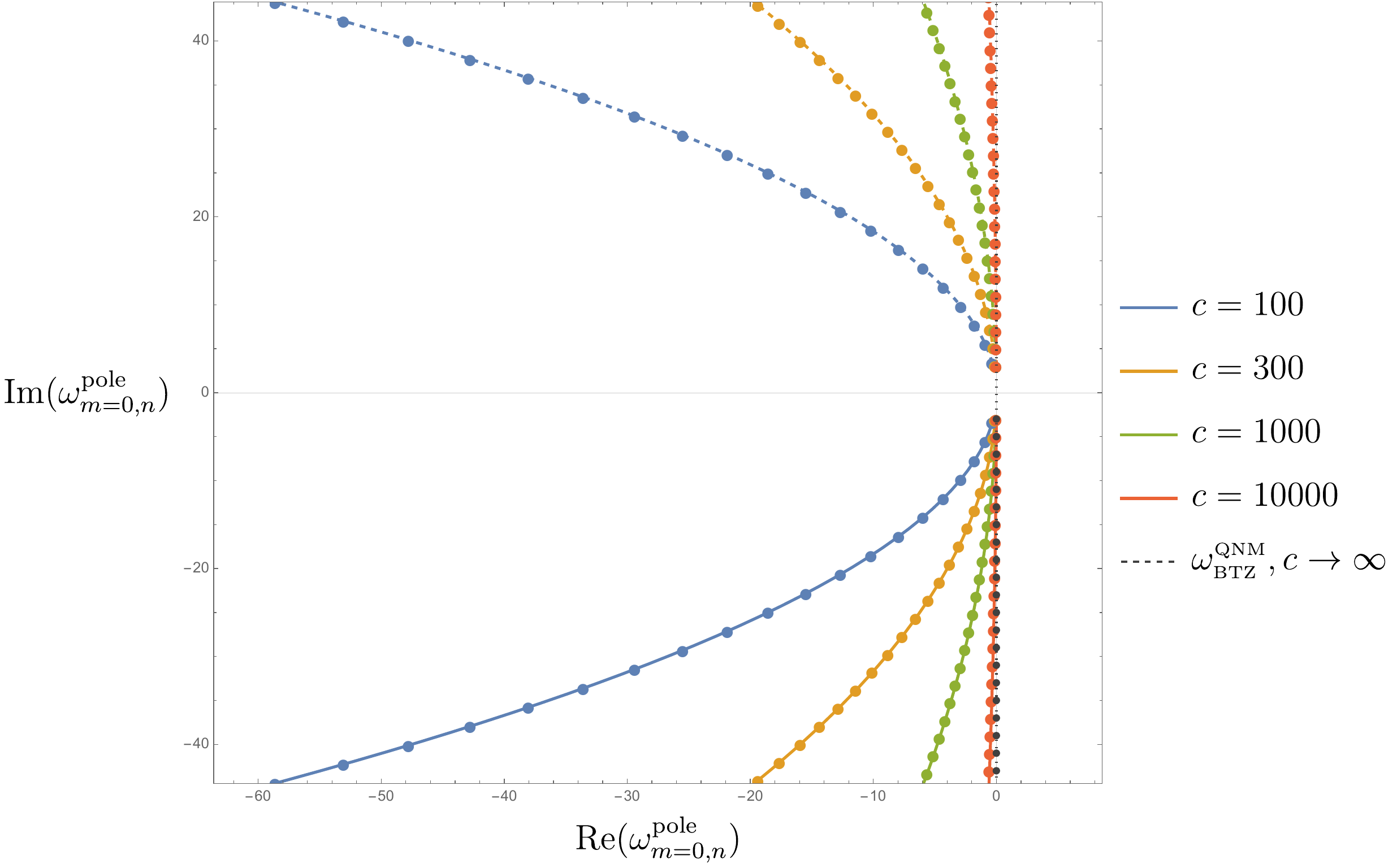}
    \caption{Pole trajectories $\omega_{m=0,n}^{\rm pole}$ of the analytically continued Liouville DOZZ coefficient in the complex $\omega$-plane for different central charges with $\Delta=3$ and $\beta_{\mt H}=2\pi$. The discrete points represent the poles $\omega_{m=0,n}^{\rm pole}$, while the colored curves illustrate their exact parabolic trajectories. As $c$ increases, the parabolas approach the imaginary-frequency axis, recovering the semiclassical BTZ QNM spectrum indicated by the black dotted line.}
    \label{fig:liouville-dozz-pole-trajectories}
\end{figure}

More precisely, we use the effective BTZ temperature $T_{\mt H}$ defined in eq.~\eqref{eq:heavy-state-BTZ-temperature},  
\begin{equation}
    \kappa=\sqrt{\frac{\Delta_{\mt H}}{2}-\frac{c-1}{24}}=\sqrt{ \frac{1}{24}+\frac{\pi^2 c}{6\beta_{\mt H}^2}}\,,
    \label{eq:liouville-momentum-temperature}
\end{equation}
and the Liouville conformal dimension,
\begin{equation}
    \alpha_1=\frac{Q-\sqrt{Q^2-2\Delta}}{2} =\frac{\sqrt{c-1}-\sqrt{c-1-12\Delta}}{2\sqrt6}\,.
    \label{eq:liouville-light-momentum-dimension}
\end{equation}
For the semiclassical branch $0<b\leq 1$, the Liouville coupling $b$ is determined by
\begin{equation}
b^2 =\frac{c-13-\sqrt{(c-1)(c-25)}}{12}\,. 
\end{equation}
To determine the leading finite-$c$ correction, substituting all these expressions into the exact pole trajectory \eqref{eq:QNM-finite-c} yields 
\begin{equation}
\begin{aligned}
    \omega_{m=0,n}^{\rm pole}
    ={}&\pm\ii\frac{2\pi}{\beta_{\mt H}}(\Delta+2n)
    -\frac{3}{c}(\Delta+2n)^2\\
    &\pm\frac{\ii}{c}\left[
    \frac{6\pi}{\beta_{\mt H}}\Delta(\Delta-2)
    +\left(\frac{13\pi}{\beta_{\mt H}}+\frac{\beta_{\mt H}}{4\pi}\right)(\Delta+2n)
    \right]+\mathcal O(c^{-2})\,.
\end{aligned}
\label{eq:liouville-finite-c-qnm-expansion}
\end{equation}
The leading finite-$c$ correction contains a quadratic real contribution, which bends the semiclassical BTZ QNM line into the complex-frequency plane. In fact, the quadratic deformation of the QNM spectrum can be seen directly from the exact pole trajectory. Since $\kappa$, $b$, and $\alpha_1$ are all real in the heavy-light configuration considered here, the real part of the pole frequencies, \ie the deformation of the linear imaginary frequency, is given exactly by
\begin{equation}
   \boxed{ \operatorname{Re}\omega_{m=0,n}^{\rm pole} =-2(\alpha_1+bn)^2 =-\frac{3}{c}(\Delta+2n)^2+\mathcal O(c^{-2})\,.}
    \label{eq:liouville-qnm-real-deformation}
\end{equation}
Thus, the semiclassical QNM spectrum, which lies along the imaginary-frequency axis, acquires a quadratic real deformation at finite $c$. This provides an explicit realization of the smoothing mechanism for bouncing-geodesic singularities discussed in the previous section. Furthermore, the exact pole trajectory admits a simple geometric interpretation, since the real and imaginary parts satisfy
\begin{equation}
    \operatorname{Re}\omega_{m,n}^{\rm pole}
    =-\frac{1}{8\kappa^2}\left(\operatorname{Im}\omega_{m,n}^{\rm pole}\right)^2\,.
    \label{eq:liouville-qnm-parabola}
\end{equation}
Thus, at finite $c$, the poles lie exactly on a parabola opening toward the negative real-frequency axis. As illustrated in figure~\ref{fig:liouville-dozz-pole-trajectories}, the parabolic trajectories become progressively less curved as $c$ increases. In the semiclassical limit, where $\kappa^2\propto c$ and $c\to\infty$, they approach the imaginary-frequency axis, recovering the linear QNM spectrum of the BTZ black hole.

\begin{figure}[t]
    \centering
    \includegraphics[width=5.5in]{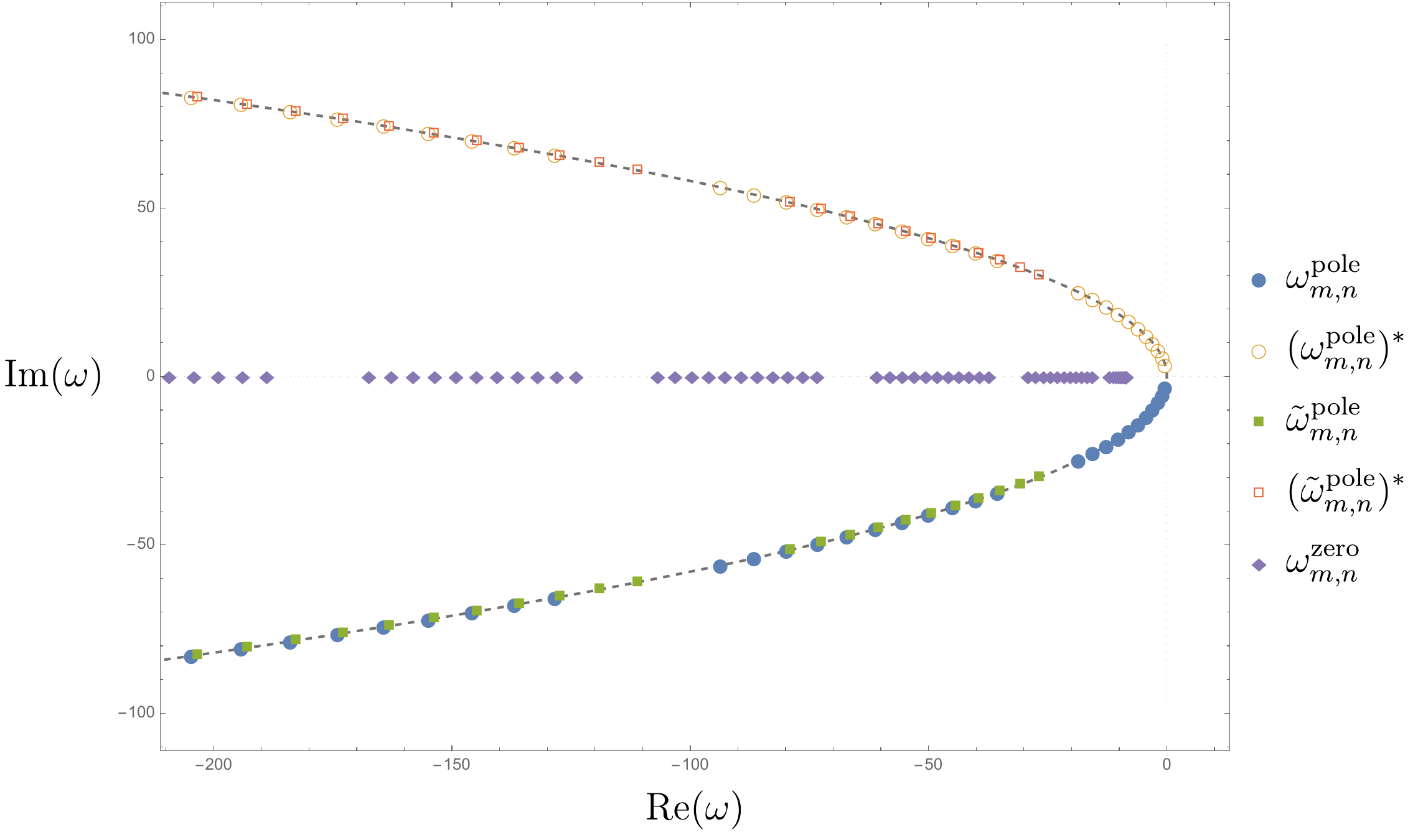}
    \caption{Pole--zero structure of the analytically continued Liouville DOZZ coefficient in the complex-frequency plane at $c=100$. The four pole branches, labeled by two integers $(m,n)$, follow the parabolic trajectory \eqref{eq:liouville-qnm-parabola}, while the zeros lie on the negative real-frequency axis, beginning at $\omega=-2\kappa^2$.}
\label{fig:liouville-dozz-pole-zero-structure}
\end{figure}

The smoothing associated with this trajectory can now be read directly
from subsection~\ref{subsec:Quadratic-QNM-Model}. For the lower-half-plane
branch of eq.~\eqref{eq:QNM-finite-c}, the semiclassically connected
inverse is
\begin{equation}
    z_{\rm sc}(\omega)
    =-\frac{\alpha_1}{b}
     +\frac{\ii\kappa}{b}
      \left(\sqrt{1+\frac{\omega}{2\kappa^2}}-1\right).
    \label{eq:liouville-inverse-quadratic}
\end{equation}
Consequently, its $\ell$th Stokes factor is
\begin{equation}
    \ee^{2\pi\ii\ell z_{\rm sc}(\omega)}
    =\ee^{-2\pi\ii\ell\alpha_1/b}
      \exp\left[
        -2\tau_\ell C
         \left(\sqrt{1+\frac{\omega}{C}}-1\right)
      \right],
    \quad
    C=2\kappa^2,\quad
    \tau_\ell=\frac{\pi\ell}{2b\kappa}\,.
    \label{eq:liouville-quadratic-model-matching}
\end{equation}
Apart from the constant phase, this is precisely
eq.~\eqref{eq:quadratic-exact-Stokes-action}, with
$c$ replaced by $C$ and $\tau_*$ by $\tau_\ell$.
Since $C=O(c)$ and $\tau_\ell\to\ell\beta_{\mt H}/2$ at fixed $\ell$,
the exact transform
\eqref{eq:quadratic-exact-transform-erfc} and its local limit
\eqref{eq:quadratic-bump-scaling} apply to this Stokes contribution
directly. At finite $c$, the square-root UV behavior leaves the would-be displaced singularity regular, whereas the strict semiclassical limit produces the corresponding displaced singularity. In the IR window, the exponent is $ \tau_\ell\omega+\tau_\ell\omega^2/(4C)+\cdots$: its second-order truncation, with appropriate contour prescription inherited from the exact continuation, reproduces the leading local profile, as in eq.~\eqref{eq:quadratic-IR-matching}. The characteristic width is $O(c^{-1/2})$. The remaining pole--zero families and conformal-block factors determine the embedding of this contribution into the full Liouville correlator.

To complete our discussion of the pole--zero structure, we note that the DOZZ coefficient also possesses zeros arising from its numerator. For generic fixed $\alpha_1$ and $\alpha_2$, the relevant $\alpha_3$-dependent factor is $\Upsilon_b(2\alpha_3)$, whose zeros are located at
\begin{equation}
    2\alpha_3=-\frac{m}{b}-nb
    \quad\text{or}\quad
    2\alpha_3=Q+\frac{m}{b}+nb\,,
    \qquad m,n\in\mathbb{Z}_{\geq0}\,.
    \label{eq:liouville-dozz-zero-families}
\end{equation}
In terms of the transition frequency $\omega=2[h(\alpha_3)-h(\alpha_2)]$, we obtain two families of zeros in the complex-frequency plane,
\begin{equation}
\begin{aligned}
\omega_{m,n}^{\rm zero}
&=-2\kappa^2
-\frac12\left(\frac{m}{b}+nb\right)^2,\\
\widetilde{\omega}_{m,n}^{\rm zero}
&=-2\kappa^2
-\frac12\left(Q+\frac{m}{b}+nb\right)^2.
\end{aligned}
\end{equation}
Since $Q=b+b^{-1}$, the second family is contained in the first through the reindexing $\widetilde{\omega}_{m,n}^{\rm zero}=\omega_{m+1,n+1}^{\rm zero}$. Consequently, all distinct zeros are captured by $\omega_{m,n}^{\rm zero}$. Unlike the complex pole trajectories, these zeros lie entirely on the negative real-frequency axis, satisfying
\begin{equation}
\operatorname{Im}\omega_{m,n}^{\mathrm{zero}}=0\,,
\qquad
\operatorname{Re}\omega_{m,n}^{\mathrm{zero}}\leq \omega_{0,0}^{\mathrm{zero}}=-2\kappa^2\,.
\end{equation}
Interestingly, the rightmost zero $\omega_{0,0}^{\mathrm{zero}}$ coincides with the lower endpoint of the transition-energy spectrum of physical Liouville primary states. Indeed, for physical Liouville momenta $\alpha_3=Q/2+\ii P$, the transition frequency satisfies
\begin{equation}
\omega=2(P^2-\kappa^2)\geq-2\kappa^2\,,
\qquad P\geq0\,.
\end{equation}
Thus, the rightmost zero lies precisely at the continuum threshold, while all remaining zeros lie below this threshold and arise from analytic continuation in the Liouville momentum. Furthermore, since $\kappa^2\propto c$ in the semiclassical limit, all these zero frequencies tend to negative infinity as $c\to\infty$. Consequently, they have no counterpart in the semiclassical holographic correlator of the BTZ black hole, consistent with the thermal product formula \eqref{eq:thermal-product-formula}, which contains only QNM poles and no zeros.

Finally, our analysis of the DOZZ coefficient in Liouville theory reveals a richer pole--zero structure at finite $c$ than that of semiclassical holographic correlators in the strict $c\to\infty$ limit. In particular, the DOZZ poles are organized into a two-dimensional lattice inherited from the zeros of $\Upsilon_b(x)$, going beyond the one-dimensional polynomial distribution $P(n)$ considered in the previous section. The DOZZ coefficients also exhibit a nontrivial distribution of zeros, suggesting that such intricate pole--zero structures may arise more generally in finite-$c$ holographic correlators. We return to these features, the generalization to more generic pole distributions, and the constraints on the associated pole--zero data in the discussion section and appendix~\ref{app:UV-IR-constraints}.

%%%%%%%%%%%%%%%%%%%%%%%%%%%%%%%%%%%%%%%%%%%%%%%%%%%%%%%%%%%%
\section{Gaussian Smoothing from the Thermal Spectral Decomposition}
\label{sec:spectral-smoothing}

In the previous sections, we investigated how finite-$c$ corrections deform the semiclassical QNM spectrum and smooth out the signals of semiclassical bouncing-geodesic singularities, \ie the complex-time singularities of thermal two-sided correlators $G_{\mt{LR}}$ and fixed-state correlators $G_{\mt H}$. However, it is natural to expect other types of quantum corrections that can also smooth the singularity structure of analytically continued holographic correlators, \eg by introducing a damping factor in frequency space.

In this section, we investigate another potential smoothing mechanism arising from finite-$\GN$ corrections by considering the thermal spectral decomposition of the holographic thermal correlator. The basic observation is that the sharp bouncing-geodesic singularity emerges in the strict semiclassical limit $\GN\to0$, or equivalently $c\to\infty$, where the thermal ensemble of black hole microstates is dominated by a saddle point at a characteristic energy. According to the eigenstate thermalization hypothesis (ETH) \cite{Srednicki:1999,DAlessio:2015qtq,Lashkari:2016vgj}, the leading thermal correlator of sufficiently simple probe operators can be approximated by the corresponding correlator in a typical black hole microstate. From the bulk perspective, this saddle-point approximation effectively replaces the thermal ensemble by a single classical black hole geometry, whose analytically continued thermal correlator $G_{\mt{LR}}$ exhibits the bouncing-geodesic singularity. At finite $\GN$, however, the thermal ensemble has a non-vanishing energy width, and the spectral decomposition receives corrections associated with transitions between microscopic states of different energies. In particular, the nonlinear energy dependence of the black hole density of states introduces an effective Gaussian correction to the frequency-space representation of the two-sided correlator $G_{\mt{LR}}(\omega)$, which is absent in the strict semiclassical limit. After transforming back to the time domain, this suppression factor gives rise to a Gaussian regularization kernel that can smooth out the divergence associated with the bouncing-geodesic singularity.

\subsection{Spectral decomposition and strict semiclassical limit}
\label{subsec:canonical-gaussian}

Let us begin by recalling the exact spectral decomposition of the two-sided thermal correlator for a Hermitian scalar operator $\cO$, \ie
\begin{equation}
    G_{\mt{LR}}(T)
    =\frac{1}{Z(\beta)}\sum_{m,n}
    \ee^{-\frac{\beta}{2}(E_m+E_n)}
    \ee^{-\ii(E_n-E_m)T}
    |\cO_{mn}|^2\,,
    \label{eq:thermal-spectral-sum-local}
\end{equation}
where $Z(\beta)=\sum_n\ee^{-\beta E_n}$ is the canonical partition function. As a holographic imprint of the black hole singularity, the analytically continued correlator exhibits an infinite family of complex-time singularities of the form\footnote{The same holographic thermal correlator, analytically continued onto a secondary Riemann sheet, exhibits a similar power-law singularity at a real Lorentzian time,
\begin{equation*}
G_{\mt{LR}}^{\rm bounce}(T,\bx=0) \sim \frac{1}{(T-T_{\rm crt})^{2\Delta}} \,.
\end{equation*}
A similar Gaussian smoothing mechanism could also apply to this real-time singularity, with $\nu=2\Delta-1$. We will not pursue this extension separately here and instead focus on the generic complex-time singularities.}
\begin{equation}
G_{\mt{LR}}(T,\bk)\supset \, \sum_{n_\pm} \frac{1}{(T-T_{n_+,n_-})^{\nu+1}} \,.  
\end{equation}
In this section, we work directly with the energy eigenstates and omit the spatial momentum, since it is implicitly included in the local operator matrix elements.

In a holographic CFT at large but finite $c$, the black hole spectrum contains an exponentially large number of closely spaced energy levels. After coarse graining over energy intervals much larger than the microscopic level spacing, the discrete spectrum can be approximated by a smooth density of states $\rho(E)$. Correspondingly, we define the microcanonical entropy,
\begin{equation}
    S(E)\equiv \log \rho(E) \sim \mathcal O\left( \frac{1}{\GN} \right)  \sim \mathcal O(c) \,,
    \label{eq:thermal-smooth-density}
\end{equation} 
with respect to a fixed coarse-graining energy scale. We further introduce the average energy and energy difference,
\begin{equation}
    E=\frac{E_m+E_n}{2}\,,\qquad
    \omega=E_n-E_m\,,
    \label{eq:thermal-average-difference}
\end{equation}
together with the coarse-grained squared matrix element $M_{\cO}(E,\omega) \equiv \overline{|\cO_{mn}|^2}$. Here, the coarse-grained average is taken over pairs of energy eigenstates satisfying $E_m\simeq E-\omega/2$ and $E_n\simeq E+\omega/2$. Replacing the discrete sums by smooth integrals over the physical energy spectrum, we obtain the smooth spectral representation of the holographic thermal correlator, \ie
\begin{equation}
    G_{\mt{LR}}(T) \simeq \frac{1}{Z(\beta)}\int \dd E\,\dd\omega\,\ee^{-\beta E-\ii\omega T}
    \rho\left(E+\frac{\omega}{2}\right) \rho\left(E-\frac{\omega}{2}\right)M_{\cO}(E,\omega)\,,
\label{eq:thermal-coarse-spectral-integral}
\end{equation}
where the partition function is given by $Z(\beta)=\int \dd E\,\ee^{S(E)-\beta E}$. Equivalently, the frequency-space correlator becomes
\begin{equation}
\begin{split}
    G_{\mt{LR}}(\omega)\simeq \frac{2\pi}{Z(\beta)} \int \dd E\,
    \ee^{-\beta E}\rho\left(E+\frac{\omega}{2}\right)\rho\left(E-\frac{\omega}{2}\right)M_{\cO}(E,\omega)\,.
\end{split}
\label{eq:thermal-coarse-frequency-integral}
\end{equation}

To characterize the matrix elements appearing in this spectral decomposition, we assume the eigenstate thermalization hypothesis (ETH), which applies to sufficiently simple operators in an appropriate chaotic energy band. According to ETH, the matrix elements of a simple operator in the energy eigenbasis take the form \cite{Srednicki:1999,DAlessio:2015qtq,Lashkari:2016vgj}
\begin{equation}
	\mathcal{O}_{mn}=O(E)\delta_{mn}+\ee^{-S(E)/2}f_{\mathcal{O}}(E,\omega)R_{mn}\,.
\end{equation}
Here, the diagonal part is controlled by the smooth ETH function $O(E)$ and diagonal fluctuations, while the off-diagonal part describes microscopic fluctuations controlled by the random variable $R_{mn}$. We focus on the off-diagonal sector relevant to the nontrivial time dependence. Since $R_{mn}$ has zero coarse-grained mean and unit variance, $\overline{|R_{mn}|^2}=1$, the coarse-grained squared matrix element can be expressed as
\begin{equation}
    M_{\cO}(E,\omega) 
    \simeq \ee^{-S(E)}F_{\cO}(E,\omega),
    \qquad
    F_{\cO}(E,\omega)\equiv |f_{\cO}(E,\omega)|^2\,.
    \label{eq:thermal-ETH-matrix-element}
\end{equation}
Combining the above ingredients, and following the ETH analysis of thermal two-point functions in \cite{Murthy:2019fgs}, we can recast the holographic thermal correlator as
\begin{equation}
   \boxed{  G_{\mt{LR}}(\omega)\simeq \frac{2\pi}{Z(\beta)} \int \dd E\, \ee^{\Phi(E,\omega)} F_{\cO}(E,\omega) \,,}
\label{eq:thermal-coarse-frequency-integral02}
\end{equation}
with 
\begin{equation}
\Phi(E,\omega) = S\left(E-\frac{\omega}{2}\right) + S\left(E+\frac{\omega}{2}\right)-S(E)-\beta E \sim \mathcal O\left( \frac{1}{\GN} \right) \,. 
\end{equation}

In the semiclassical limit $\GN\to0$, we can explicitly evaluate the mean-energy integral using the saddle-point approximation because the exponent $\Phi(E,\omega)$ becomes parametrically large. Since the characteristic energy of a macroscopic black hole also scales as $E\sim \mathcal O(1/\GN)$, we can expand the exponent for energy differences satisfying $|\omega|\ll E$, obtaining
\begin{equation}
\Phi(E,\omega)=S(E)-\beta E+\frac{S''(E)}{4}\omega^2+\mathcal O\left(S^{(4)}(E)\omega^4\right)\,.
\end{equation}
The leading contribution is thus governed by the canonical saddle $E_\beta$, determined by
\begin{equation}
\frac{\dd S(E)}{\dd E}\bigg|_{E=E_\beta}=\beta \,. 
\end{equation}
We note that the second derivative of the entropy is related to the heat capacity $C_V$ through \cite{Murthy:2019fgs}
\begin{equation}
\boxed{S''(E_\beta)=-\frac{\beta^2}{C_V}<0\,, \qquad C_V  = \left( \frac{\partial E}{\partial T}\right)_V\sim \mathcal O\left( \frac{1}{\GN} \right) \,.}
\end{equation}
The negative sign follows from the thermodynamic stability of the black hole. The coefficient $S''(E_\beta)$ also determines the variance of Gaussian fluctuations of the mean energy around the canonical saddle, since the saddle expansion is given by
\begin{equation}
\Phi(E,\omega) \approx  S(E_\beta)-\beta E_\beta-\frac{1}{2}\frac{(E-E_\beta)^2}{(\Delta E)^2}-\frac{1}{4}\frac{\omega^2}{(\Delta E)^2}\,, \quad \text{with} \quad (\Delta E)^2=-\frac{1}{S''(E_\beta)}\,. 
\end{equation}

To perform the saddle-point approximation in the strict semiclassical limit $\GN\to0$, we first note that the quadratic dependence on $\omega$ vanishes because $(\Delta E)^2\sim \mathcal O(1/\GN)\to\infty$. Performing the Gaussian integral over the mean energy then gives
\begin{equation}
G_{\mt{LR}}(\omega)\simeq 2\pi F_{\cO}(E_\beta,\omega)\,, \qquad \GN\to 0\,,
\label{eq:thermal-semiclassical-ETH}
\end{equation}
where we have also applied the same saddle-point approximation to the partition function. Thus, the leading semiclassical thermal correlator is determined by the spectral function of the probe operator evaluated at the characteristic black hole energy $E_\beta$. Intuitively, this is because the relative width $\Delta E/E_\beta\sim\sqrt{\GN}$ vanishes in the semiclassical limit. Consequently, the canonical ensemble becomes sharply localized around the canonical energy $E_\beta$, so that the dominant black hole microstates are characterized by the same macroscopic geometry. For sufficiently simple probe operators, the smooth ETH spectral function can therefore be evaluated at the canonical saddle $E_\beta$.

\subsection{Gaussian smoothing kernel}

Having derived the semiclassical relation \eqref{eq:thermal-semiclassical-ETH}, we can now connect the ETH spectral function $F_{\cO}(E_\beta,\omega)$ to the complex-time singularities of the analytically continued thermal correlator $G_{\mt{LR}}(T)$. As discussed in the previous sections, the large-frequency expansion derived from the QNM spectrum contains a family of contributions associated with an infinite set of complex-time singularities. In particular, focusing on a single sector labeled by $\alpha=(n_+,n_-)$, we have
\begin{equation}
G_{\mt{LR},\alpha}^{(0)}(T)\sim \frac{A_\alpha}{(T-T_\alpha)^{\nu+1}}\,,
\end{equation}
where $T_\alpha$ denotes the corresponding singularity location. The inverse Fourier transformation then implies the following large-frequency behavior,
\begin{equation}
G_{\mt{LR},\alpha}^{(0)}(\omega) \approx 2\pi F^{(0)}_{\cO,\alpha}(E_\beta,\omega) \sim C_\alpha\,\omega^\nu \ee^{\ii\omega T_\alpha} \,. 
\label{eq:thermal-ETH-QNM-asymptotic}
\end{equation}
Here, the superscript $(0)$ indicates the strict semiclassical limit. This relation thus provides a direct connection between the semiclassical spectral data and the complex-time singularities. In particular, the phase $\ee^{\ii\omega T_\alpha}$ encodes the location of the singularity, while the power-law factor $\omega^\nu$ determines its leading singular behavior.

However, the result \eqref{eq:thermal-ETH-QNM-asymptotic} is derived in the strict semiclassical limit,
\begin{equation}
\GN \to 0 \,, \qquad \Delta E \sim \mathcal O\left( \frac{1}{\sqrt{\GN}} \right) \,, \qquad \frac{\Delta E}{E_\beta} \sim \mathcal O(\sqrt{\GN}) \to 0 \,.   
\end{equation}
At finite $\GN$, the nontrivial energy dependence of the spectral density of black hole microstates generates a quadratic correction to the spectral exponent, \ie
\begin{equation}
\left.\frac{\rho(E+\omega/2)\rho(E-\omega/2)}{\rho(E)^2} \right|_{E=E_\beta}\simeq \exp\left[-\frac{\omega^2}{4(\Delta E)^2}\right] \,. 
\end{equation}
For a thermodynamically stable black hole, this quadratic correction is negative and introduces a Gaussian suppression kernel in the large-frequency regime. Retaining this correction in the saddle-point approximation thus gives\footnote{The thermodynamic Gaussian factor also appears in the ETH analysis of a thermally regulated two-point function in \cite{Murthy:2019fgs}. Related Gaussian corrections arising from energy fluctuations in the thermal ensemble were also discussed in \cite{Batra:2026gbq}, where the authors explicitly analyzed the gravitational backreaction of a probe worldline in two-dimensional dilaton gravity.}
\begin{equation}
\boxed{G_{\mt{LR},\alpha}(\omega)\simeq G_{\mt{LR},\alpha}^{(0)}(\omega)\exp\left[-\frac{\omega^2}{4(\Delta E)^2}\right]
\sim C_\alpha\,\omega^\nu \ee^{\ii\omega T_\alpha}\exp\left[-\frac{\omega^2}{4(\Delta E)^2}\right]\,.
}
\label{eq:thermal-gaussian-frequency}
\end{equation}
Importantly, the entropy-induced Gaussian kernel is independent of the sector $\alpha$, reflecting its universal thermodynamic origin. Since $\Delta E\sim 1/\sqrt{\GN}$, the Gaussian correction disappears at fixed $\omega$ in the strict semiclassical limit, recovering the original power-law divergences. At finite $\GN$, however, the suppression becomes significant when $\omega\sim\Delta E$. Similar to the Gaussian smoothing mechanism discussed in subsection~\ref{subsec:QNM-deformation}, this frequency-dependent suppression corresponds to a Gaussian convolution in the time domain, 
\begin{equation}
G_{\mathrm{LR},\alpha}(T) \simeq \int_{-\infty}^{+\infty}dT'\,
{\frac{\Delta E}{\sqrt{\pi}}
e^{-(\Delta E)^2(T-T')^2}}
G_{\mathrm{LR},\alpha}^{(0)}(T')\,, 
\end{equation}
with a characteristic time width $\Delta T\sim 1/\Delta E\sim\mathcal O(\sqrt{\GN})$. This Gaussian convolution at finite $\GN$ thus smooths out all corresponding complex-time singularities.   

Finally, the physical origin of this Gaussian smoothing mechanism can be understood from the thermal averaging over microscopic black hole states. The canonical ensemble involves a finite range of black hole energies around the thermal saddle, allowing transitions between microstates with different energies. Due to the concavity of the black hole entropy, the statistical weight of these transitions is increasingly suppressed as their energy difference grows. This gives rise to the Gaussian factor derived in eq.~\eqref{eq:thermal-gaussian-frequency}, whose characteristic width is determined by the canonical energy fluctuations. Our calculation here has isolated the entropy-induced contribution, while additional finite-$\GN$ corrections may also arise from the spectral function $F_{\cO}(E,\omega)$. Such corrections are closely related to the deformations of QNM trajectories investigated in the previous sections. Therefore, the thermal spectral decomposition provides a different microscopic perspective on the Gaussian smoothing of complex-time singularities in the analytically continued thermal correlator.

%%%%%%%%%%%%%%%%%%%%%%%%%%%%%%%%%%%%%%%%%%%%%%%%%%%%%%%%%%%%%%%%
\section{Discussion and Outlook}
\label{sec:discussion}
%%%%%%%%%%%%%%%%%%%%%%%%%%%%%%%%%%%%%%%%%%%%%%%%%%%%%%%%%%%%%%%%

Having established how finite-$c$ effects can smooth the bouncing-geodesic singularities, we now move to discuss several related topics and possible directions for future investigation.

\subsection{BTZ black hole}
\label{subsec:btz vs high dim bh}

\begin{figure}[t]
    \centering
    \includegraphics[width=4in]{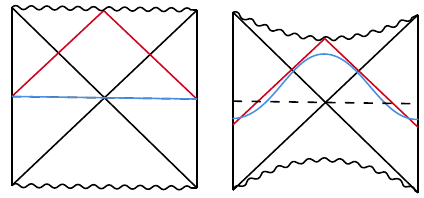}
    \caption{Penrose diagrams and geodesic configurations. Left: BTZ. The red limiting null configuration is not approached by spacelike geodesics at finite boundary time. The blue curve is the spacelike geodesic at $T=0$. Right: a higher-dimensional black hole. A family of blue spacelike geodesics approaches the bouncing geodesic presented by the red curve.}
    \label{fig:PenroseBTZ}
\end{figure}

\begin{figure}[t]
    \centering
    \includegraphics[width=1.0\linewidth]{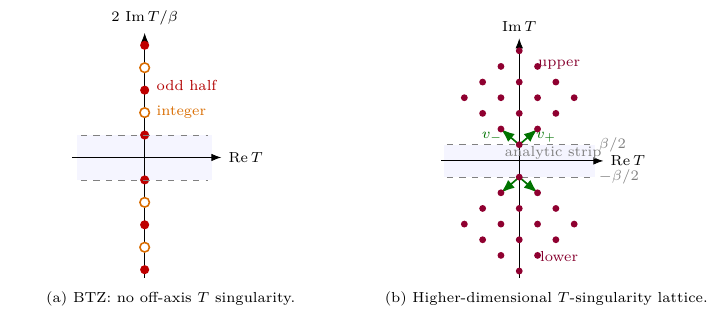}
    \caption{Locations of complex-time singularities of the analytically continued thermal correlator $G_{\mt{LR}}(T,k)$ in the complex $T$-plane.}
    \label{fig:g12_tk_lattice_classical}
\end{figure}

The BTZ black hole provides an analytically tractable example that allows us to illustrate several aspects of our analysis explicitly. However, unlike higher-dimensional AdS-Schwarzschild black holes, the BTZ geometry has no curvature singularity, and its spacelike geodesics do not approach a bouncing null configuration at finite boundary time. Formally taking $d=2$ in the critical bouncing time \eqref{eq:tstar} gives
\begin{equation}
T_{\rm crt}=-\frac{\beta}{2}\cot\frac{\pi}{d}
\overset{d=2}{\,=\,}0 \,.
\end{equation}
Naively, as shown in figure \ref{fig:PenroseBTZ}, this critical value can be understood from the square Penrose diagram of BTZ, where a pair of radial null rays originating from opposite boundaries can reach the same point on the $r=0$ boundary. However, this null configuration is not approached by a family of spacelike geodesics at any finite boundary time. Instead, it is known that the renormalized geodesic length of BTZ grows linearly at late times,
\begin{equation}
\mathcal L_{\rm ren}(T)
=2\log\left[\frac{\beta}{\pi}\cosh\left(\frac{\pi T}{\beta}\right)\right]
\approx\frac{2\pi}{\beta}T \,, \qquad T\gg\beta \,.
\end{equation}
We can see that there are similarities, as well as qualitative differences, between BTZ and higher-dimensional AdS-Schwarzschild black holes. In particular, as shown in figure~\ref{fig:g12_tk_lattice_classical}, the slope of linear QNM in BTZ is purely imaginary, causing the complex-time singularity lattice to collapse onto the imaginary axis. Notably, the resulting singularities should be interpreted as the thermal and chiral structure of the correlator, rather than a distinct family of bouncing-geodesic singularities.

However, the advantage of BTZ is that the relevant calculations can be performed explicitly. Focusing on the non-rotating BTZ black hole, we begin with the position-space thermal correlator $G_{\mt{LR}}^{\mt{BTZ}}(T,x)$ in the non-winding sector, which is known analytically. Up to an overall normalization, it takes the form
\begin{equation}\label{eq:btz-two-sided-correlator02}
G_{\mt{LR}}^{\mt{BTZ}}(T,x)\propto
\left[\cosh\left(\frac{\pi(T+x)}{\beta}\right)
\cosh\left(\frac{\pi(T-x)}{\beta}\right)
\right]^{-\Delta} \,.
\end{equation}
The correlator factorizes into two chiral contributions, whose singularities at fixed spatial separation $x$ are fully determined by the zeros of the two hyperbolic cosine factors:
\begin{equation}
T_{n}^{\pm}=\pm x+i\beta\left(n+\frac12\right),
\qquad n\in\mathbb{Z} \,, 
\end{equation}
which are the lightcone/OPE singularities and their thermal images. Analytic continuation around these singularities changes the monodromy of the chiral factors without introducing additional singularity locations on secondary sheets. By contrast, in higher-dimensional black holes, analytic continuation can access distinct geodesic branches, including bouncing geodesics that approach the spacelike singularity. These trajectories give rise to additional complex-time singularities on secondary sheets of $G_{\mt{LR}}(T,\bx)$, as discussed in section~\ref{sec:semiclassical}.

\begin{figure}[t]
    \centering
    \includegraphics[width=5in]{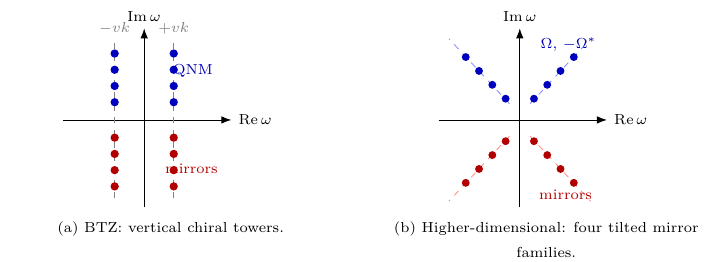}
    \caption{Pole distributions of $G_{\mt{LR}}(\omega,\bk)$. Poles in the lower half of the complex $\omega$-plane correspond to QNMs.}
    \label{fig:QNM-BTZ}
\end{figure}

On the other hand, we can evaluate the Fourier transform exactly and get
\begin{equation}
G_{\mt{LR}}^{\mt{BTZ}}(\omega,k) \propto \left|
\Gamma\left(\frac{\Delta}{2} + i\frac{\beta}{4\pi}(\omega+k)\right) \right|^2 \cdot
\left|\Gamma\left(\frac{\Delta}{2}+ i\frac{\beta}{4\pi}(\omega-k)\right)\right|^2,
\end{equation}
As shown in figure \ref{fig:QNM-BTZ}, we thus obtain four branches of poles. In particular, the poles in the lower half of the complex-frequency plane reproduce the exact QNM spectrum of the BTZ black hole \cite{Birmingham:2001pj,Birmingham:2001hc,Cardoso:2001hn,Chen:2010ik,Zhang:2011ec}, namely 
\begin{equation}
\boxed{\omega_n^{\mt{BTZ}}(k)= -\ii\frac{4\pi}{\beta}
\left(n+\frac{\Delta}{2}\right) \pm k \,, \qquad n\geq0 \,.
}\label{eq:BTZ:QNMs}
\end{equation}
In particular, the slope of this linear QNM trajectory is purely imaginary, corresponding to taking
\begin{equation}
a_1=-\ii\frac{4 \pi}{\beta} \,, \qquad \, v_+=v_-=\ii\frac{\beta}{2} \,, \qquad \text{(BTZ)}
\end{equation}
in eq.~\eqref{eq:vpm-from-QNM}. Through the Hadamard product discussed in section~\ref{subsec:qnm-singularities}, the corresponding complex-time singularities of the fixed-momentum correlator $G_{\mt{LR}}^{\mt{BTZ}}(T,k)$ can be read from the linear QNM trajectory as 
\begin{equation}\label{eq:BTZ-complex-time-lattice}
\boxed{ T_N=\ii \frac{\beta N}{2} \,, \qquad N\in\mathbb{Z}\setminus\{0\} \,, \qquad (\text{BTZ})}
\end{equation}
with the local behavior $G_{\mt{LR}}^{\mt{BTZ}}(T,k) \sim \frac{1}{(T-T_N)^{2\Delta-1}}$. 
As explained in Appendix \ref{app:four-representations}, these singularities arise from the thermal images and collisions of the chiral singularities. By contrast, higher-dimensional black holes exhibit semi-classical linear QNM trajectories with a complex slope. As derived in eq.~\eqref{eq:vpm-Tstar}, the two shifts are $v_\pm=\pm|T_{\rm crt}|+\ii\beta/2$. This gives rise to additional complex-time singularities displaced from the imaginary axis by the characteristic bouncing time $T_{\rm crt}$, \ie
\begin{equation}
    T_{n_+,n_-}=\frac{\ii\beta}{2} +(n_+-n_-)|T_{\rm crt}|
    +\frac{\ii\beta}{2}(n_++n_-),\qquad n_\pm\geq0.
\end{equation}
The off-diagonal terms $n_+\ne n_-$ lie away from the imaginary axis and carry the time shifts from bouncing geodesics. The differences between BTZ and higher-dimensional black holes from the perspective of $G_{\mt{LR}}(T, k)$ have been summarized in figure \ref{fig:g12_tk_lattice_classical}.

We can also obtain the complex-time singularities directly from the analytic expression for $G_{\mt{LR}}^{\mt{BTZ}}(T,k)$. The spatial Fourier integral can be evaluated exactly in terms of the hypergeometric function,
\begin{equation}\label{eq:BTZ-GLR-Tk}
\begin{split}
G_{\mt{LR}}^{\mt{BTZ}}(T,k)&\propto
B\left(\Delta+\frac{\ii\beta k}{2\pi},
\Delta-\frac{\ii\beta k}{2\pi}\right) \\
&\quad\times {}_2F_1\left( \Delta+\frac{\ii\beta k}{2\pi}, \Delta-\frac{\ii\beta k}{2\pi};
\Delta+\frac12;-\sinh^2\frac{\pi T}{\beta}\right) \,,
\end{split}
\end{equation}
where $B(a,b)=\Gamma(a)\Gamma(b)/\Gamma(a+b)$ is the Euler beta function. The complex-time singularity structure is encoded in the analytic properties of the hypergeometric function. However, an important subtlety is that ${}_2F_1(a,b;c;z)$ is generally multivalued, and its analytic continuation must account for nontrivial monodromy when reconstructing the complete complex-time singularity lattice. Simply evaluating the principal branch at each complex-time point does not necessarily reveal all the singularities.

To see this explicitly, recall that the hypergeometric differential equation has regular singular points at $z=0,1,\infty$. The standard solution ${}_2F_1(a,b;c;z)$ is analytic at $z=0$, but generally develops a branch point at $z=1$. For the parameters appearing in \eqref{eq:BTZ-GLR-Tk}, one has $c-a-b=1/2-\Delta$. Therefore, for generic $\Delta>1/2$, the leading singular behavior near $z=1$ is given by
\begin{equation}
{}_2F_1(a,b;c;z)\sim (1-z)^{1/2-\Delta} \,.
\end{equation}
Since $z=-\sinh^2(\pi T/\beta)$, the condition $z=1$ corresponds to
\begin{equation}
T_n=\ii\beta\left(n+\frac12\right) \,, \qquad n\in\mathbb{Z} \,.
\end{equation}
Substituting the location of complex-time singularities to eq.~\eqref{eq:BTZ-GLR-Tk}, we can immediately recover the expected local singular behavior,
\begin{equation}
G_{\mt{LR}}^{\mt{BTZ}}(T,k)\sim
\frac{1}{(T-T_n)^{2\Delta-1}} \,.
\end{equation}
However, these points account only for the odd-$N$ singularities in \eqref{eq:BTZ-complex-time-lattice}.

The remaining singularities at $T=\ii\beta n$ arise from the nontrivial monodromy of the hypergeometric function. Although the original solution is regular at $z=0$, analytic continuation around the branch point $z=1$ generally mixes the two independent local solutions near $z=0$, whose leading behaviors are $1$ and $z^{1-c}$. Consequently, the analytically continued solution can acquire a singular contribution proportional to $z^{1/2-\Delta}$, giving rise to additional complex-time singularities at
\begin{equation}
z=-\sinh^2\frac{\pi T}{\beta}=0
\qquad\Longrightarrow\qquad
T=\ii\beta n \,, \qquad n\in\mathbb{Z}\setminus\{0\} \,.
\end{equation}
Together with the singularities associated with $z=1$, these reproduce the complete complex-time singularity lattice \eqref{eq:BTZ-complex-time-lattice}.

We can also get an explicit verification by taking a fixed $\Delta$. For example, this structure becomes particularly transparent for $\Delta=1$, where the hypergeometric function reduces to an elementary expression, 
\begin{equation}
G_{\mt{LR}}^{\mt{BTZ}}(T,k) \bigg|_{\Delta=1}
\propto \frac{2\beta\sin(kT)}{\sinh(\beta k/2)\sinh(2\pi T/\beta)} \,.
\end{equation}
The complete complex-time singularity lattice \eqref{eq:BTZ-complex-time-lattice} is now manifest in the denominator, while the apparent singularity at $T=0$ is removable. This provides an explicit illustration of how the full singularity structure, including the contributions associated with hypergeometric monodromy, is captured by the analytic continuation of the thermal correlator.

The same complex-time singularities can also be understood through the contour-pinching mechanism in the spatial Fourier transform. As discussed in appendix~\ref{app:four-representations}, particularly \eqref{eq:app-BTZ-time-lattice}--\eqref{eq:app-BTZ-power}, the branch points associated with the two chiral factors of the BTZ correlator move in opposite directions in the complex spatial plane as $T$ is analytically continued. Their collisions can pinch the spatial Fourier contour, reproducing the singularity lattice \eqref{eq:BTZ-complex-time-lattice}. In particular, while the odd-$N$ singularities correspond to ordinary lightcone/OPE images, the even-$N$ singularities arise from additional collisions between branch points belonging to different chiral factors at complex spatial positions. These additional singularities are a consequence of the special chiral structure of the BTZ correlator, rather than a distinct family of bouncing-geodesic singularities (see figure~\ref{fig:BTZ pinch}).

Finally, using the explicit holographic correlation function from BTZ or CFT$_2$, we can also illustrate the subtle difference between the pole structures of the fixed-momentum correlator $G_{\mt{LR}}(T,k)$ and the one at the zero spatial separation $G_{\mt{LR}}(T,x=0)$. Taking $x=0$, the two-sided thermal correlator of BTZ reduces to 
\begin{equation}
G_{\mt{LR}}^{\mt{BTZ}}(T,x=0) \simeq \left[ \frac{\epsilon\pi}{\beta\cosh(\pi T/\beta)}\right]^{2\Delta}.
\end{equation}
Using the beta-function integral, one can get\footnote{The Gamma-function frequency dependence of spectral density is universal for a scalar primary in the high-temperature regime of a two-dimensional CFT, provided the modular-transformed torus two-point function is vacuum dominated. This universality emerges from the modular covariance of the torus two-point function, which relates the high-temperature correlator to a vacuum cylinder correlator fixed by conformal symmetry. }
\begin{equation}
G_{\mt{LR}}^{\mt{BTZ}}(\omega, x=0) \simeq \frac{\epsilon^{2\Delta}}{2\pi\Gamma(2\Delta)}
\left(\frac{2\pi}{\beta}\right)^{2\Delta-1}\left|\Gamma\left(\Delta+i\frac{\beta\omega}{2\pi}
\right)\right|^2.
\end{equation}
At coincident spatial positions $x=0$, the two types of thermal correlators in frequency are related by  
\begin{equation}
G_{\mt{LR}}(\omega,x=0)=\frac{1}{2\pi}\int_{-\infty}^{+\infty}dk \, 
G_{\mt{LR}}(\omega,k)\,, 
\end{equation}
By using Barnes' first lemma\footnote{Barnes' first lemma states that
\begin{equation}
\frac{1}{2\pi}\int_{-\infty}^{+\infty}ds\,
\Gamma(a+is)\Gamma(b+is)\Gamma(c-is)\Gamma(d-is)
=
\frac{\Gamma(a+c)\Gamma(a+d)\Gamma(b+c)\Gamma(b+d)}{\Gamma(a+b+c+d)}.
\end{equation}
}, one can explicitly get the relation as  
\begin{equation}
\frac{1}{2\pi}\int_{-\infty}^{+\infty} d k
\frac{\left|\Gamma\left(h+i\frac{\beta(\omega+k)}{4\pi}\right)\right|^2}{\Gamma(2h)}
\frac{\left|\Gamma\left(\bar h+i\frac{\beta(\omega-k)}{4\pi}\right)\right|^2}{\Gamma(2\bar h)}
=
\frac{4\pi}{\beta}\frac{\left|\Gamma\left(\Delta+i\frac{\beta\omega}{2\pi}\right)\right|^2}{\Gamma(2\Delta)}.  
\end{equation}
It is thus clear that the analytic continuation of the thermal correlator $G_{\mt{LR}}(\omega,x=0)$ on the complex-$\omega$ plane has poles at
\begin{equation}
\boxed{ \omega_n^{\rm poles}(x=0)=\pm i\frac{2\pi}{\beta}(\Delta +n)\,,}
\end{equation}
which is similar but slightly different from the standard QNM distribution of BTZ \eqref{eq:BTZ:QNMs}.

\subsection{Linear QNMs}

Since the asymptotic linear QNM trajectory is so closely tied to the AdS black hole, we can assume an exactly linear QNM family to build an analytically solvable toy model. Let us assume that the representative linear QNM family is parametrized as
\begin{equation}
\omega_n(\bk)=a_1(\bk)n+a_0(\bk) = a_1(\bk) \left(  n + h_{\rm eff}(\bk)  \right), \qquad n=0,1,2,\ldots \,, 
\end{equation}
with the effective weight
\begin{equation}
 h_{\rm eff} = \frac{a_0(\bk)}{a_1(\bk)} \,, \qquad  \bar{h}_{\rm eff} = \left( \frac{a_0(\bk)}{a_1(\bk)} \right)^\ast \,. 
 \end{equation}
For simplicity, we will also suppress the momentum dependence in the following equations.
Using the thermal product formula with these linear QNMs, one can then reconstruct the correlator as a product of Gamma functions, \ie 
\begin{equation}
\frac{G_{\rm LR}(\omega)}{G_{\rm LR}(0)}=
\frac{\Gamma\left(h_{\rm eff}+\frac{\omega}{a_1}\right)\Gamma\left(h_{\rm eff}-\frac{\omega}{a_1}\right)}{\Gamma(h_{\rm eff})^2}
\frac{\Gamma\left( \bar{h}_{\rm eff}+\frac{\omega}{a_1^\ast}\right)\Gamma\left(\bar{h}_{\rm eff}-\frac{\omega}{a_1^\ast}\right)}{\Gamma(\bar{h}_{\rm eff})^2}.
\end{equation}
Following the beta-function integral, one can write 
\begin{equation}
\int_{-\infty}^{+\infty}dT\,
e^{-i\omega T} 
\left[2\cosh\left(\frac{\lambda T}{2}\right)\right]^{-2h}
=\frac{1}{\lambda}
\frac{\Gamma(h+i\omega/\lambda)
\Gamma(h-i\omega/\lambda)}
{\Gamma(2h)}\,. 
\end{equation}
Consequently, the inverse Fourier transform of the complete QNM product yields a convolution:
\begin{equation}
G_{\rm LR}(T,\bk)=C(\bk) \int_{-\infty}^{+\infty}d\tau\,
\left[
2\cosh\left(\frac{\ii a_1}{2}\tau\right)
\right]^{-2h_{\rm eff}} \times
\left[
2\cosh\left(\frac{-\ii a_1^\ast(T-\tau)}{2}\right)
\right]^{-2\bar{h}_{\rm eff}} \,. 
\end{equation}
The complex-time singularities of the two factors are located at
\begin{equation}
\tau_{+,m}= \frac{2\pi(m+\frac{1}{2})}{a_1} , \quad  \tau_{-,n} = T + \frac{2\pi(n+\frac{1}{2})}{a_1^*},
\qquad (n,m) \in \mathbb{Z} \,.
\end{equation}
As discussed in Appendix \ref{app:four-representations}, a key feature of a convolution is that its analytic continuation may develop singularities when the singularities of the two factors pinch the integration contour. Here, the pinch condition is given by 
\begin{equation}
\tau_{+,m}= \tau_{-,n} (T) \,,
\end{equation}
which corresponds to 
\begin{equation}
 T= \ii \,\operatorname{Im}\left(\frac{2\pi}{a_1}\right)
+ m\frac{2\pi}{a_1} - n\frac{2\pi}{a_1^*}, 
\end{equation}
This result matches precisely the complex-time singularity lattice $T_{m,n}$ derived in the main text using the large-frequency Stokes expansion.

A noteworthy result is that the individual kernel is a conformal-type thermal two-point function with
\begin{equation}
h_{\rm eff}=\frac{a_0}{a_1} \,,  \qquad \beta_{\rm eff}=\frac{2\pi}{ia_1}\,. 
\end{equation}
For positive real $T$, this kernel can be decomposed as 
\begin{equation}
\left[2\cosh\left(\frac{\ii a_1T}{2}\right)\right]^{-2h}= \sum_{n=0}^{\infty}(-1)^n\frac{(2h)_n}{n!}
e^{-i\omega_nT}, \qquad \omega_n=a_0+a_1n\,,
\end{equation}
which recovers exactly the assumed linear QNM spectrum $\omega_n(\bk)$. Notably, this linear QNM spectrum realizes an explicit $\mathfrak{sl}(2,\mathbb R)$ ladder representation. From the perspective of CFT$_2$ or BTZ, this expansion constitutes a conformal descendant decomposition of the two-point function, as $| h_{\rm eff}, n \rangle$ can be identified with actual conformal descendants \cite{Zhang:2011ec,Chen:2010ik}. However, whether this effective conformal structure can be promoted to a genuine symmetry of the bulk wave equation for generic higher-dimensional AdS black holes remains an open question.

\subsection{Lowest QNM and Linear Growth of Black Hole Interior}

We have shown that the slope of the asymptotically linear QNM spectrum of AdS black holes encodes information about the black hole interior through the relation
\begin{equation}
 \frac{2\pi}{a_1} = |T_{\rm crt}| + \ii \frac{\beta}{2},
\end{equation}
which determines both the critical bouncing-geodesic time and the black hole temperature. Interestingly, the constant term in the linear QNM spectrum, or equivalently the lowest QNM $\omega_{n=0}$ also contains information about the geometry of the black hole interior. In particular, we will illustrate that the late-time linear growth of the AdS black hole interior can be related to the lowest QNM.

A simple example is first provided by the BTZ black hole. As discussed above, the lowest pole of the correlator at zero spatial separation coincides with the fundamental QNM at zero momentum,
\begin{equation}
\omega_{n=0}^{\rm{poles}}(x=0) =-\ii\frac{2\pi}{\beta}\Delta
=\omega_{n=0}^{\mt{QNM}}\big|_{k=0} \,.
\end{equation}
This lowest pole governs the late-time exponential decay of the two-sided thermal correlator. In the semiclassical geodesic approximation, the correlator is related to the renormalized geodesic length, giving
\begin{equation}
G_{\mt{LR}}^{\mt{BTZ}}(T)
\sim \ee^{-2\pi\Delta |T|/\beta},
\qquad
\mathcal L_{\rm ren}(T)
\sim \frac{2\pi}{\beta}T \,,
\end{equation}
for $T\gg\beta$. Thus, the location of the lowest QNM, \ie its dependence on the scaling dimension $\Delta$ directly determines the linear growth rate of the geodesic length. This relation also follows from the conformal structure of the high-temperature CFT$_2$ correlator, obtained from the modular transform of the vacuum contribution \cite{Birmingham:2001pj,Hartman:2013qma}.

A second example arises in JT gravity, where the Schwarzian theory provides an exact spectral description of the generating function of the geodesic-length operator. Its matrix elements take the precise form \cite{Mertens:2017mtv,Blommaert:2018oro}
\begin{equation}
\mathcal{M}_\Delta(E_1,E_2)\equiv
\langle E_1|\ee^{-\Delta\widehat{\mathcal{L}}}|E_2\rangle
=
\frac{\ee^{-S_0}}{2^{2\Delta-1}\Gamma(2\Delta)}
\left|\Gamma(\Delta+\ii k_1+\ii k_2)\right|^2
\left|\Gamma(\Delta+\ii k_1-\ii k_2)\right|^2,
\end{equation}
with $k_i=\sqrt{2CE_i}$. The Gamma-function structures appearing in the BTZ thermal correlator and the Schwarzian matrix elements share a common origin in $\mathrm{SL}(2,\mathbb{R})$ conformal representation theory. In BTZ, this structure follows from the conformally determined thermal correlator, whereas in JT gravity it arises naturally from the Schwarzian theory. Similar structures can also emerge in the conformal sectors of higher-dimensional black holes with an AdS$_2$ near-horizon region.

To make the connection with QNMs explicit, we introduce the mean energy and energy difference,
\begin{equation}
\bar E=\frac{E_1+E_2}{2},
\qquad
\omega=E_1-E_2,
\qquad
\bar k=\sqrt{2C\bar E}.
\end{equation}
The leading microcanonical entropy of JT gravity is
\begin{equation}
S_{\mathrm{JT}}(E)=S_0+2\pi\sqrt{2CE}+\cdots,
\end{equation}
which gives the inverse microcanonical temperature
\begin{equation}
\beta_{\mathrm{micro}}(\bar E)
=\left.\frac{\partial S_{\mathrm{JT}}}{\partial E}\right|_{\bar E}
\simeq\frac{2\pi C}{\bar k} \,.
\end{equation}
Consequently, at fixed high mean energy and in the near-diagonal regime, the matrix element reduces to \cite{Miyaji:2026pwc}
\begin{equation}
\mathcal{M}_\Delta(\bar E,\omega)
\simeq\mathcal N_\Delta(\bar E)
\left|\Gamma\left(\Delta+\ii\frac{\beta_{\mathrm{micro}}\omega}{2\pi}\right)\right|^2,
\qquad
\mathcal N_\Delta(\bar E)
=\frac{\ee^{-S_0}|\Gamma(\Delta+2\ii\bar k)|^2}
{2^{2\Delta-1}\Gamma(2\Delta)} \,.
\end{equation}
The analytic continuation of this expression to complex frequency exhibits poles at
\begin{equation}
\omega_n^{\mathrm{JT}}
=\pm\ii\frac{2\pi}{\beta_{\mathrm{micro}}}(\Delta+n),
\qquad n=0,1,2,\ldots.
\end{equation}
The poles in the lower half-plane reproduce the characteristic scalar QNM spectrum of an AdS$_2$ black hole, which can also be derived directly from the classical bulk wave equation \cite{Bhattacharjee:2020nul}. As discussed in \cite{Miyaji:2025yvm,Miyaji:2025jxy}, the location of the lowest pole, \ie $\omega_{n=0}^{\mathrm{JT}} \sim \Delta$ determines the late-time linear growth of the geodesic length.

These two examples suggest a more general connection between the lowest QNM and the linear growth of the black hole interior. To be more concrete, let us consider a single, exactly linear QNM family,
\begin{equation}
\omega_n=a_1(\Delta)n+\omega_{n=0}(\Delta),
\qquad n=0,1,2,\ldots.
\end{equation}
The corresponding generating function admits a spectral expansion as 
\begin{equation}
G_\Delta(T) =\sum_{n=0}^{\infty} R_n(\Delta)\ee^{-\ii\omega_n(\Delta)T},
\qquad T>0 \,,
\end{equation}
where all frequencies belong to the same linear QNM family. The geodesic length can then be extracted from the generating function through
\begin{equation}
\mathcal{L}(T)
=-\left.\partial_\Delta\log G_\Delta(T)\right|_{\Delta=0}.
\end{equation}
For a stable QNM family with $\operatorname{Im}a_1<0$, the higher overtones decay faster than the lowest mode. Furthermore, the normalization condition $G_{\Delta=0}(T)=1$ requires $\omega_{n=0}(0)=0$ and $R_0(0)=1$, while the residues of the higher QNMs vanish in this limit. As a result, it is then straightforward to show that the late-time growth rate is determined by the motion of the lowest pole,
\begin{equation}
\boxed{
\text{Linear growth rate}
=\ii\left.\frac{\partial\omega_{n=0}}
{\partial\Delta}\right|_{\Delta=0} \,.
}
\end{equation}
This relation holds explicitly in the conformal BTZ and semiclassical JT examples. Its extension to more general AdS black holes would establish a direct connection between QNM spectra and the late-time growth of the black hole interior. Furthermore, as shown in eq.~\eqref{eq:thermal-semiclassical-ETH}, the holographic thermal correlator is related to the coarse-grained matrix element in the semiclassical limit,
\begin{equation}
G_{\mt{LR}}(\omega)\simeq 2\pi F_{\cO}(E_\beta,\omega)\,, \qquad \GN\to 0\,.
\end{equation}
This identification suggests that our result provides a QNM interpretation of the findings in \cite{Miyaji:2025yvm,Miyaji:2025jxy,Miyaji:2026pwc}, where the lowest characteristic pole of the generating function was shown to determine the late-time growth rate of holographic complexity.

\subsection{Spatial momentum and Kasner singularity}

The black hole singularity leaves an imprint on the complex-time singularities of the analytically continued thermal correlator $G_{\mt{LR}}(T,\bk)$. The locations of these singularities are determined by the asymptotically linear QNM spectrum and are related to the characteristic time $T_{\rm crt}$ of almost-null bouncing geodesics. However, these locations alone do not fully characterize the local geometry near the singularity. A natural question is therefore how the more detailed information about the near-singularity geometry, particularly its Kasner exponents, can be extracted from the thermal correlator and the asymptotic QNM spectrum.

This limited information about the near-singularity geometry originates from the fact that our previous analysis has been limited to the high-frequency regime, 
\begin{equation}
|\omega|\gg |\bk| \,,
\end{equation}
relevant to the asymptotic QNM. Since the momentum-dependent terms are subleading in the bulk radial wave equation, they do not affect the leading asymptotic slope of the linear QNM trajectories. Consequently, the resulting locations of the complex-time singularities are also independent of any fixed finite $\bk$. It is thus natural to expect that the subleading momentum-dependent corrections can encode additional information about the near-singularity geometry and provide a more refined probe of the black hole interior.

To make this connection more explicit, let us consider a generic spacelike singularity whose local geometry is described by a Kasner metric,
\begin{equation}
ds^2\simeq -\dd\tau^2+a_t^2\tau^{2p_t}\dd t^2
+\sum_i a_i^2\tau^{2p_i}(\dd x^i)^2 \,.
\end{equation}
Here $\tau=0$ represents the location of the spacelike singularity, while $t$ is the Schwarzschild time coordinate, which becomes spacelike inside the horizon. The Kasner exponents $p_t$ and $p_i$ characterize the anisotropic scaling of the geometry as the singularity is approached. For a vacuum solution of the Einstein equations, they satisfy
\begin{equation}
p_t+\sum_i p_i=1 \,, \qquad
p_t^2+\sum_i p_i^2=1 \,.
\end{equation}
Such a geometry also provides a local approximation to individual Kasner epochs in the BKL description of generic spacelike singularities.

To determine the spatial momentum dependence of the thermal correlator $G_{\mt{LR}}(\omega,\bk)$ explicitly, one would need to solve the bulk scalar wave equation in the full black hole geometry with appropriate boundary conditions. Nevertheless, the characteristic momentum dependence associated with the near-singularity geometry can already be inferred from a simple scaling argument, without solving the full wave equation. To see this, let us consider a bulk scalar mode of the form
\begin{equation}
\phi(\tau,t,\bx)=\ee^{-\ii\omega t+\ii\sum_i k_i x^i}\psi(\tau) \,.
\end{equation}
An important property of the Kasner geometry is its homothetic symmetry,
\begin{equation}
\tau\rightarrow\lambda\tau \,, \qquad
t\rightarrow\lambda^{1-p_t}t \,, \qquad
x^i\rightarrow\lambda^{1-p_i}x^i \,,
\end{equation}
under which the metric transforms as $ds^2\rightarrow\lambda^2 ds^2$. In the high-frequency regime, where the scalar mass term is subleading, the bulk wave equation inherits this scaling symmetry. Consequently, the characteristic spatial momentum dependence is governed by the scaling-invariant combinations
\begin{equation}
\boxed{
\frac{k_i^2}{\omega^{2\xi_i}} \,, \qquad
\xi_i=\frac{1-p_i}{1-p_t} \,.
}
\end{equation}
Equivalently, the high-frequency expansion is controlled by the regime $|\omega|\gg |k_i|^{1/\xi_i}$. 

This observation suggests that the local Kasner geometry generates characteristic momentum-dependent corrections to the large-frequency asymptotics of the holographic correlator. Focusing on a contribution associated with a particular complex-time singularity at $T=T_\ast$, we may schematically write
\begin{equation}
G_{\mt{LR}}(\omega,\bk)\supset
A_\ast\,\ee^{\ii\omega T_\ast}\omega^\nu
\left[1+\sum_i C_i\frac{k_i^2}{\omega^{2\xi_i}}+\cdots\right] \,,
\end{equation}
where $\nu=2\Delta-d$, and the coefficients $C_i$ depend on the bulk wave equation and its matching conditions. After Fourier transformation, the corresponding singular contribution in the complex-time plane takes the form
\begin{equation}
G_{\mt{LR}}(T,\bk)\underset{T\to T_\ast}{\sim}
\frac{\widetilde A_\ast}{(T-T_\ast)^{\nu+1}}
\left[1+\sum_i\widetilde C_i k_i^2
(T-T_\ast)^{2\xi_i}+\cdots\right] \,. 
\end{equation}
Thus, while the leading singularity identifies its location $T_\ast$, the fractional powers at the order of $k_i^2$ can encode all Kasner exponents of the near-singularity geometry.

The same information should also be encoded in the asymptotic QNM spectrum \cite{Hartnoll:2026vhu}. Indeed, since the thermal product formula relates the analytic structure of $G_{\mt{LR}}(\omega,\bk)$ to the distribution of QNMs, the momentum-dependent corrections must be reflected in the large-overtone dispersion. Starting from the asymptotically linear QNM trajectory, the Kasner scaling suggests that the large-overtone expansion of QNMs should take the following schematic form:
\begin{equation}
\omega_n(\bk)=a_1n+a_0
+\sum_i k_i^2 \frac{c_i}{n^{2\xi_i}} +\cdots \,,
\qquad n\gg1 \,.
\end{equation}
The same momentum-dependent corrections to the QNMs have been explicitly derived in \cite{Hartnoll:2026vhu}. Moreover, as illustrated in \cite{Hartnoll:2026vhu}, the large-overtone QNM spectrum can also contain other momentum-dependent corrections at the order of $k_i^2/n$, which arise from the global WKB propagation and matching conditions rather than the local Kasner geometry.

These results establish explicit connections between the near-singularity Kasner geometry, the momentum dependence of holographic thermal correlators, and the asymptotic distribution of QNMs. As a summary, the relevant relations are
\begin{equation}
\boxed{
\begin{aligned}
\text{Complex-time singularity}
&\Longrightarrow
\ee^{\ii\omega T_\ast}\omega^\nu \,,
\\[1mm]
\text{Kasner geometry}
&\Longrightarrow
\frac{k_i^2}{\omega^{2\xi_i}} \,, \qquad
\xi_i=\frac{1-p_i}{1-p_t} \,,
\\[1mm]
\text{Asymptotic QNM spectrum}
&\Longrightarrow
\omega_n(\bk)=a_1n+a_0
+\sum_i c_i\frac{k_i^2}{n^{2\xi_i}}+\cdots \,.
\end{aligned}
}
\end{equation}
The momentum-dependent subleading corrections therefore provide a direct diagnostic of the near-singularity geometry beyond the leading linear QNMs. In principle, resolving their scaling along different spatial directions would allow one to reconstruct the corresponding Kasner exponents and potentially distinguish successive Kasner epochs associated with BKL dynamics. See also \cite{Ceplak:2026ard,Oh:2026dky,Xiao:2026pir} for recent related investigations about Kasner geometry, holographic correlators, QNM spectrum, and bouncing-geodesic singularities.

\subsection{IR and UV constraint on the pole-zero data}
\label{subsec:IR UV constraints}

The key result we present in the main sections is that the semiclassical linear QNM spectrum determines the locations of bouncing-geodesic singularities on the complex time plane, while nonlinear deformations of the QNM trajectories naturally lead to their smoothing. However, as indicated by the Hadamard factorization theorem \eqref{eq:hadamard-meromorphic}, the analytic structure of boundary correlators at finite $c$ or finite $\GN$ can involve both poles and zeros. Although our analysis has primarily focused on QNM poles, zeros can also play an important role in the full correlator. An explicit example is provided by Liouville theory in section~\ref{sec:finite-c-qnm-liouville}, which exhibits a richer pole-zero structure at finite $c$. In the semiclassical limit, the zeros are pushed toward infinite frequency and therefore disappear from the finite-frequency spectrum. Nevertheless, this does not imply that they can be neglected in the full finite-$c$ Hadamard representation. The key distinction is between the semiclassical and ultraviolet regimes,
\begin{equation}
\begin{aligned}
1\ll\beta|\omega|\ll c
&:\quad \text{semiclassical frequency regime},\\
\beta|\omega|\gg c
&:\quad \text{UV frequency regime}.
\end{aligned}
\end{equation}
These two regimes probe different parts of the pole-zero distribution. In particular, the UV regime at finite $c$ is sensitive to the complete pole-zero structure, including zeros that become invisible in the semiclassical regime. As we discuss below, these zeros can play an essential role in ensuring that the reconstructed correlator satisfies the UV constraints imposed by the OPE.

To understand how zeros contribute to the large-frequency behavior, recall that the Hadamard representation \eqref{eq:hadamard-meromorphic} takes the schematic form
\begin{equation}
\boxed{\log G_{\rm mero}(\omega)=Q(\omega)+\log E_{\rm zeros}(\omega)-\log E_{\rm poles}(\omega)\,,}
\end{equation}
where the pole and zero contributions enter with opposite signs. Consequently, zeros can provide cancellations of the asymptotic contributions generated by the pole distributions. A useful illustration is provided by the so-called Jensen's formula, which relates the angular average of a meromorphic function to its pole-zero counting functions,
\begin{equation}
\frac{1}{2\pi}\int_0^{2\pi}\log|G_{\rm mero}(Re^{\ii\theta})|\,\dd\theta
-\log|G_{\rm mero}(0)|
=
\int_0^R\frac{\mathscr N_{\rm zeros}(r)-\mathscr N_{\rm poles}(r)}{r}\,\dd r\,.
\end{equation}
Here $\mathscr N_{\rm zeros}(r)$ and $\mathscr N_{\rm poles}(r)$ count the zeros and poles within $|\omega|<r$, including their multiplicities. As discussed in appendix~\ref{sec:asymptotic-hadamard}, for a $k$-dimensional spectral lattice characterized by a polynomial trajectory of degree $K$, the asymptotic counting function scales as
\begin{equation}
\mathscr N(R)\sim R^{k/K}\,.
\end{equation}
In particular, a one-dimensional quadratic trajectory gives $\mathscr N(R)\sim R^{1/2}$, whereas a two-dimensional quadratic distribution yields $\mathscr N(R)\sim R$. Such multidimensional distributions can therefore generate substantially different large-frequency contributions to the Hadamard product, including powers of $\omega$ and logarithmically enhanced terms. Although Jensen's formula constrains only the angularly averaged growth, it illustrates why the combined pole-zero distribution, rather than the QNM poles alone, is relevant to the asymptotic consistency of the correlator.

However, Jensen's formula constrains only the angularly averaged growth of the correlator and does not determine its asymptotic behavior along a particular direction in the complex-frequency plane. Stronger, direction-dependent constraints arise from the short-distance behavior of boundary correlators, which is governed by the OPE. In particular, along the positive real-frequency axis, the UV asymptotics of the two-sided thermal correlator and the HHLL correlator should satisfy
\begin{equation}
\begin{aligned}
G_{\mt{LR}}(\omega,\bk)&\sim \ee^{-\beta\omega/2}\omega^{2\Delta-d}\,,\\
G_{\mt H}(\omega,\bk)&\sim\omega^{2\Delta-d}\,, 
\end{aligned}
\qquad \omega\to+\infty\,,
\end{equation}
up to overall normalization factors. However, as shown in appendix~\ref{sec:asymptotic-hadamard}, the asymptotic expansion of a generic Hadamard product can generate additional contributions involving powers of $\omega$ and logarithmically enhanced terms, such as $\omega^q\log\omega$. The OPE therefore imposes nontrivial cancellation conditions on the combined contributions from poles, zeros, and the entire Hadamard prefactor. In particular, unwanted nonpolynomial terms cannot be removed by an entire prefactor $e^{Q(\omega)}$ with polynomial $Q$ alone and must cancel among the relevant spectral contributions.

An explicit realization of these constraints is provided by the Liouville example discussed in appendix~\ref{app:UV-IR-constraints}. The finite-$c$ DOZZ coefficient contains two-dimensional quadratic distributions of poles and zeros, whose individual Hadamard products generate contributions proportional to $\omega\log\omega$ and $\sqrt{\omega}$. Requiring these unwanted terms to cancel leads to the algebraic conditions
\begin{equation}
\sum_I\frac{\epsilon_I}{a_I}=0\,,
\qquad
\sum_I\epsilon_I\frac{Q/2-d_I}{\sqrt{a_I}}=0\,,
\end{equation}
where $\epsilon_I$ denotes the signed multiplicity of each pole or zero family, and $a_I$ and $d_I$ characterize the corresponding quadratic distributions. Both conditions are explicitly satisfied by the Liouville pole-zero data. This illustrates how the OPE can impose nontrivial restrictions on the allowed spectral distributions. Nevertheless, while these cancellations already occur within the primary DOZZ coefficient, reproducing the complete ultraviolet OPE behavior of the HHLL correlator also requires the contributions of Virasoro descendants encoded in the conformal blocks. 

The distinction between the semiclassical and UV regimes can also be illustrated explicitly in Liouville theory. As shown in appendix~\ref{app:UV-IR-constraints}, taking the semiclassical limit with fixed $\beta_{\mt H}\omega$, the two-dimensional quadratic pole-zero distribution reduces to the familiar Gamma-function structure,
\begin{equation}
\mathcal W_{\rm prim}^{\rm IR}(\omega)
\propto
\left|\Gamma\left(\frac{\Delta}{2}+\ii\frac{\beta_{\mt H}\omega}{4\pi}\right)\right|^4
\sim \ee^{-\beta_{\mt H}\omega/2}\omega^{2\Delta-2}\,,
\qquad \beta_{\mt H}\omega\gg1\,,
\end{equation}
which reproduces the expected semiclassical BTZ behavior. By contrast, the UV limit at finite $c$ probes the complete two-dimensional pole-zero distribution, including the zeros and additional pole families that are pushed toward infinite frequency in the semiclassical limit. Their combined contributions are essential for satisfying the nontrivial cancellation conditions required by the ultraviolet OPE. 

These observations suggest that the consistency of finite-$c$ holographic correlators imposes strong constraints on the allowed deformations of QNM spectra, as well as on the associated zero distributions and entire Hadamard prefactors. This raises the possibility of bootstrapping the pole-zero data by combining the OPE constraints with thermal analyticity and the semiclassical limit, thereby constraining, or even partially determining, the admissible finite-$c$ QNM deformations and zero distributions. With sufficient information to fix the remaining entire prefactor, such a bootstrap approach could ultimately provide a way to reconstruct the full analytically continued boundary correlator from its pole-zero data.

\subsection{QNM Deformation and a non-Hermitian Effective Hamiltonian}
\label{subsec:QNM deformation and non-Hermitian Hamiltonian}

\begin{figure}[t]
\centering
\includegraphics[width=0.82\linewidth]{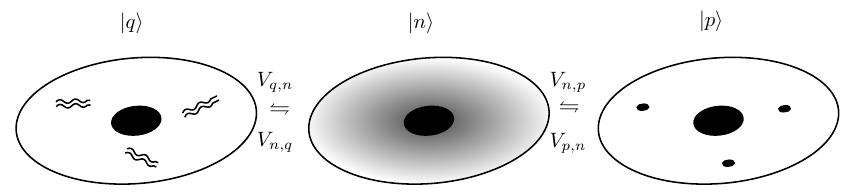}
\caption{Toy-model coupling of QNMs $|n\rangle$ to perturbative graviton states $|q\rangle$ and virtual black-hole states $|p\rangle$. The central black disk represents the fixed background $|\mathrm{H}\rangle$.}
\label{fig:discussion-qnm-gravity-coupling}
\end{figure}

To close, we consider a simple effective model to illustrate how finite-$c$ intermediate states in the $\mathrm{(HL)(HL)}$ channel can deform the semiclassical QNM spectrum. Let $|n\rangle$ be QNM resonances of the fixed black-hole background, $|q\rangle$ perturbative graviton states on that background, and $|p\rangle$ heavy states associated with other geometrical saddles, including multi-black-hole configurations. The latter appear as virtual black-hole states in the intermediate-state sum. Because the QNM frequencies $\omega_n$ are complex, the effective Hamiltonian is generally non-Hermitian:
\begin{equation}
H_{\rm eff}=\begin{pmatrix}H_Q&V_{QG}\\V_{GQ}&H_G\end{pmatrix},
\qquad H_Q=\operatorname{diag}(\omega_n),\qquad
H_G=\operatorname{diag}(\epsilon_q,E_p).
\label{eq:discussion-qnm-effective-hamiltonian}
\end{equation}
Here $V_{QG}$ and $V_{GQ}$ need not be Hermitian conjugates. Integrating out the gravitational sector gives the QNM resolvent and self-energy
\begin{equation}
G_Q(\omega)=\left[\omega-H_Q-\Sigma(\omega)\right]^{-1},\qquad
\Sigma_{nn'}(\omega)=\sum_q\frac{V_{nq}V_{qn'}}{\omega-\epsilon_q}
+\sum_p\frac{V_{np}V_{pn'}}{\omega-E_p}.
\label{eq:discussion-qnm-self-energy}
\end{equation}
Thus the dressed poles solve $\det[\omega-H_Q-\Sigma(\omega)]=0$. Even when the bare QNM spectrum is linear in $n$, the matrix elements and energy denominators can generate a nonlinear shift. 
Probe correlators can be expressed as $u^\dagger G_Q(\omega)u$, where $u$ encodes the operator information. Clearly it has the same pole structure as $G_Q(\omega)$, but the zeros or equaivalently residues of poles are generically different.

The two sectors $p,q$ can affect the overtone number differently. The weak equivalence principle  requires a universal coupling between graviton and QNM, making their matrix elements $V_{nq}$ identical for all overtone number $n$ . However, coupling to another saddle is non-perturbative and thus non-universal: schematically $V_{np}\sim\exp[-S_{\rm inst}(n,p)]$, where the instanton action can depend on the QNM label. Individual transitions are exponentially suppressed at small $G_N$, because $\exp[-S_{\rm inst}(n,p)]\sim\exp[-(S_p-S_n)]$, where the action of other saddle $S_p$ is of order $G_N^{-1}$. But a large number $N_{\rm eff}\sim\exp(G_N^{-1}+\#\log G_N)$ of contributing black-hole microstates could again enhance their sum, leaving a shift of order $G_N^\#\sim c^{-\#}$. Virtual-black-hole mixing is therefore a possible mechanism for higher-degree QNM deformation.

\section*{Acknowledgments}

We are grateful to Rong-Gen Cai, Matthew Dodelson, Ping Gao, Li Li, Robert Myers, Jun Nian, and Run-Qiu Yang for valuable discussions. We thank the organizers of Holographic Universe 2026 at YITP, Kyoto University, Quantum Information in Quantum Gravity 2026 at Tsinghua University, and 9th Holography, String Theory and Observational Cosmology, where the main results of this work were presented, as well as the many participants for stimulating discussions and valuable comments. We also acknowledge the use of various AI tools to improve the presentation of the manuscript and to assist in preparing several illustrative figures in the main text and the numerical plots in Appendix \ref{subsec:deformed-integral-EFT}. This work is supported by the National Natural Science Foundation of China (Grants No. 12675056) and Peking University under startup Grant No. 7101303985.

%%%%%%%%%%%%%%%%%%%%%%%%%%%%%%
%%%%%%%%%%%%%%%%%%%%%%%%%%%%%%

\appendix

\section{Aspects of Two-sided Correlators: $G_{\mt{LR}}(T,\mathbf{x})$ to $G_{\mt{LR}}(T,\mathbf{k})$}
\label{app:four-representations}

In the main text we used the spatial Fourier transform to relate the lightcone and bouncing singularities of $G_{\mt{LR}}(T,\bx)$ to the singularity lattice of $G_{\mt{LR}}(T,\bk)$.  Here we explain this step in more detail. Basically, when doing the $\mathbf{x}$ to $\mathbf{k}$ Fourier integral, the $\bx$ contour of the integral will decomposes into cycles surrounding poles or cuts in $\bx$ space, while the location of those $\bx$ singularities are controlled by $T$. When for certain $T$ some $\bx$ singularities collide with each other and make some pole/cut cycle pinched by two poles or branch points, the corresponding cycle can no longer avoid the $\bx$ singularity by deformation, hence its contribution diverges. That is when $T$ singularities of $G_{\mt{LR}}(T,\bk)$ occur.

\subsection{Spatial Fourier transform and the pinching mechanism}

We start from a point $T_0$ inside the first analytic strip and define
\begin{equation}
    G_{\mt{LR}}(T_0,\bk)=\int_{\mathcal{C}}\dd^{d-1}\bx\,
    \ee^{-\ii\bk\cdot\bx}G_{\mt{LR}}(T_0,\bx)=\sum_{\mathrm{poles}}I_p+\sum_{\mathrm{cuts}}I_c.
    \label{eq:app-spatial-Fourier}
\end{equation}
The defining contour $\mathcal{C}$ is a closed surface of real dimension $d-1$ in the complex coordinate space $\mathbb{C}^{d-1}$, and it is a plane glued with a large hemisphere whose direction depends on $\bk$ to make sure the contribution from the large hemisphere vanish. By Cauchy theorem, it decomposes into cycles surrounding poles and cuts, whose contributions are denoted as $I_p,I_c$. 
When $T$ is continued away from $T_0$, the singularities of the integrand
move in complex $\bx$ space. Each component cycle will be continuously drifted with the pole or cut inside it. Singularity or monodromy could happen in this process. 

\begin{figure}[ht]
    \centering
    \includegraphics[width=0.7\linewidth]{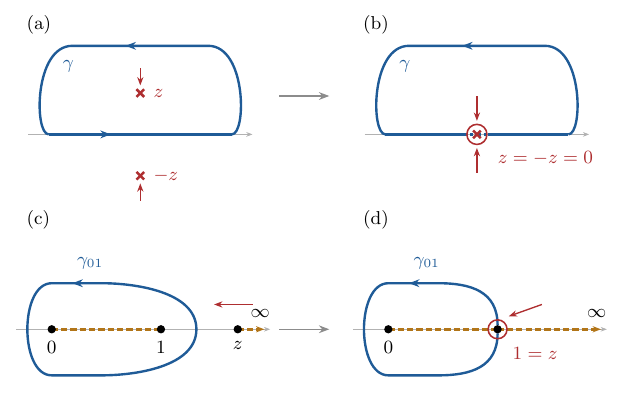}
    \caption{Two contour-pinching mechanisms in the complex $x$ plane.
    Blue curves denote integration cycles, crosses denote poles, and
    black dots denote branch points; dashed lines indicate branch cuts.
    (a,b) The cycle $\gamma$ encloses $z$ but excludes $-z$.
    Their collision at $z=0$ pinches the cycle and produces a $1/z$
    singularity. (c,d) The cycle $\gamma_{01}$ surrounds the cut $[0,1]$
    but excludes the third branch point $z$. As $z\to1$, the external
    branch point collides with the enclosed endpoint and pinches the
    cycle, producing a logarithmic singularity. Red circles mark the
    pinch locations.}
    \label{fig:pinch_toy_models}
\end{figure}

Now we use one dimensional models to illustrate one important origin of the singularity: contour pinching.  
First consider the integral with only pole cycles:
\begin{equation}
    I(z)=\int_{\mathcal{C}}\frac{\dd x}{(x+z)(x-z)}\propto\frac{N_+-N_-}{z},
    \label{eq:app-pole-pinch}
\end{equation}
where $N_\pm$ are the winding numbers of the contour around the two poles $x=\pm z$.
When $z\to0$, two poles collide with each other, giving a potential singularity at $z=0$. This singularity will indeed happen only when the two poles are captured
differently $N_+\neq N_-$ and thus pinch the contour (e.g. $N_+=1,N_-=0$, shown in figure \ref{fig:pinch_toy_models} ). When $N_+=N_-$, however, two singular contributions will cancel. In this case the contour is in fact not pinched by the poles, and one can freely deform it away from two poles, resulting in a finite integral.
Similar things happen for cut cycles and  branch points.  For example, let $\gamma_{01}$ surround
the cut $[0,1]$ but exclude $z>1$.  Then
\begin{equation}
    I(z)=\oint_{\gamma_{01}}
    \frac{\dd x}{\sqrt{x(x-1)(x-z)}}
    =C\log(z-1)+O(1),\qquad z\to1,
    \label{eq:app-external-branch-pinch}
\end{equation}
where $C\neq0$ depends on the branch and orientation.
Here the enclosed endpoint $1$ collides with the external branch point
$z$. 

In
more spatial dimensions the same statement can be written locally as
\begin{equation}
   G_{\mt{LR}}(T,\bx)\sim \Phi(T,\bx)^{-\Delta},\qquad
    \Phi(T_*,\bx_*)=0,\qquad
    \partial_{x_i}\Phi(T_*,\bx_*)=0.
    \label{eq:app-push-forward}
\end{equation}
The last condition says that the projection of the singular surface to the
$T$ plane has a critical value.  It is also necessary that the corresponding
vanishing cycle is trapped by the continued Fourier contour.  These are the
position-space version of the usual Landau conditions.

One thing need to mention is that to genuinely pinch the contour, two singular factors in the integrand have to be multiplied with each other. If one replaces the integrand of \ref{eq:app-pole-pinch} with $\frac{1}{x+z}+\frac{1}{x-z}$, then clearly the integral just gives constant results and no $z$-singularity would happen.

\subsection{Ordinary lightcone images}

Let $r=|\bx|$.  Near one of the OPE lightcones the singular part has the
universal form
\begin{equation}
    G_{\mt{LR}}(T,r)\sim
    \left[(T-T_n)^2-r^2\right]^{-\Delta},\qquad
    T_n=\ii\frac{\beta}{2}(2n+1),\quad n\in\mathbb Z,
    \label{eq:app-lightcone-image}
\end{equation}
up to a function which is regular at $T=T_n,r=0$.  The two radial branch points collide at $r=0$ when $T=T_n$.  Since the measure of the radial Fourier transform reduces to $r^{d-2}\dd r$ near the collision, setting $r=(T-T_n)y$ gives
\begin{equation}
    G_{\mt{LR}}(T,\bk)\sim \frac{1}{(T-T_n)^{2\Delta-d+1}}.
    \label{eq:app-lightcone-power}
\end{equation}
Notice that the local lightcone analysis produces the odd half-thermal images only. There is no analogous local OPE pinch at $T=\ii n\beta$ in a generic higher-dimensional thermal correlator.

There is a  subtlety about choosing the sheet. When we say the singularity of $G_{\mt{LR}}(T,\bx)$, which sheet are we talking about? In the main text we have already shown that $G_{\mt{LR}}(T,\bx)$ has infinite sheets structure in $T$, and there are different singularity structure on different sheets. Then the question is, of which sheet the singularity of $G_{\mt{LR}}(T,\bx)$ will be captured by the $\bx$ contour, and produce singularities of $G_{\mt{LR}}(T,\bk)$ by colliding and pinching. At fixed $r\neq0$, adjacent KMS strips of $G_{\mt{LR}}(T,\bx)$ can be glued through the spacelike interval between the two lightcone endpoints. The spatial Fourier integral, however, passes through $r=0$, where this interval closes. It turns out that singularity collision all happen at $r=0$, thus the main contribution to the singularities of $G_{\mt{LR}}(T,\bk)$ come from the part of $\bx$ integral that is near the $r=0$ region. We have already illustrated this point in the above discussion about light-cone singularity collision. We are going to see that similar things happen for the collision of bouncing singularities. As a result, only the cyan sheet in figure \ref{fig:BTZAdSBH} is visible to the $\bx$ integral when one is talking about singularity collision. In other word, when performing the pinch-contour analysis for the Fourier transform from $\bx$ to $\bk$, only the singularities of $G_{\mt{LR}}(T,\bx=0)$ matters.

\subsection{BTZ versus a higher dimensional black hole}
\begin{figure}[ht]
    \centering
    \includegraphics[width=0.5\linewidth]{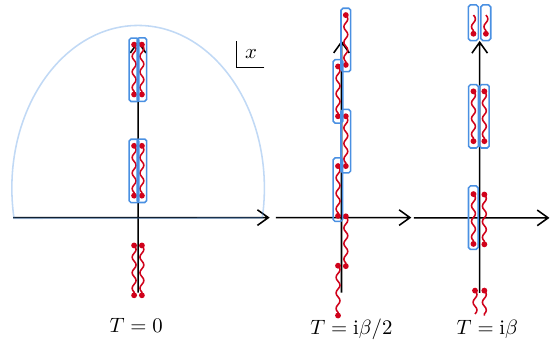}
    \caption{Illustration of singularity collision in $x$ plane for the BTZ case. Here red dots denote singularities of $G_{\mt{LR}}(T,x)$ in $x$ plane, according to \ref{eq:app-BTZ-product}, and red curly lines denote cuts between them. Light blue semicircle is the Fourier contour for the $x$ to $k$ Fourier integral \ref{eq:app-spatial-Fourier} in the case of $k<0$. It decomposes into blue cut cycles. When $T$ varies, these singularities will move in the $x$ plane, and collide with each other at $T=T_N=\ii\frac{\beta N}{2}$. Here we show the configuration of $N=0,1,2$. In the case $N=0$, the contour is actually not pinched and one can deform the contour (\eg back to the light blue one) to avoid the singularities and get a finite result. In other cases, however, the contour is genuinely pinched and gives divergent result, leading to a $T$ singularity of $G_{\mt{LR}}(T,\bk)$ at corresponding $T_N$.   }
    \label{fig:BTZ pinch}
\end{figure}

The BTZ correlator quoted in the main text is special because its left- and right-moving factors appear multiplicatively,
\begin{equation}
    G_{\mt{LR}}^{\rm BTZ}(T,x)\propto
    \left[\cosh\frac{\pi(T+x)}{\beta}
    \cosh\frac{\pi(T-x)}{\beta}\right]^{-\Delta}.
    \label{eq:app-BTZ-product}
\end{equation}
The two chiral families of branch points move in opposite directions in the complex $x$ plane.  They collide at
\begin{equation}
    T_N=\ii\frac{\beta N}{2},\qquad
    N\in\mathbb Z,
    \label{eq:app-BTZ-time-lattice}
\end{equation}
and genuinely pinch the contour for $N\neq 0$ (see figure \ref{fig:BTZ pinch}). Locally
\begin{equation}
    G_{\mt{LR}}^{\rm BTZ}(T,k)\sim(T-T_N)^{1-2\Delta}
    \label{eq:app-BTZ-power}
\end{equation}
for $N\neq 0$. In particular, the odd as well as the even half-thermal images can occur. Those odd ones are just light-cone images given in the last subsection, and they are universal for all CFTs. The even ones are specific for BTZ. They come from cross-collisions of branch points inside the chiral product \ref{eq:app-BTZ-product}.  The fact that different OPE singularities are multiplied with each other is a special property for BTZ, due to the chiral structure that only exists in 2 dimensional CFT. In higher dimensions, different OPE singularities are summed up, hence only the collision inside each OPE factor \ref{eq:app-lightcone-image} can give pinch contour singularity, that is the odd $N$ ones, while the cross-collisions between different OPE factors are not effective collisions. In the notation of \eqref{eq:complex-time-lattice}, the two lattice directions degenerate to $v_+=v_-=\ii\beta/2$ for BTZ and the fixed momentum singularities remain on the imaginary axis.

For a higher-dimensional black hole, one does not have closed expression of $G_{\mt{LR}}(T,\bx)$ as in the BTZ case. However, to do the pinch contour analysis only the local singular factors of $G_{\mt{LR}}(T,\bx)$ are required, as in \ref{eq:app-lightcone-image}. When $\bx=0$, by PL analysis with classical WKB action \cite{nextpaper2026} one can show that the $T$ singularities of $G_{\mt{LR}}(T,\bx=0)$ locate at 
\begin{equation}
    T_{m,p}=\pm m|T_{\rm crt}|+\ii\frac{\beta}{2}(2p+m+1),
    \qquad m,p\in\mathbb N_0,
    \label{eq:app-bouncing-lattice}
\end{equation}
as in eq.~\ref{eq:T_mp}. Their locations are illustrated in the last panel of figure \ref{fig:BTZAdSBH}. The $m=0$ family is the light-cone OPE singularities, and $m\neq 0$ families are bouncing-geodesic singularities. Going away from $\bx=0$, the OPE singularities have a trivial $\bx$-dependence fixed by the light-cone, as in \ref{eq:app-lightcone-image}, while other bouncing singularities could have nontrivial dependence on $\bx$. As a result, for bouncing singularities the singular factors should be modified to 
\begin{equation}
    G_{\mt{LR}}(T,r)\sim
    \left[(T-T_{m,p})^2-g(r)^2\right]^{-\Delta},
    \label{eq:app-bouncing-image}
\end{equation}
where for $m\neq0$ $g(r)$ is no longer simply $r$, but complicated function of $r$. For example, with $d=4$ and $m=1$, one has
\begin{equation}
    g(r)=\ii \frac{r^3}{3A^2} \,_2F_1\left(\frac{1}{2},\frac 3 4; \frac 7 4;\frac{r^4}{A^4}\right)=\ii\left[\frac{r^3}{3A^2}+\frac{r^7}{14A^6}+\cdots\right], \quad A=2.622....
\end{equation}
The point is $g(r)$ would still vanish at $r=0$, so for $m\neq 0$ the pinching mechanism basically remains the same as for $m=0$. Hence one gets the same $T$-singularity lattice at $T_{m,p}$ for $G_{\mt{LR}}(T,\bk)$. Equivalently, the upper-half-plane candidates form
\begin{equation}
    T_{n_+,n_-}=\ii\frac{\beta}{2}+n_+v_++n_-v_-,
    \qquad
    v_\pm=\pm|T_{\rm crt}|+\ii\frac{\beta}{2},
    \qquad n_\pm\geq0,
    \label{eq:app-lattice-vectors}
\end{equation}
as in \eqref{eq:complex-time-lattice}. The diagonal ones with $n_+=n_-$ come from the collision of OPE singularities,
whereas the off-diagonal ones encode the bouncing contributions. Hence the latter is the part of $G_{\mt{LR}}(T,\bk)$ that is expected to be smoothed when the bulk singularity is resolved at finite $c$, see the second panel of figure \ref{fig:finite-c-time-lattices}.

\section{Deformed complex contours in EFT}
\label{subsec:deformed-integral-EFT}

In this appendix, we examine the complex-contour prescription introduced in section~\ref{subsec:QNM-deformation} with more explicit details. The central question is whether an effective approximation with a truncated action can nevertheless reproduce the leading contribution of the UV-complete integral even it diverges along the original real-frequency contour. Picard--Lefschetz theory provides a useful framework for answering this question: the physical integration cycle could be decomposed into complex steepest-descent contours, 
\begin{equation}
\mathcal C_{\rm phys} \sim\sum_\sigma n_\sigma\mathcal J_\sigma.
\end{equation}
with their coefficients $n_{\sigma}$ determined by intersection numbers. The EFT contour description must reproduce the leading contributing thimbles selected by the decomposition of original physical contour, rather than simply adopt an arbitrary integral contour on which its truncated integrand converges. We illustrate this distinction using the quartic toy model and the exact quadratic QNM model discussed in section \ref{subsec:Quadratic-QNM-Model}.

\subsection{Quartic completion of an inverted Gaussian}
We first consider the quartic integral and its quadratic approximation as the toy model, \ie 
\begin{equation}
    \begin{aligned}
    I_{\mt{UV}}&=\int_{\mathbb R}\dd\omega\,
    \exp\left[-\ii(T-T_m)\omega+\frac{\alpha\omega^2}{c}-\frac{\omega^4}{c^3}\right],\\
    I_{\mt{EFT}}&=\int_{\mathcal C_{\mt{EFT}}} \dd\omega\,
    \exp\left[-\ii(T-T_m)\omega+\frac{\alpha\omega^2}{c}\right],
    \end{aligned}
    \label{eq:polynomial-prescription-example}
\end{equation}
with $\operatorname{Re}\alpha>0$. Here, we have defined the physical contour as $\mathcal{C}_{\rm phys}=\mathbb{R}$ with ignoring the power-law prefactor for simplicity. The quartic term makes $I_{\mt{UV}}$ entire in $T$ at every finite $c$, whereas the quadratic approximation diverges along the real axis. To examine the smoothing profile near $T=T_m$, we consider the regime $|T-T_m|=O(c^{-1/2})$, where the quadratic correction becomes relevant. Introducing the rescaled variables 
\begin{equation}
    u = \frac{\omega}{\sqrt{c}} \,, \qquad \xi=\ii(T-T_m)\sqrt{c} \,, 
\end{equation}
the quartic correction is suppressed by $O(c^{-1})$, while the quadratic term remains finite. This allows us to compare the leading contributions of the UV-complete integral and its quadratic approximation in the large-$c$ limit.

\begin{figure}[t]
    \centering
    \includegraphics[width=6in]{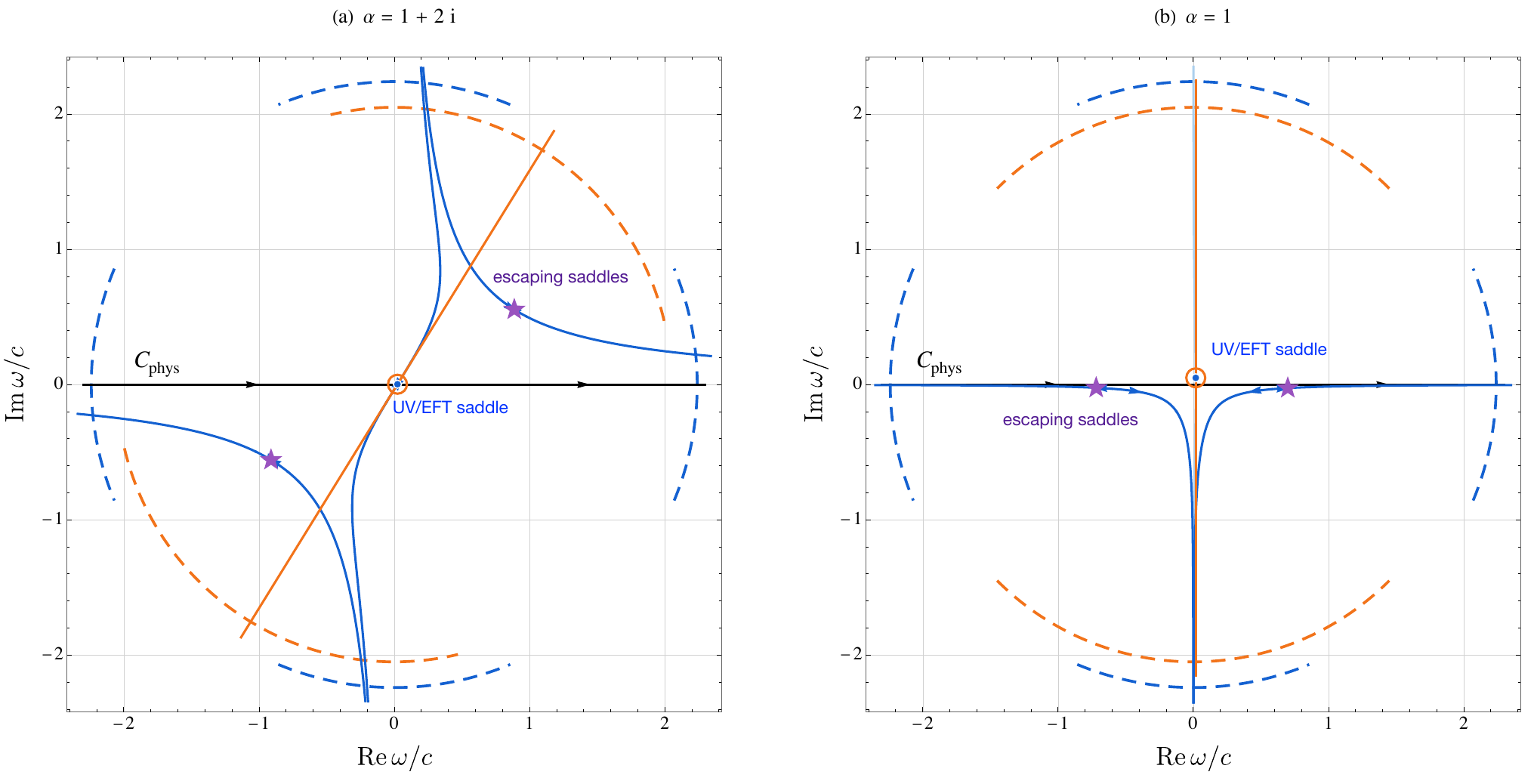}
    \caption{Picard--Lefschetz analysis of the quartic UV model~\eqref{eq:polynomial-prescription-example} and its quadratic EFT approximation, with taking $c=64$ and $\xi=\ii(T-T_m)\sqrt c=0.3+0.8\ii$. Panels (a) and (b) correspond to $\alpha=1+2\ii$ and $\alpha=1$, respectively. The black horizontal lines represent the original physical integration contours $\mathcal{C}_{\rm phys}$, while the blue and orange solid curves denote the UV Lefschetz thimbles $\mathcal{C}_{\mt{UV}}$ and the quadratic EFT contours $\mathcal{C}_{\mt{EFT}}$, respectively. The dashed arcs near the outer boundaries indicate their corresponding asymptotic convergence sectors. The blue dots and orange circles mark the local UV and EFT saddles, while the purple stars denote additional UV saddles. In panel (a), the quadratic approximation captures the dominant local saddle contribution, while the additional UV saddles are exponentially suppressed. In panel (b), the additional UV saddles instead dominate the physical integral and are not captured by the EFT contour.}
    \label{fig:eft-contours-thimbles}
\end{figure}

For $\alpha=1+2\ii$, the original contour can be rotated to $\mathcal C_{\mt{EFT}}=\ee^{\ii\theta}\mathbb R$ with $\theta=\pi/12$. This rotation remains within the quartic decay sectors, since $\cos(4\theta')>0$ throughout $0\leq\theta'\leq\pi/12$. On the final contour, the quadratic term also provides damping,
\begin{equation}
    \operatorname{Re}\left(\alpha\ee^{2\ii\theta}\right)=\frac{\sqrt3}{2}-1<0.
\end{equation}
Both integrals can therefore be evaluated on the same contour. Expanding the quartic factor gives
\begin{equation}
    \begin{aligned}
    \frac{I_{\mt{UV}}}{\sqrt c}
    &=\int_{\ee^{\ii\theta}\mathbb R}\dd u\,
    \exp\left[-\xi u+\alpha u^2-\frac{u^4}{c}\right]\\
    &=\frac{I_{\mt{EFT}}}{\sqrt c}+O(c^{-1})
    =\frac{\sqrt\pi}{\sqrt{-\alpha}}\exp\left[-\frac{\xi^2}{4\alpha}\right]+O(c^{-1}),
    \end{aligned}
    \label{eq:polynomial-prescription-expansion}
\end{equation}
where the branch of $\sqrt{-\alpha}$ is fixed by $\operatorname{Re}(\sqrt{-\alpha}\,\ee^{\ii\theta})>0$. The quadratic damping on the rotated contour makes this estimate uniform for $\xi$ in compact sets. Thus the effective Gaussian integral reproduces the leading regular profile, despite being divergent on the original real axis.

The saddle structure explains why this approximation is reliable. The local quartic saddle approaches the quadratic saddle,
\begin{equation}
    \omega_0=\frac{\xi\sqrt c}{2\alpha}+O(c^{-1/2}),
    \qquad
    \omega_\pm=\pm c\sqrt{\frac{\alpha}{2}}+O(\sqrt c).
    \label{eq:quartic-local-and-UV-saddles}
\end{equation}
The other two saddles lie at $|\omega|=O(c)$ and move outside the region described by the quadratic approximation. Their exponential weights are of order $\exp[\alpha^2c/4+O(\sqrt c)]$. For $\alpha=1+2\ii$, their magnitudes are therefore suppressed by $\exp[-3c/4+O(\sqrt c)]$. The local saddle supplies the leading contribution, while the additional contributing saddles are exponentially subleading, as illustrated in figure~\ref{fig:eft-contours-thimbles}.

Convergence of the effective integral alone does not guarantee this matching. For $\alpha=1$, the original quartic integral is instead dominated by the two saddles near $\omega=\pm c/\sqrt2$. At $T=T_m$, their contributions give
\begin{equation}
    I_{\mt{UV}}\sim 2\sqrt{\frac{\pi c}{2}}\,\ee^{c/4},
    \qquad \alpha=1,\quad T=T_m.
    \label{eq:polynomial-prescription-escape}
\end{equation}
Although rotating the quadratic contour to $-\ii\mathbb R$ makes the effective integral convergent, that contour is not equivalent to the original real line for the quartic action. It selects a different combination of UV thimbles and misses the dominant contributions, as shown in figure~\ref{fig:eft-contours-thimbles}. This example demonstrates why the effective contour must be determined from the full integral, not from the convergence of the truncated action alone.

\subsection{Exact quadratic QNM model}

We now return to the exact quadratic QNM model of section~\ref{subsec:Quadratic-QNM-Model}. Its Fourier transform is
\begin{equation}
    F_c(s)=\int_0^\infty\dd\omega\,
    \exp\left[-s\omega-2\tau_*c\left(\sqrt{1+\frac{\omega}{c}}-1\right)\right],
    \qquad s=\ii(T-T_{\rm seed}),\qquad \tau_*>0,
    \label{eq:eft-sqrt-transform}
\end{equation}
initially defined for $\operatorname{Re}s>0$. Unlike the quartic example, the UV completion is nonpolynomial. The contour prescription is fixed by continuing the exact integral through the upper half of the $s$-plane to a neighborhood of the semiclassical singularity at $s=-\tau_*$.

The substitution $u=\sqrt{1+\omega/c}$ makes the exact exponent quadratic in $u$. Under the upper continuation in $s$, the contour $[1,\infty)$ can be continued to the vertical ray from $1$ to $1-\ii\infty$. Writing $u=1-\ii\rho/(2c)$ gives the corresponding frequency-space contour,
\begin{equation}
    \omega=-\ii\rho-\frac{\rho^2}{4c},\qquad \rho\geq0,
    \label{eq:eft-sqrt-inherited-contour}
\end{equation}
which passes below the branch cut $(-\infty,-c]$. Along this contour, the exact continuation takes the form
\begin{equation}
    F_c(s)=\int_0^\infty\dd\rho\,
    \left(-\ii-\frac{\rho}{2c}\right)
    \exp\left[\frac{s\rho^2}{4c}+\ii(s+\tau_*)\rho\right],
    \qquad \operatorname{Re}s<0.
    \label{eq:eft-sqrt-continued-transform}
\end{equation}
This representation is convergent in a complex neighborhood of $s=-\tau_*$. In the smoothing region $|s+\tau_*|=O(\sqrt{\tau_*/c})$, the relevant values of $\rho$ are of order $\sqrt{c/\tau_*}$. At this scale, the contour~\eqref{eq:eft-sqrt-inherited-contour} approaches the negative imaginary axis in rescaled frequency coordinates, and eq.~\eqref{eq:eft-sqrt-continued-transform} reduces at leading order to
\begin{equation}
    -\ii\int_0^\infty\dd\rho\,
    \exp\left[-\frac{\tau_*\rho^2}{4c}+\ii(s+\tau_*)\rho\right]
    =F_c^{\mt{EFT}}(s).
    \label{eq:eft-sqrt-local-contour-limit}
\end{equation}
This is precisely the contour-defined quadratic approximation used in the main text. Its contour follows from the continuation of the exact integral, rather than from an independent choice made after truncation.

The same prescription allows us to compare higher-order approximations. Expanding the exact exponent gives
\begin{equation}
    -(s+\tau_*)\omega+\frac{\tau_*\omega^2}{4c}
    -\frac{\tau_*\omega^3}{8c^2}+\frac{5\tau_*\omega^4}{64c^3}
    +O\left(\frac{\omega^5}{c^4}\right).
    \label{eq:eft-sqrt-expansion}
\end{equation}
We denote the integral truncated at order $\omega^N$ by $F_c^{\mt{EFT},(N)}$. The quadratic and cubic truncations converge along the negative imaginary axis, but the positive quartic term makes the fourth-order truncation divergent there. This change does not invalidate the effective expansion. Near $s=-\tau_*$, the exact continued contour can also be deformed to
\begin{equation}
    \mathcal C_{\mt{EFT}}=\ee^{-2\pi\ii/3}[0,\infty),
    \label{eq:eft-sqrt-common-contour}
\end{equation}
which avoids the branch cut and provides a common convergent prescription for $N=2,3,4$. Indeed, for $\omega=\ee^{-2\pi\ii/3}\rho$, the nonlinear terms satisfy
\begin{equation}
    \operatorname{Re}\left[\frac{\tau_*\omega^2}{4c}-\frac{\tau_*\omega^3}{8c^2}+\frac{5\tau_*\omega^4}{64c^3}\right]
    =-\frac{\tau_*\rho^2}{8c}-\frac{\tau_*\rho^3}{8c^2}-\frac{5\tau_*\rho^4}{128c^3}.
    \label{eq:eft-sqrt-common-damping}
\end{equation}
Each retained nonlinear term therefore provides damping on this ray.

\begin{figure}[t]
    \centering
    \includegraphics[width=\linewidth]{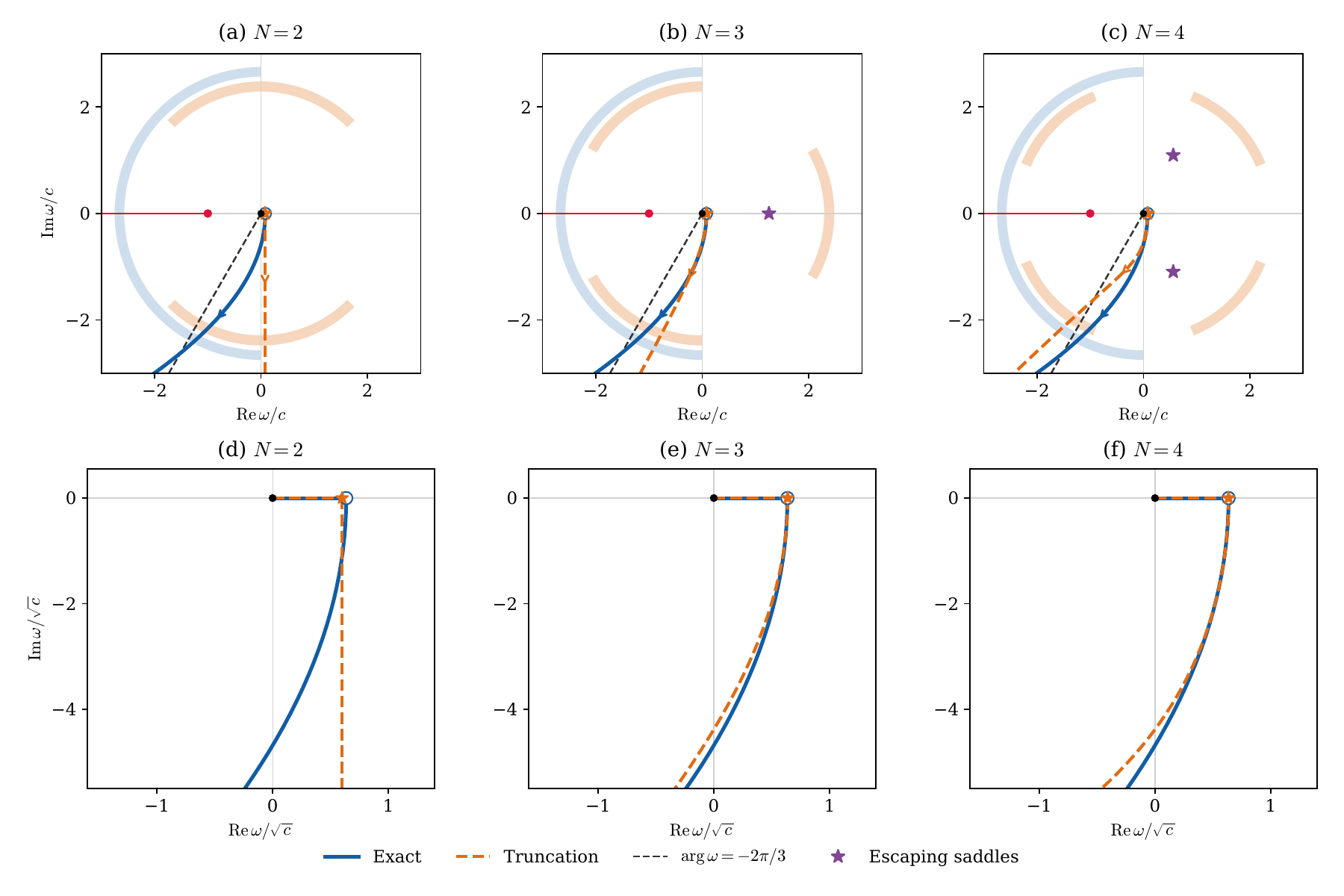}
    \caption{Contours for the exact square-root transform and its quadratic, cubic, and quartic approximations, shown in the three columns. Here $\tau_*=1$, $c=64$, and $s=-1+0.3/\sqrt c$ on the upper continuation. The upper row shows frequencies of order $c$, while the lower row resolves the smoothing region at frequencies of order $\sqrt c$. Blue solid and orange dashed curves join the endpoint to the local saddle and then follow a descending arm of the exact and truncated exponents, respectively. Blue circles and orange stars mark the local saddles, while purple stars mark additional saddles of the truncated exponents at frequencies of order $c$. Black dots mark the fixed endpoint. The outer bands indicate the exact and truncated decay sectors, and the black dashed ray $\arg\omega=-2\pi/3$ provides a common convergent prescription. The red point and line mark the branch point $\omega=-c$ and its cut.}
    \label{fig:eft-sqrt-contours}
\end{figure}

Figure~\ref{fig:eft-sqrt-contours} compares the corresponding contour geometry. The exact saddle is
\begin{equation}
    \omega_0=c\left(\frac{\tau_*^2}{s^2}-1\right),
    \label{eq:eft-sqrt-local-saddle}
\end{equation}
and approaches the fixed endpoint $\omega=0$ as $s\to-\tau_*$. The contours shown join the endpoint to the local saddle and then follow a descending arm. They are endpoint-to-infinity contours, rather than complete saddle thimbles. The coalescence of the saddle with the endpoint accounts for the error-function profile in section~\ref{subsec:Quadratic-QNM-Model}. Each truncation reproduces this local configuration, although higher truncations also introduce additional saddles at $|\omega|=O(c)$.

These additional saddles provide another illustration of the importance of the contour. For the cubic truncation at $s=-\tau_*$, there is a saddle at $\omega=4c/3$, where the exponent equals $4\tau_*c/27$. The truncated integral converges on the positive real axis, but is dominated there by this exponentially large contribution. It therefore does not approximate the continuation of the exact transform. The inherited contour~\eqref{eq:eft-sqrt-common-contour} excludes this contribution.

On the common contour, the approximation errors can be controlled directly. In the local region $|\omega|=O(\sqrt c)$, the first omitted term in an $N$th-order truncation is of order $c^{-(N-1)/2}$ in the exponent. The decay along the contour also suppresses the region $|\omega|=O(c)$, where the Taylor expansion is no longer a reliable approximation. Combining the local Taylor remainder with these tail bounds gives, for fixed $\tau_*>0$,
\begin{equation}
    \frac{F_c^{\mt{EFT},(N)}(s)-F_c(s)}{\sqrt c}
    =O\left(c^{-(N-1)/2}\right),\qquad N=2,3,4,
    \label{eq:eft-sqrt-truncation-error}
\end{equation}
uniformly for $(s+\tau_*)\sqrt{c/\tau_*}$ in compact sets on the chosen continuation. In particular, the quadratic approximation has an absolute error of order unity, while reproducing a leading profile of order $\sqrt c$, in agreement with eq.~\eqref{eq:quadratic-bump-scaling}.

Together, these examples show that the effective description requires both a local approximation to the exponent and a contour prescription inherited from the full theory. The relevant saddle and endpoint contributions must be retained, and the portions of the integral outside the effective regime must remain subleading. A convergent contour for the truncated action is not sufficient by itself, nor must it coincide geometrically with an exact UV thimble. Its role is to reproduce the leading contribution selected by the original integration cycle and its analytic continuation. In this sense, the effective integration contour retains information about the UV completion that is not contained in the truncated action alone.

\section{Asymptotics of Hadamard product}
\label{sec:asymptotic-hadamard}

Here we give an intuitive explanation of how the distribution of poles
and zeros determines the asymptotic expansion of a Hadamard product.
For lattice-labeled spectra, the discrete distribution can be decomposed
into a smooth density and oscillatory parts which retain its discreteness.
In Fourier language, these are the zero mode and the nonzero modes.
The smooth part generates the main asymptotics of $\log G(\omega)$,
including powers of $\omega$ and powers multiplied by $\log\omega$;
the nonzero modes generate exponentially small corrections in the
appropriate frequency sectors. We first describe the growth implied by
the smooth density, and then use a Fourier decomposition to identify
the origin of the two contributions.

The connection between smooth density and growth is captured by
Jensen's formula, which relates the density of poles and zeros to the radial growth the logarithm
of a meromorphic function. Let $\mathscr N(R)$ count the poles or zeros
of one family inside $|\omega|\leq R$, including multiplicities. If
$\mathscr N(R)\sim(R/|a|)^\alpha$, the corresponding canonical product
has a density-controlled growth scale of order $\alpha$, schematically
\begin{equation}
 \log G(\omega)\supset C_\alpha
 \left(\frac{\omega}{a}\right)^\alpha,
 \label{eq:dm-density-growth}
\end{equation}
with possible logarithmic factors at integer powers.

A linear pole or zero family has order one. Bending it into a
one-dimensional degree-$K$ polynomial family lowers its counting exponent
to $1/K$, since $\omega_n\sim a_Kn^K$ gives
$\mathscr N(R)\sim(R/|a_K|)^{1/K}$. For $K>1$, the same scale follows
directly from the continuum approximation:
\begin{equation}
 \sum_n\log\left(1-\frac{\omega}{a_Kn^K}\right)
 \sim\int_0^\infty dn\,
 \log\left(1-\frac{\omega}{a_Kn^K}\right)
 =C_K\left(\frac{\omega}{a_K}\right)^{1/K}.
 \label{eq:dm-continuum-product}
\end{equation}
The missing pole--zero density can again be restored by the
$k$-dimensional structure. A lattice region of linear size $L$ contains
order $L^k$ states. Thus, when the polynomial grows as $L^K$ on its
indexing cone,
\begin{equation}
 \mathscr N(R)\sim R^{k/K}.
 \label{eq:dm-counting}
\end{equation}
Higher polynomial degree dilutes the spectrum, while higher lattice
dimension increases its multiplicity, giving the smooth growth order
$k/K$. Here one can clearly see why multidimensional distribution, which can be viewed as infinite product over one-dimensional subfamilies, is fundamentally different from finitely many one dimensional families. The former will enhance the overall growth order to at most $k/K$ while the latter remains $1/K$. Essentially it is because the infinite product over one-dimensional subfamilies does not commute with the asymptotic expansion of each individual subfamilies, and it is important to perform the multidimensional product before taking
the asymptotic expansion.

To separate the smooth and oscillatory contributions explicitly, let
$P(\mathbf n)$ label one family and introduce its heat trace,
\begin{equation}
 \Theta_P(t)=\sum_{\mathbf n\in\Lambda}e^{-tP(\mathbf n)},\qquad
 \log D_P(\omega)=
 -\int_0^\infty\frac{dt}{t}\,
 e^{\omega t}\Theta_P(t).
 \label{eq:dm-heat-product}
\end{equation}
Here $D_P$ is the normalized product of
$1-\omega/P(\mathbf n)$, with any zero-frequency factor removed. $\Lambda$ is the pole or zero lattice.
For the full correlator,
\begin{equation}
 \log G(\omega)=Q(\omega)+\sum_I\epsilon_I\log D_{P_I}(\omega),
 \qquad
 \epsilon_I=+1\ \text{for zeros},\quad
 \epsilon_I=-1\ \text{for poles},
 \label{eq:dm-correlator}
\end{equation}
with multiplicities included in the family sum.

The Fourier decomposition of the distribution is implemented by
reorganizing this heat trace into modes. For a full lattice
$\Lambda=\mathbb Z^k$, with $P$ growing in every real direction,
Poisson summation gives
\begin{equation}
 \Theta_P(t)=\sum_{\mathbf m\in\mathbb Z^k}H_{\mathbf m}(t),
 \qquad
 H_{\mathbf m}(t)=\int_{\mathbb R^k}d^kx\,
 e^{-tP(x)+2\pi i\mathbf m\cdot x}.
 \label{eq:dm-poisson}
\end{equation}
The zero mode replaces the lattice sum by a continuum integral and
therefore isolates the smooth density. For a degree-$K$ polynomial,
rescaling $x=t^{-1/K}y$ shows that
\begin{equation}
 H_0(t)\sim\sum_{\ell\geq0}h_\ell t^{(\ell-k)/K}.
 \label{eq:dm-zero-mode}
\end{equation}
Its proper-time integral generates the powers and logarithms of the
large-frequency expansion, with leading scale $\omega^{k/K}$, or
$\omega^{k/K}\log\omega$ at positive integer $k/K$. This is the
heat-trace realization of the smooth density described by Jensen's
formula.
For spectra on $\mathbb{N}_0^k$, the smooth contribution also includes
terms from the faces, edges and vertex of the indexing cone. They supply
additional powers and logarithms and must be separated from the
oscillatory remainder.

The remaining nonzero modes describe the discreteness omitted by the
continuum approximation. Their Fourier oscillations produce terms
that are exponentially small at short proper time, which in turn
generate exponentially small large-frequency contributions. For example,
for the one-dimensional quadratic spectrum $P(x)=ax^2$, with $a>0$,
the Fourier integral is Gaussian:
\begin{equation}
 H_m(t)=\int_{\mathbb R}dx\,e^{-atx^2+2\pi imx}
 =\sqrt{\frac{\pi}{at}}\,
   e^{-\pi^2m^2/(at)},\qquad m\neq0.
\end{equation}
Then the proper-time integral is
\begin{equation}
 \int_0^\infty\frac{dt}{t}\,e^{\omega t}H_m(t)
 =\frac{1}{|m|}
   e^{-2\pi|m|\sqrt{-\omega/a}}.
\end{equation}
Thus the nonzero Fourier modes generate
a tower of exponentially small terms, invisible to the smooth-density
expansion.
More generally, for a nondegenerate homogeneous leading polynomial
of degree $K>1$, the combined Fourier and proper-time saddle has
$x_*\sim\omega^{1/K}$. Its exponent is
$2\pi i\mathbf m\cdot x_*$, since the proper-time saddle imposes
$P(x_*)=\omega$. The resulting contributions therefore have the
schematic form
$\exp[-A_{\mathbf m}\omega^{1/K}]$.
Picard--Lefschetz theory determines which saddles contribute and how
their contributions change across Stokes walls. Their detailed form
depends on the spectral polynomial, but their origin is the same:
they are the nonzero Fourier modes of the pole--zero distribution.

\section{IR and UV constraints for pole-zero data}
\label{app:UV-IR-constraints}

In this appendix section, we illustrate the IR and UV constraints discussed in section \ref{sec:discussion} in the case of Liouville CFT. Liouville CFT also serves as a non-trivial example for the mechanism discussed in appendix \ref{sec:asymptotic-hadamard} since its DOZZ coefficient has a two dimensional quadratic distribution. 

We retain the conventions
$\alpha_2=Q/2+i\kappa$, $\alpha_3=Q/2+iP$, and
$\omega=2(P^2-\kappa^2)$, with
$h_L=\alpha_1(Q-\alpha_1)=\Delta/2$.
The coefficient entering the primary exchange is
\begin{equation}
 \mathcal D(\omega)
 =C_L(\alpha_1,\alpha_2,\alpha_3)
  C_L(\alpha_1,\alpha_2,Q-\alpha_3).
 \label{eq:lp-dozz-pair}
\end{equation}
Using $\Upsilon_b(x)=\Upsilon_b(Q-x)$, its frequency dependence is
\begin{equation}
 \mathcal D(\omega)=\mathcal N\,
 \frac{\Upsilon_b(Q+2iP)\Upsilon_b(Q-2iP)}
 {\displaystyle\prod_{\sigma,\tau=\pm1}
  \Upsilon_b(\alpha_1+i\sigma P+i\tau\kappa)^2},
 \label{eq:lp-upsilon-product}
\end{equation}
where $\mathcal N$ is independent of $\omega$. The standard Liouville
measure is $dP$, so the primary contribution to the frequency-space
spectral weight, including the Jacobian, is
$\mathcal W_{\rm prim}(\omega)=\mathcal D(\omega)/(4P)$, up to an overall
normalization. We use this notation to distinguish the primary contribution
from the full heavy-state correlator $G_H$.

All pole and zero families depend on the two integers only through
$s_{m,n}=m/b+nb$. They therefore form a rank-one two dimensional quadratic distribution,
which can be written as
\begin{equation}
 \omega_{I;m,n}=-2\kappa^2-a_I(s_{m,n}+d_I)^2,
 \qquad m,n\geq0.
 \label{eq:lp-quadratic-families}
\end{equation}
Assigning positive multiplicity to zeros and negative multiplicity to poles,
the complete pole-zero data of $\mathcal D$ are
\begin{equation}
\begin{array}{c|c|c|c}
 \text{family}&\epsilon_I&a_I&d_I\\ \hline
 \text{zeros}&1&1/2&0\\
 \text{zeros}&1&1/2&Q\\
 \text{poles}&-2&2&\alpha_1\pm i\kappa\\
 \text{poles}&-2&2&Q-\alpha_1\pm i\kappa
\end{array}
 \label{eq:lp-family-data}
\end{equation}
Each sign denotes a separate family.

The infinite product is evaluated by first factoring the quadratic polynomial and then performing the lattice product. Each factor in the zero product is
\begin{equation}
 \omega-\omega_{m,n}
 =a\bigl[(s_{m,n}+d)^2+X_a^2\bigr]
 =a(s_{m,n}+d+iX_a)(s_{m,n}+d-iX_a),
 \qquad
 X_a=\sqrt{\frac{\omega+2\kappa^2}{a}}.
 \label{eq:lp-quadratic-factorization}
\end{equation}
The two linear factors can be summed using the Barnes double-gamma
function $\Gamma_b$. Its poles lie at $x=-s_{m,n}$, so that
$1/\Gamma_b(x)$ represents the regularized product of $s_{m,n}+x$.
We therefore define
\begin{equation}
 \mathcal P_{a,d}(\omega)
 =\frac{1}{\Gamma_b(d+iX_a)\Gamma_b(d-iX_a)}.
 \label{eq:lp-barnes-block}
\end{equation}
This function has precisely the zeros $\omega_{m,n}$, including their
lattice multiplicities.
To specify the regularization, introduce the Barnes double zeta function
\begin{equation}
 \zeta_2(s,x)=
 \sum_{m,n\geq0}(s_{m,n}+x)^{-s},
 \qquad \operatorname{Re}s>2,
 \label{eq:lp-double-zeta}
\end{equation}
initially for $\operatorname{Re}x>0$, and then analytically continue it.
With the standard normalization $\Gamma_b(Q/2)=1$,
\begin{equation}
 \log\Gamma_b(x)
 =\left.\partial_s
 \bigl[\zeta_2(s,x)-\zeta_2(s,Q/2)\bigr]\right|_{s=0}.
\end{equation}
Consequently,
\begin{equation}
 \log\mathcal P_{a,d}
 =-\left.\partial_s\bigl[
 \zeta_2(s,d+iX_a)+\zeta_2(s,d-iX_a)
 -2\zeta_2(s,Q/2)\bigr]\right|_{s=0}.
 \label{eq:lp-zeta-product}
\end{equation}
This is the regularized sum of the logarithms of the two linear
factors in \eqref{eq:lp-quadratic-factorization}. 

To recover the DOZZ pair, use
$\Upsilon_b(x)=[\Gamma_b(x)\Gamma_b(Q-x)]^{-1}$.
Since $X_{1/2}=2P$ and $X_2=P$, the numerator of
\eqref{eq:lp-upsilon-product} becomes
$\mathcal P_{1/2,0}\mathcal P_{1/2,Q}$.
Similarly, pairing the denominator factors with opposite signs of $P$
gives
\begin{equation}
 \prod_{\sigma,\tau=\pm1}
 \Upsilon_b(\alpha_1+i\sigma P+i\tau\kappa)
 =\prod_{d\in\mathcal S}\mathcal P_{2,d},
 \qquad
 \mathcal S=\{\alpha_1\pm i\kappa,\ Q-\alpha_1\pm i\kappa\}.
\end{equation}
Hence
\begin{equation}
 \mathcal D(\omega)=\mathcal N\,
 \mathcal P_{1/2,0}(\omega)\mathcal P_{1/2,Q}(\omega)
 \prod_{d\in\mathcal S}\mathcal P_{2,d}(\omega)^{-2}.
 \label{eq:lp-exact-product}
\end{equation}
Here $\mathcal N$ is independent of frequency. Positive powers supply
the zero families, while negative powers supply the pole families.

\paragraph{UV constraints}

For fixed $b$, the UV expansion follows directly from the double-gamma
asymptotics. Along the positive real-frequency ray one has
\begin{align}
 \log\mathcal P_{a,d}(\omega)
 ={}&-\frac{\omega}{2a}\log\omega
 +\frac{\log a+3}{2a}\,\omega
 +\frac{\pi(Q/2-d)}{\sqrt a}\sqrt\omega\notag\\
 &+\left(\frac{d^2-Qd}{2}+\frac{Q^2+1}{12}-\frac{\kappa^2}{a}\right)\log\omega+O(1).
 \label{eq:lp-uv-block}
\end{align}
The OPE matching conditions apply to the complete spectral function. For
the target $\log G=-\vartheta\omega+\nu\log\omega+O(1)$, the
$\omega\log\omega$ and $\sqrt\omega$ coefficients must vanish, while the
linear and logarithmic coefficients must equal $-\vartheta$ and $\nu$. Here we are considering HHLL correlator so $\vartheta=0$. 

For the primary coefficient, the first two tests reduce to the explicit
algebraic equations
\begin{equation}
 \sum_I\frac{\epsilon_I}{a_I}=0, \qquad  \sum_I\epsilon_I\frac{Q/2-d_I}{\sqrt{a_I}}=0.
 \label{eq:lp-uv-constraints}
\end{equation}
Substitution of \eqref{eq:lp-family-data} verifies both:
\begin{align}
 \sum_I\frac{\epsilon_I}{a_I}&=2+2-4=0,\notag\\
 \sum_I\epsilon_I\frac{Q/2-d_I}{\sqrt{a_I}}
 &=\sqrt2\,[Q/2+(Q/2-Q)]
   -\sqrt2\sum_{d\in\mathcal S}(Q/2-d)=0.
 \label{eq:lp-uv-verification}
\end{align}
The remaining terms are also fixed by the same data:
\begin{equation}
 \log\mathcal W_{\rm prim}(\omega)
 =-(\log16)\omega+
 \left[2\Delta+4\kappa^2-\frac{Q^2+2}{2}\right]\log\omega+O(1).
 \label{eq:lp-primary-uv}
\end{equation}
In \ref{eq:lp-dozz-pair} we only consider the contribution from primaries. 
Though two cancellations in \eqref{eq:lp-uv-constraints} are
already verified by only the primary data, exact matching of the linear and logarithmic terms in \eqref{eq:lp-primary-uv} is actually sensitive to descendants. 
For example, in position space, the summation over descendants will give an additional large internal weight factor $(16q)^{P^2}$, which cancels the
$16^{-2P^2}$ suppression of the DOZZ pair coming from the linear terms in
\eqref{eq:lp-primary-uv}. 

One can also make the Cardy density part explicit in the reflection-invariant $C_0$ normalization. Up to factors independent of $P$,
\begin{equation}
 C_0(P)\propto
 \frac{\mathcal P_{1/2,Q}}{\prod_{d\in\mathcal S}\mathcal P_{2,d}},
 \qquad
 \mathcal D\propto\rho_0(P)C_0(P)^2,
 \qquad
 \rho_0(P)=4\sqrt2\sinh(2\pi bP)\sinh(2\pi P/b).
 \label{eq:lp-density-normalization}
\end{equation}
Here $\rho_0(P)\,dP$ replaces the flat Liouville measure together with the
corresponding change of three-point normalization. Equation
\eqref{eq:lp-uv-block} now gives
\begin{equation}
 \left.\log C_0(P)^2\right|_{\sqrt\omega}
 =-\sqrt2\pi Q\sqrt\omega,
 \qquad
 \log\rho_0(P)=\sqrt2\pi Q\sqrt\omega+O(1).
 \label{eq:lp-cardy-cancellation}
\end{equation}
Thus the Cardy density shifts the square-root matching condition by
$+\sqrt2\pi Q$, exactly cancelling the coefficient contribution. 

\paragraph{IR constraints}

The IR limit resolves the same product into ordinary Gamma functions.
Take $b\to0$ with $\alpha_1=bh+O(b^3)$,
$\kappa\simeq\pi/(b\beta_H)$, and $\beta_H\omega$ fixed. The $m=0$
branches remain at finite frequency, whereas the other lattice direction
and the additional families move to the heavy scale. Their leading
frequency dependence yields
\begin{equation}
 \mathcal W_{\rm prim}^{\rm IR}(\omega)
 \propto\Gamma(h+iy)^2\Gamma(h-iy)^2,
 \qquad y=\frac{\beta_H\omega}{4\pi},\qquad h=\frac\Delta2.
 \label{eq:lp-ir-gamma}
\end{equation}
Taking $y\gg1$ after this semiclassical limit gives the symmetric BTZ
dependence $e^{-\beta_H\omega/2}\omega^{2\Delta-2}$.
Equivalently, the two linearized roots have
$r_\pm=-h$, $v_\pm=\pm i\beta_H/(4\pi)$, and
$\epsilon_\pm=-2$. Since
$\sum_{n\geq0}^{\rm reg}\log(n-z)=-\log\Gamma(-z)+\text{constant}$,
the IR constraints and their verification are
\begin{equation}
 \sum_{\pm}\epsilon_\pm v_\pm=0,
 \qquad
 \sum_{\pm}\epsilon_\pm(r_\pm+\tfrac12)=2\Delta-2,
 \qquad
 \sum_{\pm}\epsilon_\pm v_\pm[\log(-v_\pm)-1]
 =-\frac{\beta_H}{2}.
 \label{eq:lp-ir-constraints}
\end{equation}
The logarithms use the conjugate Stirling branches.

\paragraph{Discussion}

These calculations make clear the roles of zeros and of the two-dimensional
lattice.  In the $C_0^2$
normalization, the four pole families have zero net square-root
coefficient; the zero family of $C_0^2$ supplies the negative term in
\eqref{eq:lp-cardy-cancellation} while the zeros of $\rho_0$ give exactly the Cardy like behaviour. Meanwhile, counting lattice points with
$|\omega_{m,n}|<R$ gives $N(R)\sim R$, rather than $R^{1/2}$ for a
single quadratic sequence. This restores the order-one growth needed for
the linear term in \eqref{eq:lp-primary-uv}.

%%%%%%%%%%%%%%%%%%%%%%%%%%%%%%%%%%%%%%%%%%%%%%%%
%%%%%%%%%%%%%%%%%%%%%%%%%%%%%%%%%%%%%%%%%%%%%%%%

%%%%%%%%%%%%%%%%%%%%%%%%%%%%%%%%%%%%%%%%%%%%%%%%%%%%%%%%%%%%%%%%%%%%%%
%%%%%%%%%%%%%%%%%%%%%%%%%%%%%%%%%%%%%%%%%%%%%%%%%%%%%%%%%%%%%%%%%%%%%%
\bibliographystyle{jhep}
\bibliography{references}

@article{Blommaert:2018oro,
    author = "Blommaert, Andreas and Mertens, Thomas G. and Verschelde, Henri",
    title = "{The Schwarzian Theory - A Wilson Line Perspective}",
    eprint = "1806.07765",
    archivePrefix = "arXiv",
    primaryClass = "hep-th",
    doi = "10.1007/JHEP12(2018)022",
    journal = "JHEP",
    volume = "12",
    pages = "022",
    year = "2018"
}

@article{Hartman:2013qma,
    author = "Hartman, Thomas and Maldacena, Juan",
    title = "{Time Evolution of Entanglement Entropy from Black Hole Interiors}",
    eprint = "1303.1080",
    archivePrefix = "arXiv",
    primaryClass = "hep-th",
    doi = "10.1007/JHEP05(2013)014",
    journal = "JHEP",
    volume = "05",
    pages = "014",
    year = "2013"
}

@article{Mertens:2017mtv,
    author = "Mertens, Thomas G. and Turiaci, Gustavo J. and Verlinde, Herman L.",
    title = "{Solving the Schwarzian via the Conformal Bootstrap}",
    eprint = "1705.08408",
    archivePrefix = "arXiv",
    primaryClass = "hep-th",
    doi = "10.1007/JHEP08(2017)136",
    journal = "JHEP",
    volume = "08",
    pages = "136",
    year = "2017"
}

@article{Xiao:2026pir,
    author = "Xiao, Zi-Qing and Nian, Jun and Li, Li",
    title = "{Channel-Resolved High-Overtone Quasinormal Modes Decode Kasner Scaling inside Black Holes}",
    eprint = "2609.17426",
    archivePrefix = "arXiv",
    primaryClass = "hep-th",
    month = "9",
    year = "2026"
}

@article{Miyaji:2025yvm,
    author = "Miyaji, Masamichi and Ruan, Shan-Ming and Shibuya, Shono and Yano, Kazuyoshi",
    title = "{Non-perturbative overlaps in JT gravity: from spectral form factor to generating functions of complexity}",
    eprint = "2502.12266",
    archivePrefix = "arXiv",
    primaryClass = "hep-th",
    reportNumber = "YITP-25-28",
    doi = "10.1007/JHEP06(2025)251",
    journal = "JHEP",
    volume = "06",
    pages = "251",
    year = "2025"
}

@article{Miyaji:2025jxy,
    author = "Miyaji, Masamichi and Ruan, Shan-Ming and Shibuya, Shono and Yano, Kazuyoshi",
    title = "{Universal Time Evolution of Holographic and Quantum Complexity}",
    eprint = "2507.23667",
    archivePrefix = "arXiv",
    primaryClass = "hep-th",
    reportNumber = "RIKEN-iTHEMS-Report-25",
    doi = "10.1103/fsdt-d3p9",
    journal = "Phys. Rev. Lett.",
    volume = "136",
    number = "15",
    pages = "151602",
    year = "2026"
}

@article{Miyaji:2026pwc,
    author = "Miyaji, Masamichi and Ruan, Shan-Ming and Shibuya, Shono and Yano, Kazuyoshi",
    title = "{Codimension-Zero Holographic Complexity and Spectral Three-Point Correlators}",
    eprint = "2610.13289",
    archivePrefix = "arXiv",
    primaryClass = "hep-th",
    reportNumber = "RIKEN-iTHEMS-Report-26",
    month = "10",
    year = "2026"
}

@article{Bhattacharjee:2020nul,
    author = "Bhattacharjee, Srijit and Sarkar, Subhodeep and Bhattacharyya, Arpan",
    title = "{Scalar perturbations of black holes in Jackiw-Teitelboim gravity}",
    eprint = "2011.08179",
    archivePrefix = "arXiv",
    primaryClass = "gr-qc",
    doi = "10.1103/PhysRevD.103.024008",
    journal = "Phys. Rev. D",
    volume = "103",
    number = "2",
    pages = "024008",
    year = "2021"
}

@article{Chen:2010ik,
    author = "Chen, Bin and Long, Jiang",
    title = "{Hidden Conformal Symmetry and Quasi-normal Modes}",
    eprint = "1009.1010",
    archivePrefix = "arXiv",
    primaryClass = "hep-th",
    doi = "10.1103/PhysRevD.82.126013",
    journal = "Phys. Rev. D",
    volume = "82",
    pages = "126013",
    year = "2010"
}

@article{Zhang:2011ec,
    author = "Zhang, Hong-bao",
    title = "{$SL(2,R)$ symmetry and quasi-normal modes in the BTZ black hole}",
    eprint = "1102.4721",
    archivePrefix = "arXiv",
    primaryClass = "hep-th",
    reportNumber = "CCTP-2011-06",
    doi = "10.1007/JHEP03(2011)009",
    journal = "JHEP",
    volume = "03",
    pages = "009",
    year = "2011"
}

@article{Cardoso:2001hn,
    author = "Cardoso, Vitor and Lemos, Jose P. S.",
    title = "{Scalar, electromagnetic and Weyl perturbations of BTZ black holes: Quasinormal modes}",
    eprint = "gr-qc/0101052",
    archivePrefix = "arXiv",
    doi = "10.1103/PhysRevD.63.124015",
    journal = "Phys. Rev. D",
    volume = "63",
    pages = "124015",
    year = "2001"
}

@article{Birmingham:2001hc,
    author = "Birmingham, Danny",
    title = "{Choptuik scaling and quasinormal modes in the AdS / CFT correspondence}",
    eprint = "hep-th/0101194",
    archivePrefix = "arXiv",
    doi = "10.1103/PhysRevD.64.064024",
    journal = "Phys. Rev. D",
    volume = "64",
    pages = "064024",
    year = "2001"
}

@article{Oh:2026dky,
    author = "Oh, Hyunwoo",
    title = "{Kasner Exponents from the Bouncing Singularity}",
    eprint = "2610.31699",
    archivePrefix = "arXiv",
    primaryClass = "hep-th",
    month = "10",
    year = "2026"
}

@article{Chakravarty:2026fdm,
    author = "Chakravarty, Joydeep and Zigdon, Yoav",
    title = "{A signature of the timelike singularity in holography}",
    eprint = "2610.30134",
    archivePrefix = "arXiv",
    primaryClass = "hep-th",
    month = "10",
    year = "2026"
}

@article{Chen:2026efa,
    author = "Chen, Heng-Yu and Hikida, Yasuaki and Koga, Yasutaka",
    title = "{Quasinormal modes, complex null geodesics and singularity resolution in regular AdS black holes}",
    eprint = "2610.30915",
    archivePrefix = "arXiv",
    primaryClass = "hep-th",
    reportNumber = "RUP-26-31",
    month = "10",
    year = "2026"
}

@article{Komargodski:2026oob,
    author = "Komargodski, Zohar",
    title = "{Fate of Bouncing Singularities at Finite $G_N$}",
    eprint = "2609.31309",
    archivePrefix = "arXiv",
    primaryClass = "hep-th",
    month = "9",
    year = "2026"
}

@article{Ceplak:2026ard,
    author = "{\v{C}}eplak, Nejc and Valach, Samuel",
    title = "{Where's Kasner? OPE Signature of Kasner Exponents}",
    eprint = "2609.23776",
    archivePrefix = "arXiv",
    primaryClass = "hep-th",
    month = "9",
    year = "2026"
}

@article{Batra:2026gbq,
    author = "Batra, Gauri and Levine, Adam and Shenker, Stephen H.",
    title = "{A backreacted bouncing geodesic in 2D}",
    eprint = "2609.32071",
    archivePrefix = "arXiv",
    primaryClass = "hep-th",
    month = "9",
    year = "2026"
}

@article{Buric:2026sit,
    author = "Buri{\'c}, Ilija and Gusev, Ivan and Parnachev, Andrei",
    title = "{Stringy resolution of bouncing singularities}",
    eprint = "2610.12311",
    archivePrefix = "arXiv",
    primaryClass = "hep-th",
    month = "10",
    year = "2026"
}

@article{Hartnoll:2026vhu,
    author = "Hartnoll, Sean A. and Zhiboedov, Alexander",
    title = "{Can one hear the shape of a black hole singularity?}",
    eprint = "2609.07514",
    archivePrefix = "arXiv",
    primaryClass = "hep-th",
    reportNumber = "CERN-TH-2026-217",
    month = "9",
    year = "2026"
}

@article{Horowitz:2023ury,
    author = "Horowitz, Gary T. and Leung, Henry and Queimada, Leonel and Zhao, Ying",
    title = "{Boundary signature of singularity in the presence of a shock wave}",
    eprint = "2310.03076",
    archivePrefix = "arXiv",
    primaryClass = "hep-th",
    doi = "10.21468/SciPostPhys.16.2.060",
    journal = "SciPost Phys.",
    volume = "16",
    number = "2",
    pages = "060",
    year = "2024"
}

@article{Bhattacharya:2025vyi,
    author = "Bhattacharya, Jyotirmoy and Padhi, Nibedita and Sharma, Aditya and Singha, Sourav",
    title = "{Thermal product formula for shear modes}",
    eprint = "2504.17781",
    archivePrefix = "arXiv",
    primaryClass = "hep-th",
    doi = "10.1007/JHEP08(2025)170",
    journal = "JHEP",
    volume = "08",
    pages = "170",
    year = "2025"
}

@article{Grozdanov:2025ulc,
    author = "Grozdanov, Sa{\v{s}}o and Vrbica, Mile",
    title = "{Thermal field theory correlators in the large-N limit and the spectral duality relation}",
    eprint = "2509.18074",
    archivePrefix = "arXiv",
    primaryClass = "hep-th",
    doi = "10.1007/JHEP02(2026)106",
    journal = "JHEP",
    volume = "02",
    pages = "106",
    year = "2026"
}

@article{Grozdanov:2024wgo,
    author = "Grozdanov, Sa{\v{s}}o and Vrbica, Mile",
    title = "{Duality Constraints on Thermal Spectra of 3D Conformal Field Theories and 4D Quasinormal Modes}",
    eprint = "2406.19790",
    archivePrefix = "arXiv",
    primaryClass = "hep-th",
    doi = "10.1103/PhysRevLett.133.211601",
    journal = "Phys. Rev. Lett.",
    volume = "133",
    number = "21",
    pages = "211601",
    year = "2024"
}

@article{DAlessio:2015qtq,
    author = "D'Alessio, Luca and Kafri, Yariv and Polkovnikov, Anatoli and Rigol, Marcos",
    title = "{From quantum chaos and eigenstate thermalization to statistical mechanics and thermodynamics}",
    eprint = "1509.06411",
    archivePrefix = "arXiv",
    primaryClass = "cond-mat.stat-mech",
    doi = "10.1080/00018732.2016.1198134",
    journal = "Adv. Phys.",
    volume = "65",
    number = "3",
    pages = "239--362",
    year = "2016"
}

@article{Srednicki:1999,
   title={The approach to thermal equilibrium in quantized chaotic systems},
   volume={32},
   ISSN={1361-6447},
   url={http://dx.doi.org/10.1088/0305-4470/32/7/007},
   DOI={10.1088/0305-4470/32/7/007},
   number={7},
   journal={Journal of Physics A: Mathematical and General},
   publisher={IOP Publishing},
   author={Srednicki, Mark},
   year={1999},
   month=Jan, pages={1163–1175} }

@article{Murthy:2019fgs,
    author = "Murthy, Chaitanya and Srednicki, Mark",
    title = "{Bounds on chaos from the eigenstate thermalization hypothesis}",
    eprint = "1906.10808",
    archivePrefix = "arXiv",
    primaryClass = "cond-mat.stat-mech",
    doi = "10.1103/PhysRevLett.123.230606",
    journal = "Phys. Rev. Lett.",
    volume = "123",
    number = "23",
    pages = "230606",
    year = "2019"
}

@article{Lashkari:2016vgj,
    author = "Lashkari, Nima and Dymarsky, Anatoly and Liu, Hong",
    title = "{Eigenstate Thermalization Hypothesis in Conformal Field Theory}",
    eprint = "1610.00302",
    archivePrefix = "arXiv",
    primaryClass = "hep-th",
    doi = "10.1088/1742-5468/aab020",
    journal = "J. Stat. Mech.",
    volume = "1803",
    number = "3",
    pages = "033101",
    year = "2018"
}

@article{Balasubramanian:2017fan,
    author = "Balasubramanian, Vijay and Bernamonti, Alice and Craps, Ben and De Jonckheere, Tim and Galli, Federico",
    title = "{Heavy-Heavy-Light-Light correlators in Liouville theory}",
    eprint = "1705.08004",
    archivePrefix = "arXiv",
    primaryClass = "hep-th",
    doi = "10.1007/JHEP08(2017)045",
    journal = "JHEP",
    volume = "08",
    pages = "045",
    year = "2017"
}

@article{Grabovsky:2024jwf,
    author = "Grabovsky, David",
    title = "{Heavy states in 3d gravity and 2d CFT}",
    eprint = "2403.13757",
    archivePrefix = "arXiv",
    primaryClass = "hep-th",
    doi = "10.1007/JHEP07(2024)287",
    journal = "JHEP",
    volume = "07",
    pages = "287",
    year = "2024"
}

@article{Berenstein:2022ico,
    author = "Berenstein, David and Grabovsky, David and Li, Ziyi",
    title = "{Aspects of holography in conical AdS$_{3}$}",
    eprint = "2205.02256",
    archivePrefix = "arXiv",
    primaryClass = "hep-th",
    doi = "10.1007/JHEP04(2023)029",
    journal = "JHEP",
    volume = "04",
    pages = "029",
    year = "2023"
}

@article{Fitzpatrick:2014vua,
    author = "Fitzpatrick, A. Liam and Kaplan, Jared and Walters, Matthew T.",
    title = "{Universality of Long-Distance AdS Physics from the CFT Bootstrap}",
    eprint = "1403.6829",
    archivePrefix = "arXiv",
    primaryClass = "hep-th",
    doi = "10.1007/JHEP08(2014)145",
    journal = "JHEP",
    volume = "08",
    pages = "145",
    year = "2014"
}

@article{Dorn:1994xn,
    author = "Dorn, Harald and Otto, H. J.",
    title = "{Two and three point functions in Liouville theory}",
    eprint = "hep-th/9403141",
    archivePrefix = "arXiv",
    reportNumber = "HU-BERLIN-IEP-94-2, DESY-94-066",
    doi = "10.1016/0550-3213(94)00352-1",
    journal = "Nucl. Phys. B",
    volume = "429",
    pages = "375--388",
    year = "1994"
}

@article{Teschner:2001rv,
    author = "Teschner, Jorg",
    title = "{Liouville theory revisited}",
    eprint = "hep-th/0104158",
    archivePrefix = "arXiv",
    doi = "10.1088/0264-9381/18/23/201",
    journal = "Class. Quant. Grav.",
    volume = "18",
    pages = "R153--R222",
    year = "2001"
}

@article{Ponsot:1999uf,
    author = "Ponsot, B. and Teschner, J.",
    title = "{Liouville bootstrap via harmonic analysis on a noncompact quantum group}",
    eprint = "hep-th/9911110",
    archivePrefix = "arXiv",
    year = "1999"
}

@article{Fitzpatrick:2015dlt,
    author = "Fitzpatrick, A. Liam and Kaplan, Jared",
    title = "{Conformal Blocks Beyond the Semi-Classical Limit}",
    eprint = "1512.03052",
    archivePrefix = "arXiv",
    primaryClass = "hep-th",
    doi = "10.1007/JHEP05(2016)075",
    journal = "JHEP",
    volume = "05",
    pages = "075",
    year = "2016"
}

@article{Chen:2017yze,
    author = "Chen, Hongbin and Hussong, Charles and Kaplan, Jared and Li, Daliang",
    title = "{A Numerical Approach to Virasoro Blocks and the Information Paradox}",
    eprint = "1703.09727",
    archivePrefix = "arXiv",
    primaryClass = "hep-th",
    doi = "10.1007/JHEP09(2017)102",
    journal = "JHEP",
    volume = "09",
    pages = "102",
    year = "2017"
}

@article{Fitzpatrick:2016ive,
    author = "Fitzpatrick, A. Liam and Kaplan, Jared and Li, Daliang and Wang, Junpu",
    title = "{On information loss in AdS$_{3}$/CFT$_{2}$}",
    eprint = "1603.08925",
    archivePrefix = "arXiv",
    primaryClass = "hep-th",
    doi = "10.1007/JHEP05(2016)109",
    journal = "JHEP",
    volume = "05",
    pages = "109",
    year = "2016"
}

@article{Cotler:2016fpe,
    author = "Cotler, Jordan S. and Gur-Ari, Guy and Hanada, Masanori and Polchinski, Joseph and Saad, Phil and Shenker, Stephen H. and Stanford, Douglas and Streicher, Alexandre and Tezuka, Masaki",
    title = "{Black Holes and Random Matrices}",
    eprint = "1611.04650",
    archivePrefix = "arXiv",
    primaryClass = "hep-th",
    reportNumber = "SU-ITP-16-19, SU-ITP-16/19, YITP-16-124",
    doi = "10.1007/JHEP05(2017)118",
    journal = "JHEP",
    volume = "05",
    pages = "118",
    year = "2017",
    note = "[Erratum: JHEP 09, 002 (2018)]"
}

@article{Saad:2018bqo,
    author = "Saad, Phil and Shenker, Stephen H. and Stanford, Douglas",
    title = "{A semiclassical ramp in SYK and in gravity}",
    eprint = "1806.06840",
    archivePrefix = "arXiv",
    primaryClass = "hep-th",
    month = "6",
    year = "2018"
}

@article{Saad:2019lba,
    author = "Saad, Phil and Shenker, Stephen H. and Stanford, Douglas",
    title = "{JT gravity as a matrix integral}",
    eprint = "1903.11115",
    archivePrefix = "arXiv",
    primaryClass = "hep-th",
    month = "3",
    year = "2019"
}

@article{Balasubramanian:1999zv,
    author = "Balasubramanian, Vijay and Ross, Simon F.",
    title = "{Holographic particle detection}",
    eprint = "hep-th/9906226",
    archivePrefix = "arXiv",
    reportNumber = "HUTP-99-A032, UCSBTH-99-1",
    doi = "10.1103/PhysRevD.61.044007",
    journal = "Phys. Rev. D",
    volume = "61",
    pages = "044007",
    year = "2000"
}

@article{Louko:2000tp,
    author = "Louko, Jorma and Marolf, Donald and Ross, Simon F.",
    title = "{On geodesic propagators and black hole holography}",
    eprint = "hep-th/0002111",
    archivePrefix = "arXiv",
    doi = "10.1103/PhysRevD.62.044041",
    journal = "Phys. Rev. D",
    volume = "62",
    pages = "044041",
    year = "2000"
}

@article{Aparicio:2011zy,
    author = "Aparicio, Joao and Lopez, Esperanza",
    title = "{Evolution of Two-Point Functions from Holography}",
    eprint = "1109.3571",
    archivePrefix = "arXiv",
    primaryClass = "hep-th",
    reportNumber = "IFT-UAM-CSIC-11-59, IFT-UAM/CSIC-11-59",
    doi = "10.1007/JHEP12(2011)082",
    journal = "JHEP",
    volume = "12",
    pages = "082",
    year = "2011"
}

@article{Balasubramanian:2012tu,
    author = {Balasubramanian, V. and Bernamonti, A. and Craps, B. and Ker{\"a}nen, V. and Keski-Vakkuri, E. and M{\"u}ller, B. and Thorlacius, L. and Vanhoof, J.},
    title = "{Thermalization of the spectral function in strongly coupled two dimensional conformal field theories}",
    eprint = "1212.6066",
    archivePrefix = "arXiv",
    primaryClass = "hep-th",
    doi = "10.1007/JHEP04(2013)069",
    journal = "JHEP",
    volume = "04",
    pages = "069",
    year = "2013"
}

@article{Witten:2010cx,
    author = "Witten, Edward",
    editor = "Andersen, Joergen E. and Boden, Hans U. and Hahn, Atle and Himpel, Benjamin",
    title = "{Analytic Continuation Of Chern-Simons Theory}",
    eprint = "1001.2933",
    archivePrefix = "arXiv",
    primaryClass = "hep-th",
    journal = "AMS/IP Stud. Adv. Math.",
    volume = "50",
    pages = "347--446",
    year = "2011"
}

@article{Penrose:1965Singularity,
    author = "Penrose, Roger",
    title = "{Gravitational Collapse and Space-Time Singularities}",
    doi = "10.1103/PhysRevLett.14.57",
    journal = "Phys. Rev. Lett.",
    volume = "14",
    pages = "57--59",
    year = "1965"
}

@article{HawkingPenrose:1970,
    author = "Hawking, S. W. and Penrose, R.",
    title = "{The Singularities of Gravitational Collapse and Cosmology}",
    doi = "10.1098/rspa.1970.0021",
    journal = "Proc. Roy. Soc. Lond. A",
    volume = "314",
    pages = "529--548",
    year = "1970"
}

@book{Belinski:2017awb,
    author = "Belinski, Vladimir and Henneaux, Marc",
    title = "{The Cosmological Singularity}",
    isbn = "978-1-107-04747-1, 978-1-108-54799-4",
    publisher = "Cambridge University Press",
    month = "11",
    year = "2017"
}

@article{Belinsky:1970ew,
    author = "Belinsky, V. A. and Khalatnikov, I. M. and Lifshitz, E. M.",
    title = "{Oscillatory approach to a singular point in the relativistic cosmology}",
    doi = "10.1080/00018737000101171",
    journal = "Adv. Phys.",
    volume = "19",
    pages = "525--573",
    year = "1970"
}

@article{Belinsky:1982pk,
    author = "Belinsky, V. a. and Khalatnikov, I. m. and Lifshitz, E. m.",
    title = "{A General Solution of the Einstein Equations with a Time Singularity}",
    doi = "10.1080/00018738200101428",
    journal = "Adv. Phys.",
    volume = "31",
    pages = "639--667",
    year = "1982"
}

@article{Damour:2002et,
    author = "Damour, T. and Henneaux, M. and Nicolai, H.",
    title = "{Cosmological billiards}",
    eprint = "hep-th/0212256",
    archivePrefix = "arXiv",
    reportNumber = "IHES-P-02-80, AEI-2002-092, ULB-TH-02-33, IHES-P-02-08",
    doi = "10.1088/0264-9381/20/9/201",
    journal = "Class. Quant. Grav.",
    volume = "20",
    pages = "R145--R200",
    year = "2003"
}

@article{Jia:2025jbi,
    author = "Jia, Hewei Frederic and Rangamani, Mukund",
    title = "{Thermal spectral function asymptotics and black hole singularity in holography}",
    eprint = "2512.15114",
    archivePrefix = "arXiv",
    primaryClass = "hep-th",
    doi = "10.1103/tzm2-jjvg",
    journal = "Phys. Rev. D",
    volume = "114",
    number = "4",
    pages = "046025",
    year = "2026"
}

@article{Giombi:2026kdz,
    author = "Giombi, Simone and Li, Yue-Zhou and Shan, Jieru",
    title = "{Bouncing singularities and thermal correlators on line defects}",
    eprint = "2603.11012",
    archivePrefix = "arXiv",
    primaryClass = "hep-th",
    doi = "10.1007/JHEP07(2026)170",
    journal = "JHEP",
    volume = "07",
    pages = "170",
    year = "2026"
}

@article{Grinberg:2020fdj,
    author = "Grinberg, Matan and Maldacena, Juan",
    title = "{Proper time to the black hole singularity from thermal one-point functions}",
    eprint = "2011.01004",
    archivePrefix = "arXiv",
    primaryClass = "hep-th",
    doi = "10.1007/JHEP03(2021)131",
    journal = "JHEP",
    volume = "03",
    pages = "131",
    year = "2021"
}

@article{Grozdanov:2026ktq,
    author = "Grozdanov, Sa{\v{s}}o and Movrin, Vita and Valach, Samuel",
    title = "{Bouncing Geodesics, Singularities, and the Cavity Thermal Product Formula in Asymptotically Flat and de Sitter Black Holes}",
    eprint = "2606.11297",
    archivePrefix = "arXiv",
    primaryClass = "hep-th",
    month = "6",
    year = "2026"
}

@article{Festuccia:2005pi,
    author = "Festuccia, Guido and Liu, Hong",
    title = "{Excursions beyond the horizon: Black hole singularities in Yang-Mills theories. I.}",
    eprint = "hep-th/0506202",
    archivePrefix = "arXiv",
    reportNumber = "MIT-CTP-3641",
    doi = "10.1088/1126-6708/2006/04/044",
    journal = "JHEP",
    volume = "04",
    pages = "044",
    year = "2006"
}

@article{Ceplak:2024bja,
    author = "{\v{C}}eplak, Nejc and Liu, Hong and Parnachev, Andrei and Valach, Samuel",
    title = "{Black hole singularity from OPE}",
    eprint = "2404.17286",
    archivePrefix = "arXiv",
    primaryClass = "hep-th",
    doi = "10.1007/JHEP10(2024)105",
    journal = "JHEP",
    volume = "10",
    pages = "105",
    year = "2024"
}

@article{Afkhami-Jeddi:2025wra,
    author = "Afkhami-Jeddi, Nima and Caron-Huot, Simon and Chakravarty, Joydeep and Maloney, Alexander",
    title = "{Imprint of the black hole singularity on thermal two-point functions}",
    eprint = "2510.21673",
    archivePrefix = "arXiv",
    primaryClass = "hep-th",
    doi = "10.1103/1j3s-fzjl",
    journal = "Phys. Rev. D",
    volume = "114",
    number = "2",
    pages = "025005",
    year = "2026"
}

@article{Grozdanov:2026cut,
    author = "Grozdanov, Sa{\v{s}}o and Valach, Samuel and Vrbica, Mile",
    title = "{Bouncing geodesics, black hole singularities, and singularities of thermal correlators}",
    eprint = "2603.15598",
    archivePrefix = "arXiv",
    primaryClass = "hep-th",
    doi = "10.1007/JHEP08(2026)022",
    journal = "JHEP",
    volume = "08",
    pages = "022",
    year = "2026"
}

@article{Nunez:2003eq,
    author = "Nunez, Alvaro and Starinets, Andrei O.",
    title = "{AdS/CFT Correspondence, Quasinormal Modes, and Thermal Correlators in N=4 SYM}",
    eprint = "hep-th/0302026",
    archivePrefix = "arXiv",
    doi = "10.1103/PhysRevD.67.124013",
    journal = "Phys. Rev. D",
    volume = "67",
    pages = "124013",
    year = "2003"
}

@article{Kovtun:2005ev,
    author = "Kovtun, Pavel K. and Starinets, Andrei O.",
    title = "{Quasinormal Modes and Holography}",
    eprint = "hep-th/0506184",
    archivePrefix = "arXiv",
    reportNumber = "NSF-KITP-05-41",
    doi = "10.1103/PhysRevD.72.086009",
    journal = "Phys. Rev. D",
    volume = "72",
    pages = "086009",
    year = "2005"
}

@article{Dodelson:2023vrw,
    author = "Dodelson, Matthew and Iossa, Cristoforo and Karlsson, Robin and Zhiboedov, Alexander",
    title = "{A Thermal Product Formula}",
    eprint = "2304.12339",
    archivePrefix = "arXiv",
    primaryClass = "hep-th",
    reportNumber = "CERN-TH-2023-062",
    doi = "10.1007/JHEP01(2024)036",
    journal = "JHEP",
    volume = "01",
    pages = "036",
    year = "2024"
}

@article{Maldacena:2001kr,
    author = "Maldacena, Juan Martin",
    title = "{Eternal black holes in anti-de Sitter}",
    eprint = "hep-th/0106112",
    archivePrefix = "arXiv",
    doi = "10.1088/1126-6708/2003/04/021",
    journal = "JHEP",
    volume = "04",
    pages = "021",
    year = "2003"
}

@article{Herzog:2002pc,
    author = "Herzog, C. P. and Son, D. T.",
    title = "{Schwinger-Keldysh propagators from AdS/CFT correspondence}",
    eprint = "hep-th/0212072",
    archivePrefix = "arXiv",
    reportNumber = "NSF-ITP-02-175, INT-PUB-02-53",
    doi = "10.1088/1126-6708/2003/03/046",
    journal = "JHEP",
    volume = "03",
    pages = "046",
    year = "2003"
}

@article{Skenderis:2008dg,
    author = "Skenderis, Kostas and van Rees, Balt C.",
    title = "{Real-time gauge/gravity duality}",
    eprint = "0805.0150",
    archivePrefix = "arXiv",
    primaryClass = "hep-th",
    doi = "10.1103/PhysRevLett.101.081601",
    journal = "Phys. Rev. Lett.",
    volume = "101",
    pages = "081601",
    year = "2008"
}

@article{Skenderis:2008dh,
    author = "Skenderis, Kostas and van Rees, Balt C.",
    title = "{Real-time gauge/gravity duality: Prescription, Renormalization and Examples}",
    eprint = "0812.2909",
    archivePrefix = "arXiv",
    primaryClass = "hep-th",
    reportNumber = "ITFA-2008-50",
    doi = "10.1088/1126-6708/2009/05/085",
    journal = "JHEP",
    volume = "05",
    pages = "085",
    year = "2009"
}

@article{Fidkowski:2003nf,
    author = "Fidkowski, Lukasz and Hubeny, Veronika and Kleban, Matthew and Shenker, Stephen",
    title = "{The Black hole singularity in AdS/CFT}",
    eprint = "hep-th/0306170",
    archivePrefix = "arXiv",
    doi = "10.1088/1126-6708/2004/02/014",
    journal = "JHEP",
    volume = "02",
    pages = "014",
    year = "2004"
}

@article{Birmingham:2001pj,
    author = "Birmingham, Danny and Sachs, Ivo and Solodukhin, Sergey N.",
    title = "{Conformal field theory interpretation of black hole quasi-normal modes}",
    eprint = "hep-th/0112055",
    archivePrefix = "arXiv",
    doi = "10.1103/PhysRevLett.88.151301",
    journal = "Phys. Rev. Lett.",
    volume = "88",
    pages = "151301",
    year = "2002"
}

@article{Zamolodchikov:1995aa,
    author = "Zamolodchikov, Alexander B. and Zamolodchikov, Alexei B.",
    title = "{Structure constants and conformal bootstrap in Liouville field theory}",
    eprint = "hep-th/9506136",
    archivePrefix = "arXiv",
    doi = "10.1016/0550-3213(96)00351-3",
    journal = "Nucl. Phys. B",
    volume = "477",
    pages = "577--605",
    year = "1996"
}

@article{Teschner:1995yf,
    author = "Teschner, Jorg",
    title = "{On the Liouville three point function}",
    eprint = "hep-th/9507109",
    archivePrefix = "arXiv",
    doi = "10.1016/0370-2693(95)01200-A",
    journal = "Phys. Lett. B",
    volume = "363",
    pages = "65--70",
    year = "1995"
}

@article{Harlow:2011ny,
    author = "Harlow, Daniel and Maltz, Jonathan and Witten, Edward",
    title = "{Analytic Continuation of Liouville Theory}",
    eprint = "1108.4417",
    archivePrefix = "arXiv",
    primaryClass = "hep-th",
    doi = "10.1007/JHEP12(2011)071",
    journal = "JHEP",
    volume = "12",
    pages = "071",
    year = "2011"
}

@article{Fitzpatrick:2015zha,
    author = "Fitzpatrick, A. Liam and Kaplan, Jared and Walters, Matthew T.",
    title = "{Virasoro Conformal Blocks and Thermality from Classical Background Fields}",
    eprint = "1501.05315",
    archivePrefix = "arXiv",
    primaryClass = "hep-th",
    doi = "10.1007/JHEP11(2015)200",
    journal = "JHEP",
    volume = "11",
    pages = "200",
    year = "2015"
}

@article{Dodelson:2025jff,
    author = "Dodelson, Matthew and Iossa, Cristoforo and Karlsson, Robin",
    title = "{Bouncing off a stringy singularity}",
    eprint = "2511.09616",
    archivePrefix = "arXiv",
    primaryClass = "hep-th",
    doi = "10.1007/JHEP07(2026)161",
    journal = "JHEP",
    volume = "07",
    pages = "161",
    year = "2026"
}

@article{Turiaci:2016cvo,
    author = "Turiaci, Gustavo J. and Verlinde, Herman",
    title = "{On CFT and Quantum Chaos}",
    eprint = "1603.03020",
    archivePrefix = "arXiv",
    primaryClass = "hep-th",
    doi = "10.1007/JHEP12(2016)110",
    journal = "JHEP",
    volume = "12",
    pages = "110",
    year = "2016"
}

@article{Festuccia:2008zx,
    author = "Festuccia, Guido and Liu, Hong",
    title = "{A Bohr-Sommerfeld quantization formula for quasinormal frequencies of AdS black holes}",
    eprint = "0811.1033",
    archivePrefix = "arXiv",
    primaryClass = "gr-qc",
    reportNumber = "MIT-CTP-3995, SCIPP-08-11",
    doi = "10.1166/asl.2009.1029",
    journal = "Adv. Sci. Lett.",
    volume = "2",
    pages = "221--235",
    year = "2009"
}

@article{AliAhmad:2026wem,
    author = "Ali Ahmad, Shadi and Almheiri, Ahmed and Lin, Simon",
    title = "{Continuing past the inner horizon using WKB}",
    eprint = "2601.02354",
    archivePrefix = "arXiv",
    primaryClass = "hep-th",
    month = "1",
    year = "2026"
}

@article{nextpaper2026,
    author = "Liu, Ming-Xuan. and Ruan, Shan-Ming",
    title = "{From Picard–Lefschetz theory to Holographic Correlators and Singularity}",
    archivePrefix = "arXiv",
    primaryClass = "hep-th",
    journal = "in preparation",
    year = "2026"
}
%\bibliographystyle{utphys}
%%%%%%%%%%%%%%%%%%%%%%%%%%%%%%%%%%%%%%%%%%%%%%%%%%%%%%%%%%%%%%%%%%%%%%
%%%%%%%%%%%%%%%%%%%%%%%%%%%%%%%%%%%%%%%%%%%%%%%%%%%%%%%%%%%%%%%%%%%%%%

\end{document}